\documentclass[11pt]{article}
\usepackage{jheppub}
\pdfoutput=1
\usepackage{epstopdf}

\usepackage{mathrsfs}
\usepackage{psfrag}
\usepackage{color}

\usepackage{subfig}
\usepackage{slashed}

\usepackage{todonotes}

\usepackage{amsmath}
\usepackage{amsfonts}
\usepackage{graphicx}
\usepackage{amssymb}
\usepackage{xcolor}

\usepackage{mathtools}

\usepackage{kbordermatrix}

\newcommand{\eq}{\begin{equation}}
\newcommand{\eqe}{\end{equation}}
\newcommand{\eqa}{\begin{eqnarray}}
\newcommand{\eqae}{\end{eqnarray}}

\title{On the Positive Geometry of Conformal Field Theory}

\author{Nima Arkani-Hamed,$^1$}
\author{Yu-tin Huang,$^{2,3}$}
\author{Shu-Heng Shao$^1$}

\affiliation{$^1$ School of Natural Sciences, Institute for Advanced Study, Princeton, NJ 08540, USA}

\affiliation{$^2$ Department of Physics and Astronomy, National Taiwan University, Taipei 10617, Taiwan}
\affiliation{$^3$ Physics Division, National Center for Theoretical Sciences, National Tsing-Hua University,
No.101, Section 2, Kuang-Fu Road, Hsinchu, Taiwan}

\abstract{It has long been clear that the conformal bootstrap is associated with a rich geometry. In this paper we undertake a systematic exploration of this geometric structure as an object of study in its own right. We study conformal blocks for the minimal $SL(2,R)$  symmetry present in conformal field theories in all dimensions. Unitarity demands that the Taylor coefficients of the four-point function lie inside a polytope ${\bf U}$ determined by the operator spectrum,  while crossing demands they lie on a  plane ${\bf X}$.  The conformal bootstrap is then geometrically interpreted as demanding a non-empty intersection of ${\bf U} \cap {\bf X}$.
We find that the conformal blocks enjoy a surprising  positive determinant property. This implies that $\bf U$ is an example of a famous polytope -- the cyclic polytope. The face structure of cyclic polytopes is completely understood. This lets us fully characterize the intersection ${\bf U} \cap {\bf X}$ by a simple combinatorial rule, leading to a number of new exact statements about the spectrum and four-point function in any conformal field theory.}

\begin{document}

\begin{flushright}
\vspace{10pt} \hfill{NCTS-TH/1816} \vspace{20mm}
\end{flushright}
\maketitle

\section{Introduction}
Conformal field theories are characterized by a set of operators $O_i$
with dimensions $\Delta_i$ and spins $s_i$, together with the three-point functions coefficients
given by  $c_{ijk}$. Polyakov's dream \cite{Polyakov:1974gs} of the conformal bootstrap
program was to determine the space of allowed $\left\{ \Delta_i, s_i,
c_{ijk} \right\}$ from locality and unitarity, via the consistency of
the operator product expansion (OPE)  for four-point correlators in two different
channels. This puts a quadratic constraint on the $c_{ijk}$ depending
on the putative spectrum for $\Delta_i, s_i$, somewhat analogous to
the Jacobi identity for the structure constants of a Lie group. Indeed
Polyakov's vision was that the conformal bootstrap would lead to a
classification of possible conformal field theories (CFT) mirroring the classification of Lie
groups.

Unlike the older S-matrix bootstrap program--where the constraints from locality and unitarity on the analytic structure of four-particle scattering amplitudes were not well-understood (and are indeed still not well-understood to this day)--the conformal bootstrap poses a completely well-defined
mathematical problem. Given some  putative CFT data $\left \{\Delta_i,
s_i, c_{ijk} \right \}$, it is possible to unambiguously check whether
it does or does not define a unitary CFT. The fundamental challenge of
determining the space of consistent CFTs in this way lies in the
infinite nature of the problem. The four-point functions depends
continuously on conformal cross-ratios, and we must contend with the
infinite number of operators and the continuous spectrum of
dimensions.  For instance, if we are only given operators with
dimensions $\Delta_i < 100$, say, how can we check whether just this
spectrum of low-lying dimensions is consistent? How can we make
definite statements about them that are completely robust and
independent of the details associated with all the even higher-dimensional operators?

To make progress, it is necessary to render this infinite problem
finite in some reliable way. For instance, instead of dealing with the
full four-point function, we can take a finite approximation to it by
considering a truncation of its Taylor expansion to $n$ terms, around
some convenient kinematic point. Schematically this gives us an
$n$-dimensional vector ${\bf F}$, and the conformal bootstrap turns
into a well-defined geometry problem in this $n$-dimensional space.
The modern revival of the conformal bootstrap \cite{Rattazzi:2008pe} began by numerically
studying this geometry problem using techniques from linear
programming, and both this numerical approach as well as various
analytic avatars of this program had some spectacular successes  over
the past decade.   See \cite{Rychkov:2016iqz,Simmons-Duffin:2016gjk,Poland:2016chs,Poland:2018epd} for reviews on this subject.

Stimulated by these developments, in this paper we initiate a
systematic study of the geometry associated with the conformal
bootstrap. Working with some fixed number $n$ of Taylor coefficients
is analogous to using a microscope with fixed resolution to look at
both the four-point function and the space of CFTs. We would then
like to ask: how is the space of consistent operator dimensions
$\left \{ \Delta_i , s_i\right\}$ ``carved out" by the bootstrap at this
resolution?  Given some set of consistent $\left \{ \Delta_i ,s_i\right\}$
at this order, what constraints do we find on the Taylor coefficients
${\bf F}$ of the four-point function? And finally, clearly as we
increase the resolution by increasing $n$, the allowed space of
$\left\{ \Delta_i ,s_i\right\}$ is further refined. How can we
systematically see this refinement as we increase $n$ one step at a
time?

We begin by giving a trivial  interpretation of the conformal bootstrap
equations using the language of polytopes in projective space. Given a
putative spectrum $\left\{\Delta_i, s_i \right\}$, unitarity implies
that $\bf{F}$  must be in the interior of the convex hull
${\bf U}(\Delta_i,s_i)$ of a collection of points associated with the
Taylor coefficient of conformal blocks ${\bf G}(\Delta_i,s_i)$, while
crossing symmetry implies that ${\bf F}$ must lie on a fixed
hyperplane ${\bf X}$ depending only on the external operator
dimensions.  The bootstrap then demands that the crossing hyperplane
${\bf X}$ must intersect the unitarity polytope $\bf{U}$. If this
intersection is non-trivial, the Taylor coefficients ${\bf F}$ must
lie inside it, i.e.\ ${\bf F} \subset {\bf U} \cap {\bf X}$.

In this language, the non-trivial aspect of the problem is to
determine what the polytope ${\bf U}$ (and ultimately the intersection
${\bf U} \cap {\bf X}$) ``looks like". This is a standard challenge in
the study of polytopes. A polytope can be defined as the convex hull
of a collection of points, and this is precisely how it arises in our
example. But this definition does not allow us to easily check whether
some given point  ${\bf F}$ is or isn't inside the polytope. For this,
we need a dual definition, where the polytope is cut out by giving a
collection of inequalities ${\bf W}_{a} \cdot {\bf F} \geq 0$, where $a$ is the index that runs over the boundaries of the polytope.

In general, given a random collection of vertices, the only systematic
method for determining the facets of their convex hull is by
exhaustion: simply trying all possible $(n-1)$-tuples of vertices as
candidates for a face, and seeing whether the remaining vertices all
lie on the same side of this plane. Thus if the vertices ${\bf V}_i$
are given as $n$ dimensional vectors, the facet
structure is entirely determined by specifying, for all $n$-tuples
$\left \{i_1, \cdots, i_n \right \}$, whether the determinants
$\langle {\bf V}_{i_1} \cdots {\bf V}_{i_n} \rangle$ are positive,
negative, or zero. This data is known as the \textit{oriented matroid}
associated with the configuration of vectors ${\bf V}_i$. Of course,
for a random collection of ${\bf V}_i$, this method is hopelessly
impractical, and linear programming provides a more effective way of
determining the face structure numerically.

From a mathematical point of view, however, it is often possible to do
much better than this, and various infinite classes of polytopes have
been studied and ``understood" by mathematicians over the past
century. This requires some special structure for the vertices, which
allows  the faces of all co-dimensions to be
fully understood analytically. 
From the point of view of physics,
we should not expect geometric problems handed to us by quantum field
theory to be random and unstructured, so it is natural to ask whether
there is something special about the polytopes associated with the
conformal bootstrap problem, reflecting some special structure of the
conformal block vectors ${\bf G}(\Delta_i,s_i)$, which allows the face structure of
these polytopes to be understood analytically.

One famous class of polytopes that have been ``understood" directly
generalize convex polygons in two dimensions, and are known as
\textit{cyclic polytopes}. Here the vertices have a natural ordering ${\bf
V}_1, {\bf V}_2, \cdots$, and the special property they enjoy is that
all the ordered determinants are positive, $\langle {\bf V}_{i_1} \cdots
{\bf V}_{i_n} \rangle > 0$ for $i_1<i_2<..<i_n$. The convex hull of
these vertices is known as a cyclic polytope, and the positivity
allows a beautiful characterization of their full face structure.
This notion of positivity (and the closely related ideas of the \textit{positive Grassmannian}) has played a prominent role in the development of the \textit{Amplituhedron} \cite{Arkani-Hamed:2013jha} approach to the reformulating the physics of scattering amplitudes starting from primitive combinatoric-geometric ideas of \textit{positive geometry}.
We will see that precisely this same ``positive" structure makes a
striking appearance in the geometry of the conformal bootstrap.

We will work in the simplest possible setting, by exploring the
geometry of the conformal bootstrap with the minimal degree of
conformal symmetry possible, associated with the $SL(2,\mathbb{R})$ subgroup
corresponding to conformal transformations in $d=1$ dimension. The
$SL(2,\mathbb{R})$ conformal blocks are labeled only by the dimension $\Delta$,
and so the conformal block vectors ${\bf G}_\Delta$ also only depend
on $\Delta$. Note that the $\Delta$'s have a natural ordering from
small to large dimensions. Remarkably, we find, with a small but
fascinating caveat, that the conformal block vectors enjoy the
positivity property that $\langle {\bf G}_{\Delta_1} \cdots {\bf
G}_{\Delta_n} \rangle > 0$, and thus the unitarity polytope ${\bf U}$
is a cyclic polytope!

We will explore the conformal bootstrap from this geometric point of
view. As an illustration of the main ideas, we will fully solve the
geometry problem leading to the determination of the intersection
${\bf U} \cap {\bf X}$, when keeping up to six terms in the Taylor
expansion of the four-point function, corresponding to a
two-dimensional geometry for ${\bf U} \cap {\bf X}$. This will let us
transparently see how the spectrum of consistent operator dimensions
is ``carved out" by unitarity and crossing symmetry, and will lead to
a number of new, exact results for the spectrum and four-point
functions of any CFT.
In particular, the intersection  ${\bf U}\cap {\bf X}$, which bounds the value of the four-point function $\bf F$,  obeys  an interesting combinatoric rule, and its shape undergoes various intricate ``phase transitions" as the spectrum is continuously varied.
 We will also sketch the nature of the recursive
method that allows us to ``increase resolution" as we keep more terms
in the Taylor expansion and consider higher-dimensional geometries.
Looking ahead, we make some preliminary comments on the more general
geometries appropriate to higher-dimensional conformal symmetries.

The positive geometry associated with the conformal bootstrap is
relevant to the physics of conformal field theory, while the
positivity seen in the conformal blocks, and the problem of
determining the intersection of cyclic polytopes with the crossing
plane, is mathematically interesting in its own right. We have thus
endeavored to present our results in a self-contained way, not
assuming any prior knowledge of either conformal field theory nor
positive geometry, keeping both physicists and mathematician readers
in mind.  All the necessary background on polytopes and positivity is
included in the main text. In Appendix \ref{app:block}, we also give an introduction
to conformal blocks, starting with a Lorentzian integral
representation going back to Polyakov's pioneering paper \cite{Polyakov:1974gs,Czech:2016xec,Kravchuk:2018htv}. The
necessary integrals are trivial to carry out in this representation,
leading to a conceptually transparent and rapid computation for
conformal blocks in even dimensions.

\section{Conformal Bootstrap in One Dimension}\label{sec:bootstrap}

In this section we set up the physical problem -- the conformal bootstrap of  four-point functions in one dimension \cite{Gaiotto:2013nva,Paulos:2016fap,Maldacena:2016hyu,Hogervorst:2017sfd,Rychkov:2017tpc,Qiao:2017xif,Mazac:2018mdx,Mazac:2018ycv,Mazac:2018qmi,Liendo:2018ukf}.  Unitary  $1d$ CFTs arise in various physical contexts.
 For example, the theory on a conformal line defect in a higher dimensional CFT can be described by a 1$d$ CFT \cite{Gaiotto:2013nva} (see also \cite{Billo:2016cpy}).    They can also arise as the boundary theory of a QFT in a classical $AdS_2$ background \cite{Paulos:2016fap}.
 Moreover, since the 1$d$ conformal group $SL(2,\mathbb{R})$ is a subgroup of the higher-dimensional conformal group, every four-point function in higher-dimensional CFT can be thought of as a 1$d$ conformal four-point function and can be decomposed into the $SL(2,\mathbb{R})$ conformal blocks.
 See Appendix A of \cite{Qiao:2017xif} for a comprehensive overview of 1$d$ CFTs.

In one dimension, the conformal symmetry is $SL(2,\mathbb{R})$, which acts on the 1$d$ spacetime coordinate $x\in \mathbb{R}$ as
\begin{align}
x\to x' (x) = {ax+b\over cx+d}\,,
\end{align}
where $a,b,c,d\in \mathbb{R}$ with $ad-bc=1$. A conformal primary operator $\phi(x)$ with scaling dimension $\Delta_\phi$  transforms as $\phi(x) \to \phi'(x)$ under $SL(2,\mathbb{R})$,  where
\begin{align}
\left| {\partial x' \over \partial x }\right|^{\Delta_\phi}   \phi'(x')  =  \phi(x)\,.
\end{align}
The physical observables in a unitary $1d$ CFT includes the correlation functions of conformal primary operators.  The correlation functions transform covariantly under $SL(2,\mathbb{R})$, obey reflection positivity, and factorize via the operator product expansion (OPE).

Let us consider the four-point function $\langle \phi(x_1)\phi(x_2)\phi(x_3)\phi(x_4)\rangle$ of identical, real conformal primary operators $\phi$ with scaling dimension $\Delta_\phi$.  We will assume $x_1<x_2<x_3<x_4$ and define the $SL(2,\mathbb{R})$ invariant cross ratio $z$ as
\begin{align}
z  = {x_{12} x_{34} \over x_{13}x_{24}} \in (0,1)\,,
\end{align}
where $x_{ij}=x_i - x_j$.  We note that $1-z= {x_{14} x_{23}\over x_{13}x_{24}} $.
The  $SL(2,\mathbb{R})$ covariance of the four-point function implies that, up to an overall factor, it can be written as a function of the cross ratio
\begin{align}\label{pre4pt}
\langle \phi(x_1)\phi(x_2)\phi(x_3)\phi(x_4)\rangle  = {1\over |x_{12}|^{2\Delta_\phi} |x_{34}|^{2\Delta_\phi} }
F(z)\,.
\end{align}
Using the OPE between $\phi(x_1)$ and $\phi(x_2)$,  the four-point function $F(z)$ can be written as a sum over all the intermediate conformal primary operators ${\cal O}_i$ in the $\phi\times \phi$ OPE channel
\begin{align}\label{unitarity}
\text{Unitarity}:~~F(z)  = \sum_{i }  p_{i}   G_{\Delta_{i}}(z)\,,~~~~p_i>0\,,
\end{align}
where  the $SL(2,\mathbb{R})$ conformal block $G_\Delta(z)$  is \cite{Dolan:2011dv}
\begin{align}\label{block}
G_\Delta(z) = z^\Delta\,_2F_1(\Delta,\Delta ,2\Delta,z)\,.
\end{align}
Here $p_{i}$ is the three-point function coefficient square between $\phi,\phi,{\cal O}_i$.
Unitarity implies that the $p_{i}$'s are positive.
Hence in a unitary $1d$ CFT, the four-point function $F(z)$ can be expanded on the conformal blocks with positive coefficients.  In Appendix \ref{app:block} we review the derivation of the conformal blocks in general dimensions.

Alternatively,  we can use the OPE between $\phi(x_2)$ and $\phi(x_3)$ to expand the four-point function and express the four-point function $\langle \phi(x_1)\phi(x_2)\phi(x_3)\phi(x_4)\rangle$ as in \eqref{pre4pt} and \eqref{unitarity}, but only with $1\leftrightarrow 3$ exchange.  Since the OPE channels are identical in the two expansions, the coefficients $p_i$'s are the same as well.  Comparing the two expansions, we conclude that
\begin{align}\label{crossing}
\text{Crossing}:~~F(z)  =  \left( {z\over 1-z}\right)^{2\Delta_\phi } F(1-z)\,.
\end{align}
This is known as the crossing symmetry of the four-point function.  The prefactor $\left( {z\over 1-z}\right)^{2\Delta_\phi } $ originates from the ratios of prefactors in \eqref{pre4pt} in the two OPE channels.

To summarize, a four-point function $F(z)$ in a unitary 1$d$ CFT satisfies the unitarity condition \eqref{unitarity} and the crossing equation \eqref{crossing}.  The central goal in the conformal bootstrap program is to study the solution space of $\Delta_i$ and $p_{i}$.

The simplest analytic examples of the unitary $1d$ four-point function are the generalized free boson and fermion.  The generalized free field theory contains a fundamental bosonic/fermionic conformal primary operator $\phi(x)$, together with a tower of double-trace operators that are quadratic in $\phi$.  The theory is free in the sense that all correlation functions in the generalized free field theory are computed by Wick contractions.

 The four-point function of $\phi$ in the generalized free boson ($+$) and fermion ($-$) theory is simply given by the sum of three different Wick contractions (with signs):
 \begin{align}
&\langle \phi(x_1)\phi(x_2) \phi(x_3)\phi(x_4) \rangle\notag\\
&= \langle  \phi(x_1)\phi(x_2) \rangle \langle\phi(x_3)\phi(x_4) \rangle
\pm  \langle  \phi(x_1)\phi(x_3) \rangle \langle\phi(x_2)\phi(x_4) \rangle
+ \langle  \phi(x_1)\phi(x_4) \rangle \langle\phi(x_2)\phi(x_3) \rangle\notag\\
&=  {1\over (x_{12}^2x_{34}^2)^{\Delta_\phi} }
\pm {1\over (x_{13}^2x_{24}^2)^{\Delta_\phi} }
+ {1\over (x_{14}^2x_{23}^2)^{\Delta_\phi} } \,,
\end{align}
which gives
\begin{align}
F^\pm(z) =
1 \pm (z\bar z)^{\Delta_\phi}  +  \left(  {z\bar z\over (1-z)(1-\bar z)}\right)^{\Delta_\phi}
\,.
\end{align}
$F^\pm(z)$ manifestly satisfies the crossing symmetry \eqref{crossing}.

The conformal block decomposition of the generalized free boson four-point function is
\begin{align}\label{4pt}
F^+(z) =  1  + \sum_{n=0}^\infty
p^+_{n} G_{2\Delta_\phi+2n }(z) \,,
\end{align}
with positive coefficients (see, for example, \cite{Gaiotto:2013nva})
\begin{align}
p^+_n  =  {(2\Delta_\phi) _n \over 2^{2n-1} (2n)!(2\Delta_\phi+n -1/2)_n }\,,
\end{align}
where $(a)_b = \Gamma(a+b)/\Gamma(a)$.
The $\phi\times \phi$ OPE channel contains the identity plus an infinite tower of double-trace operators of the form $\phi \partial^{2n} \phi +\cdots$ where the $\cdots$ represents other distributions of the derivative such that the operator is  a conformal primary.
The scaling dimensions of these intermediate operators are
\begin{align}
\Delta = 2\Delta_\phi + 2n  \,,~~~~n \in \mathbb{Z}_{\ge0}\,.
\end{align}

On the other hand, the block decomposition of the four-point function in the generalized free fermion is
\begin{align}\label{4pt}
F^-(z) =  1  + \sum_{n=0}^\infty
p^-_{n} G_{2\Delta_\phi+2n+1 }(z)\,,
\end{align}
where
\begin{align}
p^-_n   = { (2\Delta_\phi)_n (2\Delta_\phi)_{2n+1}\over
4^n (2n+1)!(2\Delta_\phi+n+1/2)_n}\,.
\end{align}
The intermediate operators in the $\phi\times \phi$ OPE are schematically of the form $\phi \partial^{2n+1} \phi +\cdots$ with scaling dimension
\begin{align}
\Delta = 2\Delta_\phi +2n +1\,,~~~~~n\in \mathbb{Z}_{\ge0}\,.
\end{align}
  Note that in the fermionic theory, $\phi\phi$ vanishes due to Fermi statistics, so the lowest non-identity  dimension operator is $\phi \partial \phi$ with $\Delta = 2\Delta_\phi+1$.  In fact, given the external operator dimension $\Delta_\phi$, $\Delta_{gap}=  2\Delta_\phi+1$ is the maximum possible gap in a unitary $1d$ CFT four-point function \cite{Gaiotto:2013nva,Mazac:2016qev,Mazac:2018mdx}.

Finding a solution to the conformal bootstrap equations \eqref{unitarity} and \eqref{crossing} is quite challenging in general.  To tame this infinite-dimensional problem in the space of all four-point functions, let us discretize the problem to start with.  For example, we can discretize the four-point function $F(z)$ by its first $2N+1$ derivatives around $z=1/2$:
\begin{align}
F(z) \rightarrow {\bf F}=
 \left(\begin{array}{c}F^0 \\F^1 \\ \vdots \\F^{2N+1}\end{array}\right) \in \mathbb{P}^{2N+1} \,,
\end{align}
where
\begin{align}
F^I \equiv {1\over I!} \partial_z^I F(z)\Big|_{z=1/2}\,,~~~I=0,1,\cdots, 2N+1\,.
\end{align}
Let us now ask what are the constraints on the $(2N+2)$-dimensional vector $\bf F$ from unitarity \eqref{unitarity} and crossing \eqref{crossing}.  Given any solution $\bf F$ to this truncated problem, $\lambda {\bf F}$ is again a solution for any real $\lambda$. Hence it is convenient to think of $\bf F$ as a vector in $\mathbb{P}^{2N+1}$.

The constraint from crossing symmetry \eqref{crossing} is easy to state in the finite-dimensional problem.  Expanding \eqref{crossing} around $z=1/2$, we have
\begin{align}
F^0 + F^1 y +  F^2 y^2 +\cdots = \left( {1+2y \over 1-2y} \right)^{2\Delta_\phi}   \left[ F^0 -F^1 y +F^2 y^2+\cdots\right] \,,
\end{align}
where $y\equiv z-1/2$.  Matching the coefficients of powers of $y$, we obtain  $N+1$ linear relations (that depend on $\Delta_\phi$) among the $F^I$'s.  We can use these relations to solve $F^{odd}$ in terms of $F^{even}$.  For example, when $N=2$, we obtain the following 3 linear relations:
\begin{align}
\begin{split}
&F^1 = 4\Delta_\phi F^0\,,~~~~\\
&F^3=  {16\over 3} (\Delta_\phi-4\Delta_\phi^3) F^0 +4\Delta_\phi F^2\, \\
&F^5 =   {64\over 15} \Delta_\phi (32\Delta_\phi^4 -20 \Delta_\phi^2 +3 )F^0
- {16\over3} \Delta_\phi (4\Delta_\phi^2 -1) F^2
+4\Delta_\phi F^4 \,.
\end{split}
\end{align}
To say it more geometrically, the crossing equation restricts the four-point function $\bf F$ to lie on an $N$-dimensional plane, denoted as ${\bf X}[\Delta_\phi]$,  in $\mathbb{P}^{2N{+}1}$.  We will call  this plane the  \textit{crossing plane}.

Let us switch gear to the unitarity constraint \eqref{unitarity}.  We first Taylor expand the conformal block $G_\Delta(z)$ around $z=1/2$ to forma  a $(2N+2)$-dimensional \textit{block vector} ${\bf G}_\Delta$:
\begin{align}\label{prebv}
G_\Delta(z) \rightarrow  {\bf G}_\Delta=
 \left(\begin{array}{c}G_\Delta^0 \\ G_\Delta^1 \\ \vdots \\G_\Delta^{2N+1}\end{array}\right) \in \mathbb{P}^{2N+1}\,.
\end{align}
The unitarity constraint \eqref{unitarity} demands the four-point function vector $\bf F$ to lie in the positive span  of the block vectors ${\bf G}_\Delta$:
\begin{align}
{\bf F}  =  \sum_\Delta p_\Delta {\bf  G}_\Delta\,,
\end{align}
with $p_\Delta$ all positive.

\section{Polytopes and Positive Geometry}\label{sec:polytope}
\begin{figure}
\begin{center}
\includegraphics[scale=0.6]{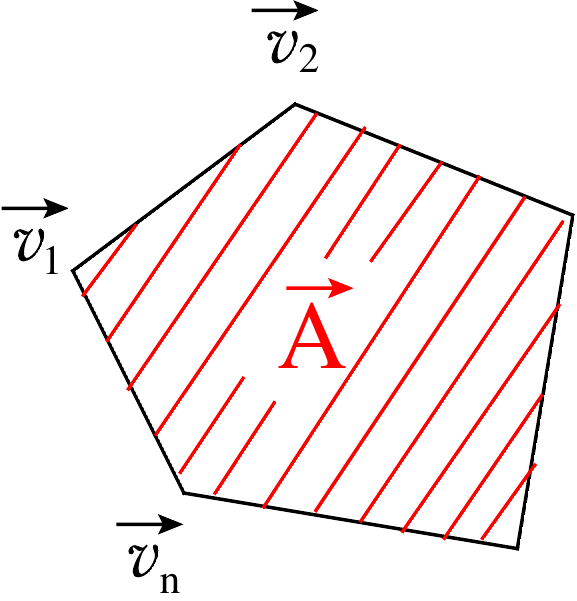}
\caption{A point $\vec A$ inside a polytope is given by a non-negative sum of the vertices $\vec v_i$.}
\label{PolyGon}
\end{center}
\end{figure}

Geometric problems of the character of equation (2.21), asking for a vector to be expressible as a {\it positive} linear combination of some fixed set of vectors, are ubiquitous in mathematics and physics. In this section we give a brief introduction to the elementary ideas of projective polytopes, which is a useful language for thinking about such problems. A nice reference for discussing polytopes from this perspective (usually couched in terms of ``cones") can be found in \cite{grunbaum2003convex}.

\begin{figure}
\centering
\includegraphics[width=.3\textwidth]{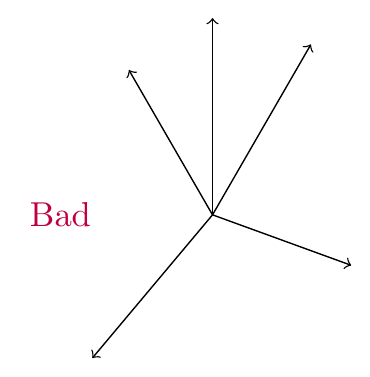}~~~~~~~~~~~~~~~
\includegraphics[width=.4\textwidth]{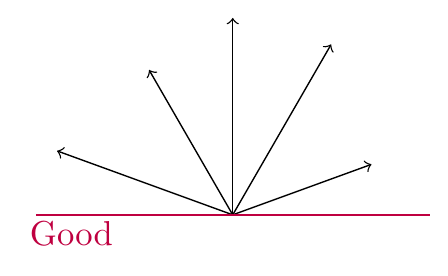}
\caption{The ``bad" and ``good" configurations.}\label{fig:badgood}
\end{figure}

To emphasize the generality of the discussion, let's switch notation, and ask to characterize the space of $(d+1)$ dimensional vectors ${\bf A}$ which can be expressed as a positive linear combination of a given set of vectors ${\bf V}_i$:
\begin{equation}
{\bf A} = \sum_i c_i {\bf V}_i \, ,  \, {\rm where} ~~c_i > 0
\end{equation}
We assume for simplicity that the number of vectors is greater than or equal to $(d+1)$, so that the space of all ${\bf A}$'s is top-dimensional.
Clearly, for this equation to put any interesting constraint on the space of allows ${\bf A}$'s, all the vectors ${\bf V}_i$ must lie on the same side of some hyperplane. If this is not the case, then {\it any} vector can be expressed as a linear combination of the ${\bf V}_i$ and there are no constraints on ${\bf A}$ at all. But if the ${\bf V}_i$ are all the same side of some hyperplane, ${\bf A}$ is constrained to lie inside a cone spanned by the ${\bf V}_i$. Examples of ``bad" and ``good" configurations of vectors in $(d+1)=2$ dimensions are shown in Figure \ref{fig:badgood}.

It is convenient to think of this cone projectively in the following way. Clearly, the space of all possible ${\bf A}$'s is invariant under ${\bf A} \to t {\bf A}$ for any (positive) rescaling $t>0$. Similarly, the cone spanned by vectors ${\bf V}_i$ is exactly the same as that spanned by the vectors ${\bf V}^\prime_i = t_i {\bf V}_i$ for all $t_i > 0$. Thus, the data associated with this problem is naturally projective: we can think of the equivalence class of all ${\bf A} \sim t {\bf A}$. All these vectors are on the same side of some hyperplane. Thus we can write
\eq\label{projective}
{\bf V}_i=t_i \left(\begin{array}{c}1 \\ \vec{V}_i\end{array}\right), \quad {\bf A}=t \left(\begin{array}{c}1 \\ \vec{A}\end{array}\right)\,.
\eqe
where $t_i>0,t>0$. Note that we have arranged the top component to be positive for all the vectors, reflecting the fact that they are on the same side of the hyperplane $(1,\vec{0})$, i.e.\ $(1,\vec{0}) \cdot {\bf A} = t > 0$. Because of the projective invariance, the data of the problem is given by $\vec{V}_i$ and we want to determine the space of allowed $\vec{A}$. Since
\begin{equation}
\sum_i c_i {\bf V}_i = \left(\begin{array}{c} \sum t_i c_i \\ \sum_i t_i c_i \vec{V}_i\end{array}\right) =
\left(\begin{array}{c} \sum_i c^\prime_i \\ \sum_i c^\prime_i \vec{V}_i\end{array}\right)\,
\end{equation}
where $c^\prime_i = t_i c_i$ are also positive, the space of $\vec{A}$ is given by
\begin{equation}
\vec{A} = \frac{\sum_i c^\prime_i \vec{V}_i}{\sum_i c^\prime_i}\,, \, {\rm with} ~~ c^\prime_i > 0
\end{equation}
This is a weighted sum of the vectors ${\vec V}_i$, familiar for instance from the notion of ``center of mass". This defines the {\it convex hull} of the vectors $\vec{V}_i$, which is a {\it polytope} in $\mathbb{R}^d$.

 More directly, we can think of the equation ${\bf A} = \sum_i c_i {\bf V}_i$ as a {\it projective polytope} in $\mathbb{P}^d$. This is a special case of the well-known general fact, that any geometric questions in $\mathbb{R}^d$ that do not involve a notion of a metric are more usefully posed in $\mathbb{P}^d$. The description in $\mathbb{R}^d$ manifests the affine symmetries of translations and  linear transformations, $T_d \times SL(d)$,  but these are just a subgroup of the larger group $SL(d+1)$ of symmetries that map linear spaces to linear spaces, which is made manifest in the projective space.

 Being explicit with the indices, we denote ${ A}^I$ with an upstairs index $I=0,\cdots, d$ to be a point in the projective space, with the symmetry under ${A}^I \to L^I_J { A}^J$. A hyperplane is denoted by ${ W}_I$ with a lower $SL(d+1)$ index. The $T_d \times SL(d)$ subgroup of $SL(d+1)$ is the one that keeps the hyperplane at infinity--in our notation above the co-vector $(1,\vec{0})$, fixed, while the full $SL(d+1)$ allows the most general transformations that also move infinity.
Since the projective symmetry is $SL(d+1)$, the only invariant tensor we have is the antisymmetric tensor. Any invariant must involve the antisymmetric contraction of $(d+1)$ vectors which we will denote by $\langle {\bf V}_1,{\bf V}_2, \cdots, {\bf V}_{d+1} \rangle$. This allows us to easily characterize the notions of ``incidence". For instance, if these $d+1$ vectors are coplanar, then this bracket must vanish.  Note that this statement is invariant under arbitrary (positive or not) rescaling of the vectors. It is also easy to describe linear subspaces of any dimensionality $p$: they are simply antisymmetric tensors with either $(p+1)$ upstairs indices or $(d+1) - (p+1)$ downstairs indices.

There are further advantages of the projective picture. One rather trivial one is that we don't need to awkwardly normalize the weights to add up to one, as we do when talking about the usual convex hull in $\mathbb{R}^d$. But the projective language is most useful for thinking about the boundary structure of the polytope.
Indeed so far we have defined a polytope by the convex hull of its vertices. It is also possible to define the polytope by a collection of hyperplanes $W_{a I}$, the facets, that cut out the polytope by the linear inequalities
\eq\label{Boundary}
{\bf A} \cdot {\bf W}_a = A^IW_{aI}>0 \,,\quad \forall \, a\,.
\eqe
where the index $a$ labels the facet.

This is useful, since given the convex hull definition, it is not easy to check whether a given point ${\bf A}$ is or isn't inside the polytope. The reason is that the expression ${\bf A} = \sum_i c_i {\bf V}_i$ is highly redundant--there are many more $c_i$ than the dimensionality of the space, and hence there is no unique representation of ${\bf A}$ in this form. All we ask that is that there is exist {\it some} representation for which all the $c_i$ are positive, but there are also (infinitely) many others for which some of the coefficients can be negative. By contrast the facet definition does allow us to practically check whether or not a given ${\bf A}$ is in the polytope--we simply check whether it satisfies all the inequalities defined by the boundary hyperplanes.

Given the convex hull definition of a polytope, how can we determine the facets? Let's see how the projective picture helps us do this in the simplest case of a 2$d$ polygon, where the answer is pictorially obvious: the facets are edges connecting consecutive vertices $({\bf V}_i, {\bf V}_{i+1})$, so $W_{i I} = \epsilon_{I J K} V_{i}^J V_{i+1}^K$.  To begin with, consider points $a,b,c$ in the two-dimensional plane, projectively associated with three-vectors ${\bf a}, {\bf b}, {\bf c}$. We can distinguish three situations, where $c$ is on one side on the line $(ab)$, on the line $(ab)$, and on the other side of $(ab)$, as having $\langle {\bf a}, {\bf b}, {\bf c} \rangle > 0, \langle {\bf a}, {\bf b}, {\bf c} \rangle = 0, \langle {\bf a}, {\bf b}, {\bf c} \rangle <0$ respectively. Note that these signs (and zeros) are invariant under the {\it positive} rescaling of all the vectors we allow for our projective polytopes.

\begin{figure}
\begin{center}
\includegraphics[scale=0.6]{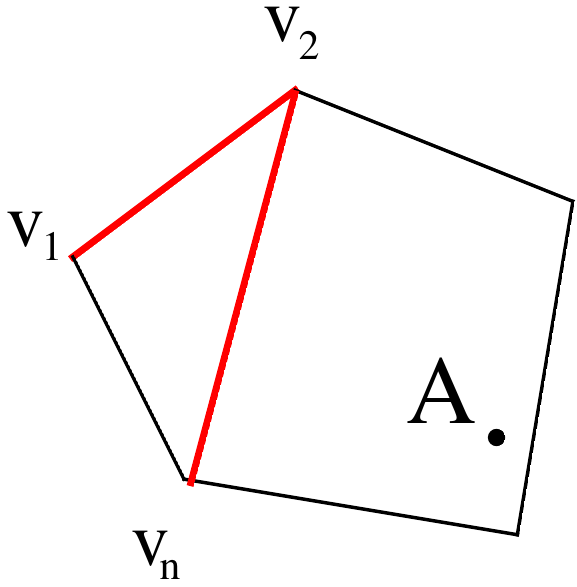}
\caption{A polygon. While the line segment $({\bf V}_n{\bf V}_2)$ is not an edge of the polygon, $({\bf V}_1{\bf V}_2)$ is.}
\label{BoundaryG}
\end{center}
\end{figure}

Let's now discuss how to understand the boundaries of a 2$d$ polygon, which are just edges specified by the vertices at their end points.
But not every pair of points gives rise to an edge.
As shown in Figure \ref{BoundaryG}, while the line segment $({\bf V}_1,{\bf V}_2)$ is  an edge of the polygon, $({\bf V}_2,{\bf V}_n)$ is not.
What is the condition on a pair of points such that they give rise to an edge?
Since the line segment $({\bf V}_2,{\bf V}_n)$ is in the interior of the polygon, there are points on its either side.  It follows that the determinant $\epsilon_{I_1I_2I_3}A^{I_1}V_{i_1}^{I_2}V_{i_2}^{I_3}$ can take either sign in the interior of the polygon, where $\epsilon$ is the three-dimensional Levi-Cevita tensor.  Thus,  for $({\bf V}_{i_1},{\bf V}_{i_2})$ to be an edge, we must have
\eq\label{Vab}
\langle {\bf A},{\bf  V}_{i_1},{\bf V}_{i_2}\rangle \equiv \epsilon_{I_1I_2I_3}A^{I_1}V_{i_1}^{I_2}V_{i_2}^{I_3}\;\;{\rm have\;the\;same\;sign} \,,
\eqe
for all $\bf A$ in the interior of the polytope.
 Since $\bf A$ can be any point in the convex hull of $\{{\bf V}_i\}$, it follows  that (\ref{Vab}) amounts to
\eq
 \langle {\bf  V}_i ,{\bf V}_{i_1},{\bf V}_{i_2}\rangle\;\;{\rm have\;the\;same\;sign} ~~~~~\forall \, i\,.
\eqe

This straightforwardly generalizes to higher dimensions. In $\mathbb{P}^d$, the facets for the convex hull of a set of vectors $\{{\bf V}_i\}$ are given by the set of $d$ vectors $({\bf V}_{i_1},\cdots,{\bf V}_{i_d})$ such that
\eq
 \langle {\bf V}_i ,{\bf V}_{i_1},{\bf V}_{i_2},\cdots, {\bf V}_{i_d}\rangle
 \equiv \epsilon_{I_0\cdots I_{d}}V_i^{I_0} V_ {i_1}^{I_1}\cdots V_{i_d}^{I_d}~
 \;\;{\rm have\;the\;same\;sign} ,~~~~~\forall \, i\,.
\eqe
This means that the points ${\bf V}_{i_1}, \cdots, {\bf V}_{i_d}$ lie on a facet of the polytope, as
\eq\label{DualVec}
W_{aI}=\epsilon_{II_1\cdots I_d}V^{I_1}_{i_1}\cdots V^{I_d}_{i_d} \,.
\eqe
Note that in general dimensions, a facet may well have more than $d$ vertices on it. Any $(d+1)$ vertices ${\bf V}_{i_1}, \cdots, {\bf V}_{i_{d_+1}}$ lying on the same facet will then satisfy $\langle {\bf V}_{i_1}, \cdots, {\bf V}_{i_{d+1}} \rangle = 0$, and any $d$ of them will define  exactly the same facet $W_{a I}$ projectively (i.e.\ up to rescaling).

We see that the boundary structure of a polytope, defined as the convex hull of $n$ points $\{{\bf V}_i\}_{i=1}^n$, is fully fixed by specifying the zeroes and signs of all ${n \choose d+1}$ brackets $\langle {\bf V}_{i_1}, \cdots, {\bf V}_{i_{d+1}} \rangle$. This is known as the {\it oriented matroid} of the configuration of vectors $\{ {\bf V}_i \}$. Thus, to determine the face structure of a polytope, one needs to compute the signs  of the determinants for all possible collections of $d{+}1$ vectors.
The computational complexity grows polynomially with $n$.

The projective language also lets us easily determine the intersection of planes of various dimensions; we simply write down the various planes as antisymmetric tensors and contract the indices with the epsilon tensor in the only way possible.  For instance,  a $k$-plane in $\mathbb{P}^{d}$ is defined by  $k{+}1$ vectors $\{ {\bf Z}_1,\cdots, {\bf Z}_k\}$.
In $d$ dimensions, in general a $k$-plane (defined by $k{+}1$ vectors) intersects with $p$-plane (defined by $p{+}1$ vectors) on a $(p{+}k{-}d)$-plane. For example, a 2-plane  $({\bf Z}_1,\,{\bf Z}_2,\,{\bf Z}_3)$ intersects a line  $({\bf Z}_a,\,{\bf Z}_b)$ in $\mathbb{P}^{3}$ at the point
\eq
{\bf Z}_a\langle {\bf Z}_b,{\bf Z}_1,{\bf Z}_2,{\bf Z}_3\rangle- {\bf Z}_b\langle {\bf Z}_a,{\bf Z}_1, {\bf Z}_2,{\bf Z}_3\rangle\,.
\eqe
For general $(p,k)$, the intersection is given as:
\eqa
&&Z^{[I_1}_{j_1}{\cdots} Z^{I_{d'}]}_{j_{d'}}
\langle {\bf Z}_{ j_{d'{+}1}},{\cdots}, {\bf Z}_{j_{p{+}1}},{\bf Z}_{i_1},{\cdots},{\bf Z}_{ i_{k{+}1}}\rangle
{+}({-})^{p}Z^{[I_1}_{j_2}\cdots Z^{I_{d'}]}_{j_{d'}{+}1}
\langle {\bf Z}_{j_{d'{+}2}},{\cdots},{\bf Z}_{j_1},{\bf Z}_{i_1},{\cdots},{\bf Z}_{ i_{k{+}1}}\rangle\nonumber\\
&&{+}({-})^{p+1}Z^{[I_1}_{j_3}{\cdots} Z^{I_{d'}]}_{j_{d'}{+}2}\langle {\bf Z}_{j_{d'{+}3}},{\cdots}, {\bf Z}_{j_2}, {\bf Z}_{i_1},{\cdots}, {\bf Z}_{i_{k{+}1}}\rangle{+}{\cdots}\,
\eqae
where $d'=p+1+k-d$. In this paper, we will be interested in the intersection of a $k$-plane $\bf X$, with a $d$-dimensional polytope with vertices $\{{\bf V}_i\}$. From the above, we see that it intersects with a $(d{-}k)$-face at a point, which is given as
\eq
{\bf V}_1\langle {\bf V}_2,{\bf V}_3,{\bf V}_4,\cdots,{\bf V}_{d{-}k},{\bf X}\rangle
-{\bf V}_2\langle  {\bf V}_1,{\bf V}_3,{\bf V}_4,\cdots,{\bf V}_{d{-}k},{\bf X}\rangle
+{\bf V}_3\langle {\bf  V}_1,{\bf  V}_2,{\bf  V}_4,\cdots,{\bf V}_{d{-}k},{\bf X}\rangle+\cdots\,.
\eqe
By definition, this point is in the interior of the polytope if all the coefficients of ${\bf V}_i$'s
\eq\label{Intersect}
\langle {\bf V}_2,{\bf V}_3,{\bf V}_4,\cdots,{\bf V}_{d{-}k},{\bf X}\rangle,\;
-\langle  {\bf V}_1,{\bf V}_3,{\bf V}_4,\cdots,{\bf V}_{d{-}k},{\bf X}\rangle,\;
\langle {\bf V}_1, {\bf V}_2, {\bf  V}_4,\cdots,{\bf V}_{d{-}k},{\bf X}\rangle,\quad {\rm e.t.c}\,,
\eqe
 have the same sign. Thus if the $k$-plane $\bf X$ intersects the polytope, we must be able to find  $d{-}k{+}1$ vertices such that the signs of the determinants with $\bf X$ satisfies the above pattern.

Finally, we will often reduce the dimension of the problem by projecting the geometry through fixed vectors or planes in the problem.
The former include the identity block vector ${\bf G}_0=(1,0,\cdots,0)$, the infinity vector ${\bf G}_\infty=(0,\cdots,0,1)$, and the latter involve the crossing-plane $\bf X$. Projecting through ${\bf G}_0$ means that we are considering
\eq
\langle 0,\cdots\rangle=\det\left(\begin{array}{cccc}1 & * & *& * \\0 & * & * & * \\\vdots & * & * & * \\0 & * & * & *\end{array}\right)\,,
\eqe
where $\langle i,\cdots\rangle$ will be a short hand notation for $\langle {\bf G}_{\Delta_i},\cdots\rangle$. We see that projection through ${\bf G}_0$  lops off the first component of each vector, and reducing the dimension by 1. Similarly for the projection through ${\bf G}_\infty$ we  simply lob off the last component, which  corresponds to the geometry of the previous dimension. For the projection through $\bf X$, we can understand as  implemented by performing some $GL$($d{+}1$) transformation such that $\bf X$ takes the form
\eq
k=0:\;\left(\begin{array}{c}1 \\0 \\ \vdots \\0\end{array}\right),\quad k=1,\quad \left(\begin{array}{cc}1 & 0   \\0 & 1   \\ \vdots & \vdots   \\0 & 0    \end{array}\right),\;{\rm e.t.c.}
\eqe
Then one considers the geometry defined through the determinant $\langle {\bf X},\cdots \rangle$.
The net effect is that the first $(k{+}1)$-components of all vectors are chopped off, leaving behind a $(d{-}k{+}1)$-dimensional space.

\section{Cyclic Polytopes}\label{sec:cyclic}
As mentioned in Section \ref{sec:polytope}, the complexity of computing the convex hull of a set of vectors grows polynomially with the number of vectors. For our purpose the number of vectors is associated with the number of conformal primaries, which is infinite. Thus a priori it would appear that the problem is intractable.

However as alluded to in the introduction, the science of studying polytopes is precisely about finding situations where the face structure can be fully determined analytically, which is possible when the data of the oriented matroid -- all the zeroes and signs of the $\langle{\bf V}_{i_1}, \cdots, {\bf V}_{i_{d+1}}\rangle$ -- have some nice properties. Indeed, in two dimensions, all polytopes are simply convex polygons. Note that in this case, the vertices have a natural ordering $\{{\bf V}_1,\cdots, {\bf V}_n \}$, and the oriented matroid is simply that all the $\langle {\bf V}_{i_1} {\bf V}_{i_2} {\bf V}_{i_3} \rangle > 0$ for $i_1<i_2<i_3$. Observe that since the brackets are antisymmetric, the ordering on vertices was crucial to make this statement. When this is satisfied, we immediately conclude that the facets are $({\bf V}_i {\bf V}_{i+1})_I$.

\begin{figure}
\centering
\includegraphics[width=.25\textwidth]{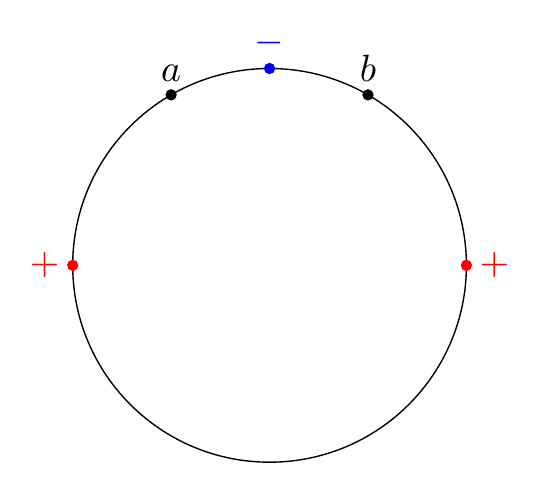}~~~~
\includegraphics[width=.25\textwidth]{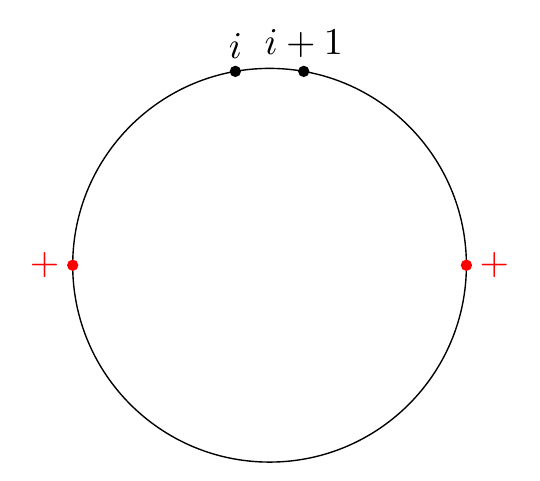}\\
\includegraphics[width=.25\textwidth]{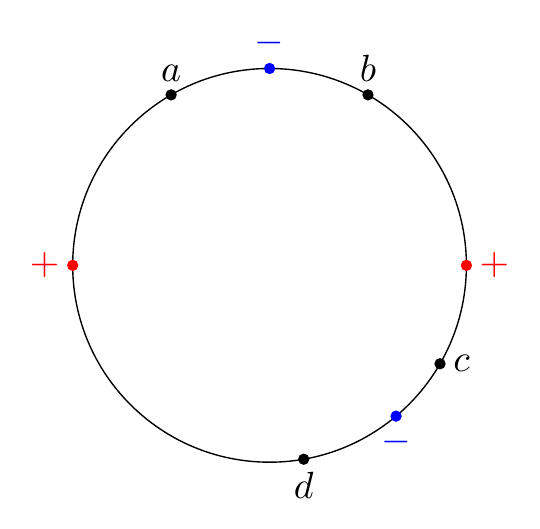}~~~~
\includegraphics[width=.25\textwidth]{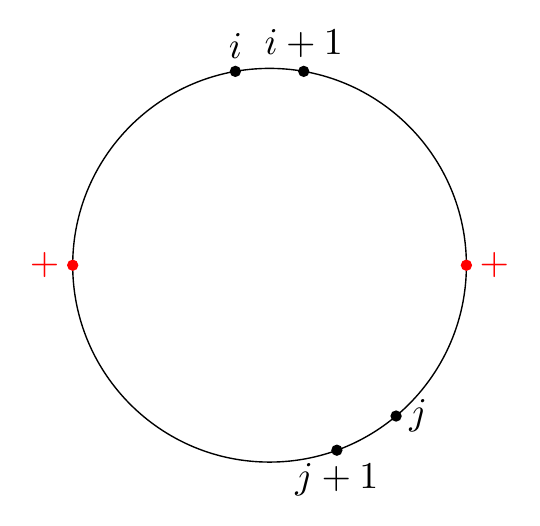}\\
\includegraphics[width=.25\textwidth]{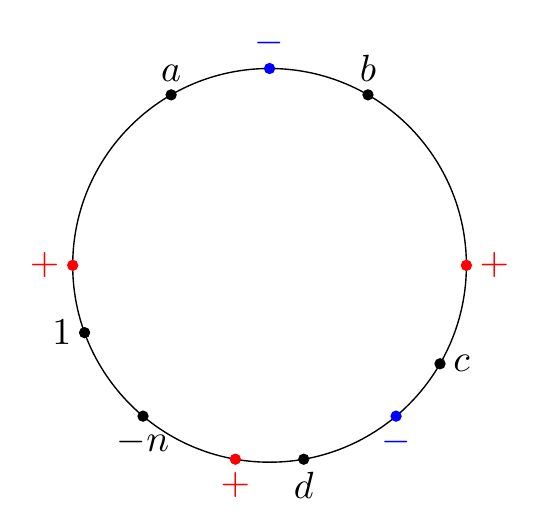}~~~~
\includegraphics[width=.25\textwidth]{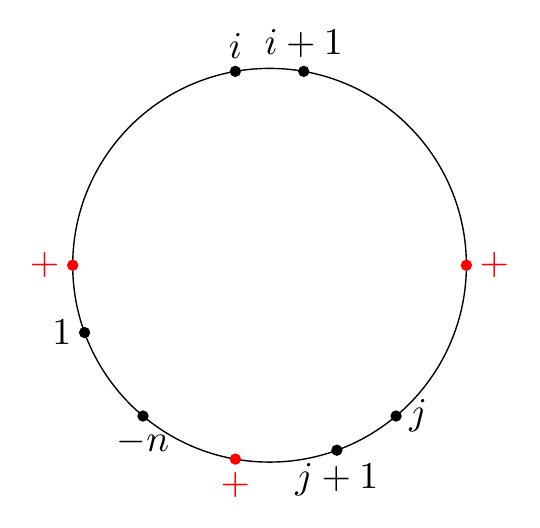}
\caption{Signs of brackets involving the point ${\bf V}_i$ depend on its position relative to the other indices. In the top figure we see that for $d=2$, the signs can be positive or negative depending on whether there is a gap between $a,b$. When there is no gap, the signs are always positive. The middle figure shows the same phenomenon for $d=4$. The bottom figure shows the novelty for odd $d$, where ${\bf V}_1$ and -${\bf V}_n$ must appear.} \label{fig:proofcyclic}
\end{figure}

There is a beautiful class of polytopes in any number of dimensions, known as cyclic polytopes, that directly generalizes this structure for convex polygons. The vertices have an ordering ${\bf V}_1, \cdots, {\bf V}_n$, where $n$ can be arbitrarily large, and obey
\eq\label{PosDet}
\langle{\bf  V}_{i_1} , {\bf V}_{i_2},\cdots  , {\bf V}_{i_d} \rangle >0,\quad \forall \;  i_1<i_2<\cdots<i_d   \,.
\eqe
Remarkably, this allows us to determine the face structure analytically. To see how this works, we return to the case of the  polygon, but now understand the facets purely algebraically without drawing a picture, since pictures will not be available to us in general dimensions. Recall that in order for $({\bf V}_a,{\bf V}_b)$ to be a facet,  $\langle {\bf V}_i, {\bf V}_a,{\bf V}_b \rangle$ must have the same sign for all $i$ different than $a,b$. Without loss of generality we can take $a<b$. Then, we see that for $i<a<b$ the bracket is positive, as well as for $a<b<i$, since in this case we simply pass $i$ through two vectors to arrange it in manifestly positive (increasing) ordering in the bracket. But if there is any gap between $a,b$, so that we can have some $i$ with $a<i<b$, this bracket will be negative. We conclude that $(a,b)$ can not have a gap between them, and the so the facets must be of the form $({\bf V}_i {\bf V}_{i+1})$. This argument generalizes for all even $d$. For instance as seen in the Figure \ref{fig:proofcyclic}, consider the case of $d=4$. Here a putative facet is $({\bf V}_{a} {\bf V}_b {\bf V}_c {\bf V}_d)$, but we can easily see that they must pair up into two consecutive sets $(i,i+1,j,j+1)$ in order for all the brackets to have the same sign.

For odd $d$ there is a slight difference. Consider $d=3$. Here it is easy to see that all the facets must have either the first vector $``1"$ or the last vector $``n"$; and $\langle {\bf V}_1 {\bf V}_i {\bf V}_{i+1} {\bf V}_j \rangle > 0$ for all $j$, and also $-\langle {\bf V}_i {\bf V}_{i+1} {\bf V}_n {\bf V}\rangle_j>0$ for all $j$ (note the extra minus sign).

The reason for the difference between even and odd $d$ is also related to the name of these objects. Given ${\bf V}_1, \cdots, {\bf V}_n$ that are ``positive" in the sense that all the ordered brackets are positive, it is natural to look at what happens to this data under the cyclic shift ${\bf V}_1 \to {\bf V}_2, {\bf V}_2 \to {\bf V}_3, \cdots, {\bf V}_n \to {\bf V}_1$. This preserves the ordering of everything except between $n$ and $1$, so all the positive brackets involving $n$ will pick up a factor of $(-1)^{d+1 - 1} = (-1)^d$. Thus, for all the brackets to stay positive, we must have a {\it twisted} cyclic symmetry under which ${\bf V}_n \to (-1)^d {\bf V}_1$. For $d$ even this is an honest cyclic symmetry, and so cyclic polytopes are really cyclically invariant (starting with the familiar polygon example in $d=2$). But for $d$ odd, cyclic polytopes are not literally cyclically invariant, and the vertices ${\bf V}_1$ and ${\bf V}_n$ play a special role in the boundary structure.

For general $d$, the facets of the cyclic polytope are

\eqa\label{Boundaries}
d \in {\rm odd}:&&\quad \{{\bf W}_a\}=\begin{array}{c} \left({\bf V}_1,\, {\bf V}_{i_1},\,{\bf V}_{i_1{+}1}, \,{\bf V}_{i_2},\,{\bf V}_{i_2{+}1}, \cdots,\,{\bf V}_{i_{\frac{d{-}1}{2}}},\,{\bf V}_{i_{\frac{d{-}1}{2}}{+}1}\right) \\ \bigcup \left( -1\right) \left( {\bf V}_{i_1},\,{\bf V}_{i_1{+}1},\,{\bf V}_{i_2},\,{\bf V}_{i_2{+}1},\cdots,\, {\bf V}_{i_{\frac{d-1}{2}}},\,{\bf V}_{i_{\frac{d-1}{2}}{+}1},\,{\bf V}_n\right)\end{array}\,,\nonumber \\
d \in {\rm even}:&&\quad  \{{\bf W}_a\}=\left( {\bf V}_{i_1},\,{\bf V}_{i_1+1},\,{\bf V}_{i_2},\,{\bf V}_{i_2{+}1}, \cdots, {\bf V}_{i_{\frac{d}{2}}},\,{\bf V}_{i_{\frac{d}{2}}+1}\right)\,.
\eqae
Note in particular that all the vectors ${\bf V}_i$ are vertices.
Thus if \eqref{PosDet} holds true, the task of computing the convex hull is already completed!
 Furthermore, all the lower dimensional faces simply follows from removing one vertex from the facet at a time.

Cyclic polytopes have a number of extremal properties that make them special amongst all polytopes. For instance, for a fixed number of vertices in any dimension, cyclic polytopes have the maximum number of possible facets, of all dimensionality! Cyclic polytopes have also made an important appearance in the physics of scattering amplitudes. If we group the vectors ${\bf V}_1, \cdots, {\bf V}_{n}$ into a $(d+1) \times n$ matrix, all the minors of this matrix are positive--the matrix gives a point in the {\it positive Grassmannian}, which has figured prominently in the study of on-shell diagrams \cite{ArkaniHamed:2012nw} and the Amplituhedron \cite{Arkani-Hamed:2013jha}. 
Indeed, while for general $N^k MHV$ amplitudes with $k>1$, the Amplituhedron represents a Grassmannian generalization of polytopes,  for $k=1$   it reduces to  the cyclic polytope   \cite{Hodges:2009hk}.

There are many natural questions we can ask about general polytopes
that cannot be readily answered, for which the results are instead
trivially available in the case of cyclic polytopes. 
Let us begin with a particularly simple example. 
 Given any polytope with vertices $({\bf V}_1,
\cdots, {\bf V}_n)$, we can choose to look at the convex hull of any subset
of these vertices, to get a new polytope that naturally sits ``inside"
the old one. In general we cannot make any predictions for the face
structure of the new polytope.  However given a cyclic polytope, if we
choose any subset of the vertices $({\bf V}_{a_1}, \cdots, {\bf V}_{a_m})$, we
trivially get another cyclic polytope (with fewer vertices),
since all the ordered determinants are trivially a subset of the old
ones and are still positive.

As another example,  as we have already
discussed, it is often useful to project a space of interest through
some vector. In particular it is often interesting to a take a
$d$-dimensional polytope and project through one of its vertices $\bf V$
to go down to a $(d-1)$-dimensional polytope. For general
polytopes, we cannot easily predict the face structure of the projected
polytope. 
Indeed,  the vertices of the higher
dimensional polytope can end up inside the lower dimensional one. However, again everything can be understood for cyclic
polytopes. 
Given some $d$-dimensional cyclic polytope with vertices
$({\bf V}_1, \cdots, {\bf V}_n)$, if we project through ${\bf V}_1$, the projected
vertices $({\bf V}_2,\cdots,{\bf V}_n)$ form a cyclic polytope in $(d-1)$
dimensions. This is because the projected minors 
\begin{align}
\langle
{\bf V}_{a_1}, \cdots, {\bf V}_{a_d} \rangle_{\rm proj} = \langle {\bf V}_1 {\bf V}_{a_1}
\cdots {\bf V}_{a_d} \rangle
\end{align}
are positive when the $a_i$ are ordered.  
The same is similarly true if we project through the last vertex ${\bf V}_n$.  
Thus, for instance, if we take a $d=3$ cyclic polytope and project
through the vertex ${\bf V}_1$, we are left with a $d=2$ cyclic polytope,
which is just the polygon with vertices $({\bf V}_2, \cdots, {\bf V}_n)$. In this
simple example it is easy to see how the faces of the ``upstairs"
polytope project down to those of the projected one. The
(two-dimensional) faces of the upstairs polytope are $(1, i, i+1)$ and
$(i, i+1, n)$; the (one-dimensional) edges of these are $(i, i+1)$,
together with $(1, i), (1,i+1)$ and $(n,i), (n, i+1)$. Of these, after
projection, $(i, i+1)$ end up as the edges of the polygon,
$(1,i),(1,i+1)$ get projected down to the vertices $i,i+1$, while
$(n,i),(n,i+1)$ end up inside the polygon.

 \begin{figure}
\centering
\subfloat[]{
\includegraphics[width=.25\textwidth]{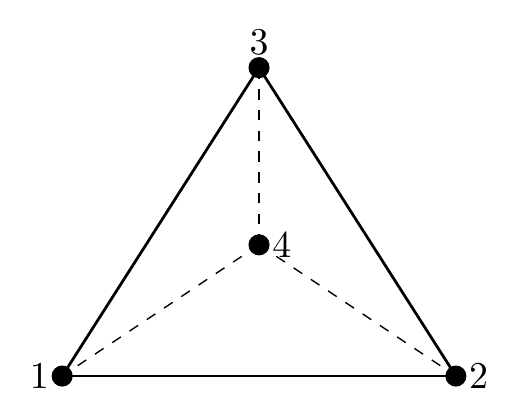}
}
~~~~~~~~~~~
\subfloat[]{
\includegraphics[width=.15\textwidth]{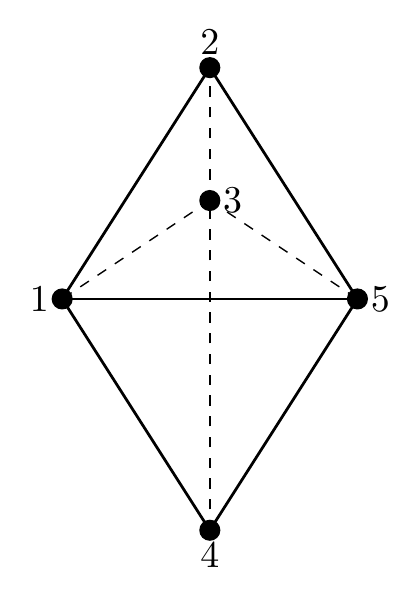}
}
\caption{Both the tetrahedron and the triangular bipyramid are cyclic polytopes.}\label{fig:45}
\end{figure}

\begin{figure}
\centering
\subfloat[]{\label{fig:6va}
\includegraphics[width=.25\textwidth]{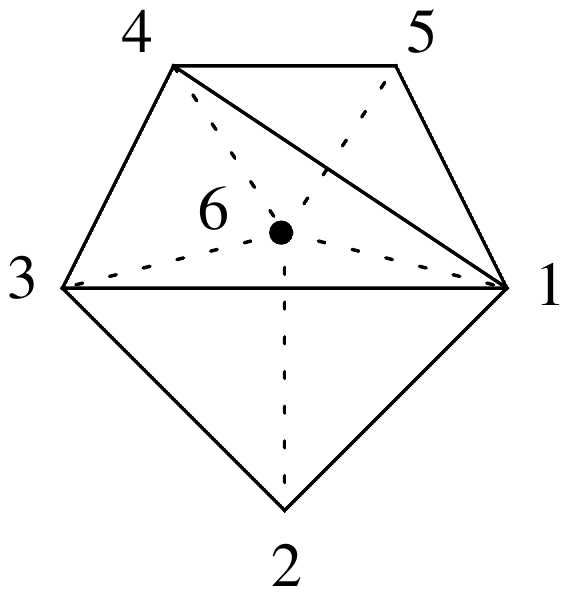}
}
~~~~~~~~~~~
\subfloat[]{\label{fig:6vb}
\includegraphics[width=.15\textwidth]{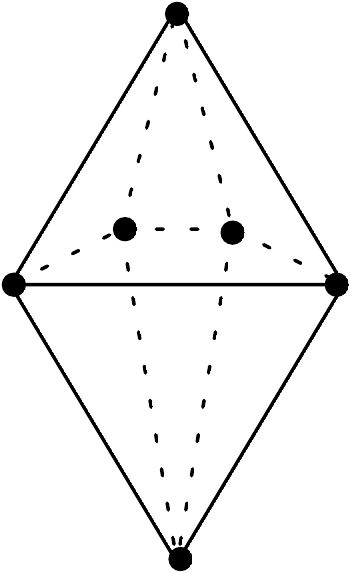}
}
\caption{Two $3d$ polytopes with six vertices. (a) Cyclic polytope. (b) Non-cyclic polytope.}
\end{figure}

Let us give a few examples of cyclic polytopes in two and three dimensions.
For $d=2$, any convex polygon is a cyclic polytope. Indeed, the faces (to be more precise, edges) are all the adjacent pairs $({\bf V}_i ,{\bf V}_{i+1})$, as in \eqref{Boundaries}.

For $d=3$,  the  polytopes with four (tetrahedron) and five (triangular bipyramid) are both cyclic (see Figure \ref{fig:45}).  Moving on to  polytopes with six vertices, the one in Figure \ref{fig:6va} is cyclic with the ordering of vertices given there.  On the other hand, the square bipyramid in Figure \ref{fig:6vb} is the simplest example of a non-cyclic simplicial polytope.
To see this, let us note that in a $3d$ cyclic polytope with six vertices, the first vertex 1 is on 5 different faces, i.e.\ $({\bf V}_1,{\bf V}_2,{\bf V}_3), ({\bf V}_1,{\bf V}_3,{\bf V}_4),({\bf V}_1,{\bf V}_4,{\bf V}_5), ({\bf V}_1,{\bf V}_5,{\bf V}_6), ({\bf V}_6,{\bf V}_1,{\bf V}_2)$ according to \eqref{Boundaries}.  However, the maximum number of faces that share a vertex is 4 in Figure \ref{fig:6vb}, hence the latter cannot be cyclic.

The canonical example for cyclic polytopes in general dimensions is that constructed from a moment curve. A moment curve is the following algebraic curve in $\mathbb{P}^d$:
\eq\label{momentcurve}
\left( \begin{array}{c}
1 \\x \\ x^2 \\\vdots \\ x^d \end{array}\right)
\in \mathbb{P}^d, \quad x \in \mathbb{R} \,.
\eqe
Consider any $d+1$  points ${\bf V}_i = (1,x_i,\,x_i^2,\,x_i^3,\,\cdots\,,\,x_i^d)^T$ labeled by $x_i$ ($i=1,\cdots, n$) on the curve.  We will order them such that $x_1<x_2<\cdots <x_{d+1}$.
We have
\eq
\langle {\bf V}_1,{\bf V}_2,\cdots, {\bf V}_{d{+}1}\rangle=\prod_{i<j}(x_j-x_i) \,.
\eqe
To see this, we note that the RHS  vanishes if $x_i=x_j$, and is a polynomial of degree $\frac{d(d-1)}{2}$ in the $x_i$'s.
It follows that as long as we order $x_1<x_2<\cdots <x_n$, the determinant will be positive. Hence, the convex hull of any $n$ points ${\bf V}_i$ on the moment curve is a cyclic polytope.

Another example of cyclic polytope comes from the  monomial $z^{\Delta_i}$ with a collection of  $\Delta_i$'s. This is the contribution to the CFT four-point function from a single operator (not necessarily primary) of scaling dimension $\Delta_i$.  Let us first consider the vector generated by Taylor expanding $z^{\Delta_i}$ around some point  $z_0>0$:
\eq
{\bf V}_i=
\left( \begin{array}{c}
z_0^{\Delta_i} \\ \Delta_i z_0^{\Delta_i{-}1} \\\frac{\Delta_i(\Delta_i{-}1)}{2} z_0^{\Delta_i{-}2} \\ \vdots \\\frac{\prod_{a=0}^{d{-}1}(\Delta_i-a)}{d!}z_0^{\Delta_i{-}d} \end{array}\right)\,.
\eqe
We will often refer to this as the Taylor scheme.
Again since $\langle {\bf V}_1,{\bf V}_2,\cdots, {\bf V}_{d{+}1}\rangle$ must vanish whenever $\Delta_i=\Delta_j$, it is straightforward to see that
\eq\label{Scaling}
\langle {\bf V}_1,{\bf V}_2,\cdots,{\bf V}_{d{+}1}\rangle=\left(\prod_{a=1}^{d}\frac{1}{a!}\right)\prod_{i<j}(\Delta_j-\Delta_i)z_0^{\sum_{i}\Delta_i-\frac{d(d-1)}{2}}\,,
\eqe
which is positive for ordered $\Delta_i$'s. We conclude that the convex hull of vectors from the Taylor coefficients of $z^{\Delta_i}$ forms a cyclic polytope.
There is in fact a enlightening way of understanding this positivity. Instead of the Taylor expansion, we will consider the vector constructed by evaluating $z^{\Delta_i}$ at distinct but ordered $z_i>0$:
\eq\label{posdet}
{\bf V}_i=
\left( \begin{array}{c}
z_0^{\Delta_i} \\ z_1^{\Delta_i}  \\ \vdots \\ z_d^{\Delta_i}\end{array}\right)\,,\quad 0<z_0<z_1<\cdots<z_d \,.
\eqe
This will be referred to as the position scheme.
We will prove that the determinants of ordered vectors in the position scheme are also positive. Note that this the positivity in the position scheme implies that in the Taylor scheme, since we can take the $z_i$s arbitrary close to $z_0$ and the determinants of the two schemes are equivalent.

We will proceed by proving that $\langle {\bf V}_1,\cdots , {\bf V}_{d+1}\rangle$ in the position scheme \eqref{posdet} must have the same sign for any $z_i$'s with $0<z_0<z_1<\cdots<z_d$ and $0<\Delta_1<\Delta_2<\cdots<\Delta_{d{+}1}$. The overall sign can be fixed later by considering an arbitrary example. Now the statement that the determinants \eqref{posdet} have the same sign is equivalent to that the determinant $\langle {\bf  V}_1,\cdots , {\bf V}_{d+1}\rangle$ can never be zero. In other words,  for any real $c_i$'s, there is no solution to
\eq
\sum_i c_i \left( \begin{array}{c}
z_0^{\Delta_i} \\ z_1^{\Delta_i}  \\ \vdots \\ z_d^{\Delta_i}\end{array}\right)=0\,.
\eqe
 Said it in yet another way, the function
\eq
g_{d{+}1}(z)=c_1 z^{\Delta_1}+c_2 z^{\Delta_2}+\cdots +c_{d{+}1} z^{\Delta_{d+1}}
\eqe
cannot have $d{+}1$ positive roots for any choice of $c_i$'s.
We will  prove by induction.
For $d=0$, this is obviously true because $g_1(z)=c_1 z^{\Delta_1}$  does not have any root for $z>0$.
Let us assume $g_n(z)$ has at most  $n{-}1$ positive roots for any choice of the $c_i$'s.
Now if we further assume $g_{n+1}$ has  $n{+}1$ positive roots for some choices of the $c_i$'s, we will show that this leads to a contradiction.
 Since $g_{n+1}$ has $n{+}1$ positive roots so does the product $z^{-\Delta_{n+1}}g_{n+1}(z)$. Now take the derivative of this product:
\eq
\left(z^{-\Delta_{n+1}}g_{n+1}(z)\right)'=c_1 (\Delta_1-\Delta_{n+1})z^{\Delta_1}+c_2 (\Delta_2-\Delta_{n+1})z^{\Delta_2}+\cdots +c_{n}(\Delta_n-\Delta_{n+1})z^{\Delta_{n}}\,.
\eqe
From calculus (Descartes' rule of signs) we know that if a smooth function has $n{+}1$ positive roots, then its derivative must have at least $n$ positive roots.  However the right hand side of the above is special case of $g_{n}(z)$, which cannot have $n$ positive roots, and thus a contradiction.
We conclude that $g_{d+1}(z)$ cannot have $d+1$ positive roots for all $n$, which implies that $\langle {\bf V}_1,\cdots ,{\bf V}_{d+1}\rangle$ cannot vanish and has a fixed sign. This completes the proof that the convex hull of vectors in the position scheme of $z^\Delta$ form a cyclic polytope.

We can phrase our two examples in the following more general way. We are interested in characterizing functions of two variables $F(x,y)$, such that in some suitable ranges for $x,y$, and for any ordered $x_1<x_2<\cdots<x_n$ and $y_1<y_2<\cdots<y_n$, the matrix $M_{a A} = F(x_a,y_A)$ has positive determinant det$(M)>0$. Such classes of functions are known as {\it Tchebycheff systems}, and were first studied by Tchebycheff in the context of the theory of interpolating functions. The technique  above, using Descartes' rule of signs, is one of several interesting approaches to understanding these functions. We refer the interested reader to the classic text on this subject  \cite{karlin1966tchebycheff}.

In the following section, we will apply similar arguments to the full conformal block in the Taylor scheme.

\section{Positivity of the Conformal Block}
We now return to the block vectors $G^I_{\Delta}$ \eqref{prebv}, obtained by Taylor expanding the conformal block around $z=\frac{1}{2}$. Since the block is positive for $0<z<1$, we will define the block vector ${\bf G}_{\Delta}$ projectively by normalizing the first component to be $1$, i.e.\ $G^0_\Delta=1$. We  have
\eqa\label{TaylorBlock}
{\bf G}_{\Delta}=\left(\begin{array}{c}1 \\ 2 \Delta\alpha \\ 4\Delta(\Delta-1)+2\Delta\alpha \\ \frac{8}{3}\Delta\alpha(\Delta^2{-}\Delta{+}1) \\ \frac{8}{3} \Delta(\Delta{-}1)(\Delta^2{-}\Delta{+}3) +4\Delta\alpha \\ \frac{16}{15}\left[\Delta\alpha(\Delta^2{-}\Delta{+}1)(\Delta^2{-}\Delta{+}6)-2\Delta^2(\Delta{-}1)^2\right] \\ \vdots\end{array}\right)\,,
\eqae
where
\eq\label{alphaeq}
\alpha\equiv  \frac{_2F_1\left(\Delta,\Delta{+}1,2\Delta,\frac{1}{2}\right)}{_2F_1\left(\Delta,\Delta,2\Delta,\frac{1}{2}\right)}\,.
\eqe
The polynomial dependence of the block vector on $\Delta$ and $\alpha$ follows from the Gauss contiguous relation of the hypergeometric function.
\begin{figure}
\begin{center}
\includegraphics[scale=0.5]{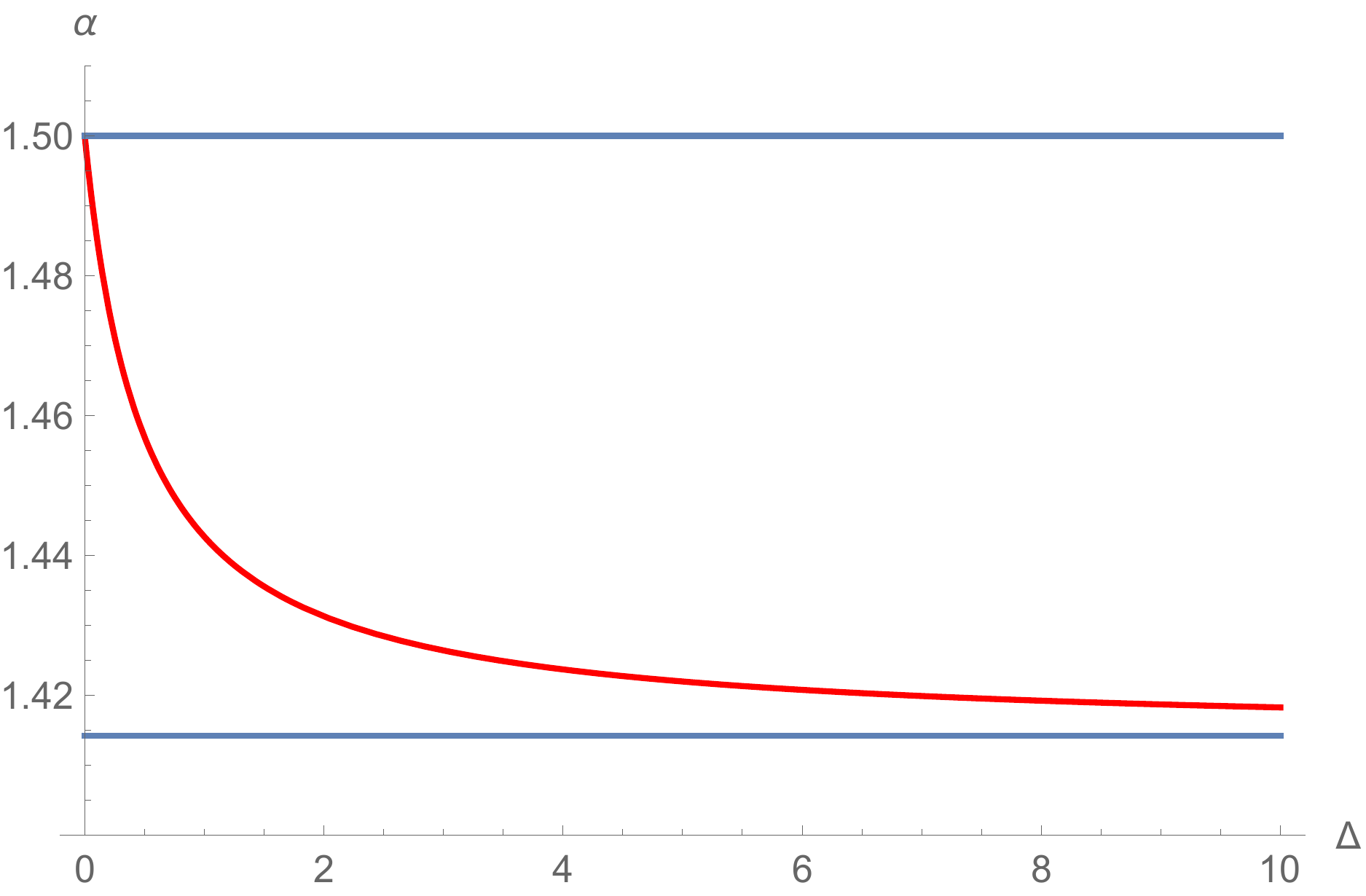}
\caption{The red curve is $\alpha$ \eqref{alphaeq} as a function of $\Delta$.  The value of $\alpha$ is bounded in a narrow region with the upper bound given by $3/2$ ($\Delta\rightarrow 0$) and the lower bound given by $\sqrt{2}$ ($\Delta\rightarrow \infty$).}
\label{AlphaPlot}
\end{center}
\end{figure}
While $\alpha$ is a non-polynomial function of $\Delta$, numerically it is approximately a constant.
As shown in Figure \ref{AlphaPlot},  the value of  $\alpha$ decreases monotonically from $3/2$ to $\sqrt{2}$ as we vary $\Delta$:
\eq
\frac{3}{2}>\alpha>\sqrt{2},\quad \Delta \in [0, \infty)\,.
\eqe
If we approximate $\alpha$ by a constant, the Taylor block vector in (\ref{TaylorBlock}) becomes equivalent to $(1,\Delta,\Delta^2,\cdots)$ up to a $\Delta$-independent $GL$ transformation.
For example, in the large $\Delta$ limit $\alpha\to \sqrt{2}$, and  the 3-dimensional block vector approaches
\eqa
{\bf G}_{\Delta \gg1}=\left(\begin{array}{c}1 \\ 2\sqrt{2}\Delta \\ 4\Delta^2-2(2-\sqrt{2})\Delta \\ \frac{8\sqrt{2}}{3}(\Delta^3-\Delta^2+\Delta) \end{array}\right)\,.
\eqae
Applying a $\Delta$-independent $GL(3)$ rotation on the last three rows, we can convert it to a moment curve \eqref{momentcurve}:
\eq
\left(\begin{array}{ccc}\frac{1}{2\sqrt{2}} & 0 & 0 \\ \frac{\sqrt{2}-1}{4}& \frac{1}{4} & 0 \\ - \frac{1}{4} & \frac{1}{4}  &  \frac{3}{8\sqrt{2}} \end{array}\right)G^I_{\Delta \gg1}=\left(\begin{array}{c}  1 \\ \Delta  \\ \Delta^2\\ \Delta^3 \end{array}\right)\,.
\eqe
Note that since the $GL(3)$ matrix is a constant matrix, it only changes the determinant by an overall constant.
 Thus for large scaling dimensions we see that the block vectors form a cyclic polytope as discussed in Section \ref{sec:polytope}.

\begin{figure}
\begin{center}
\includegraphics[scale=0.25]{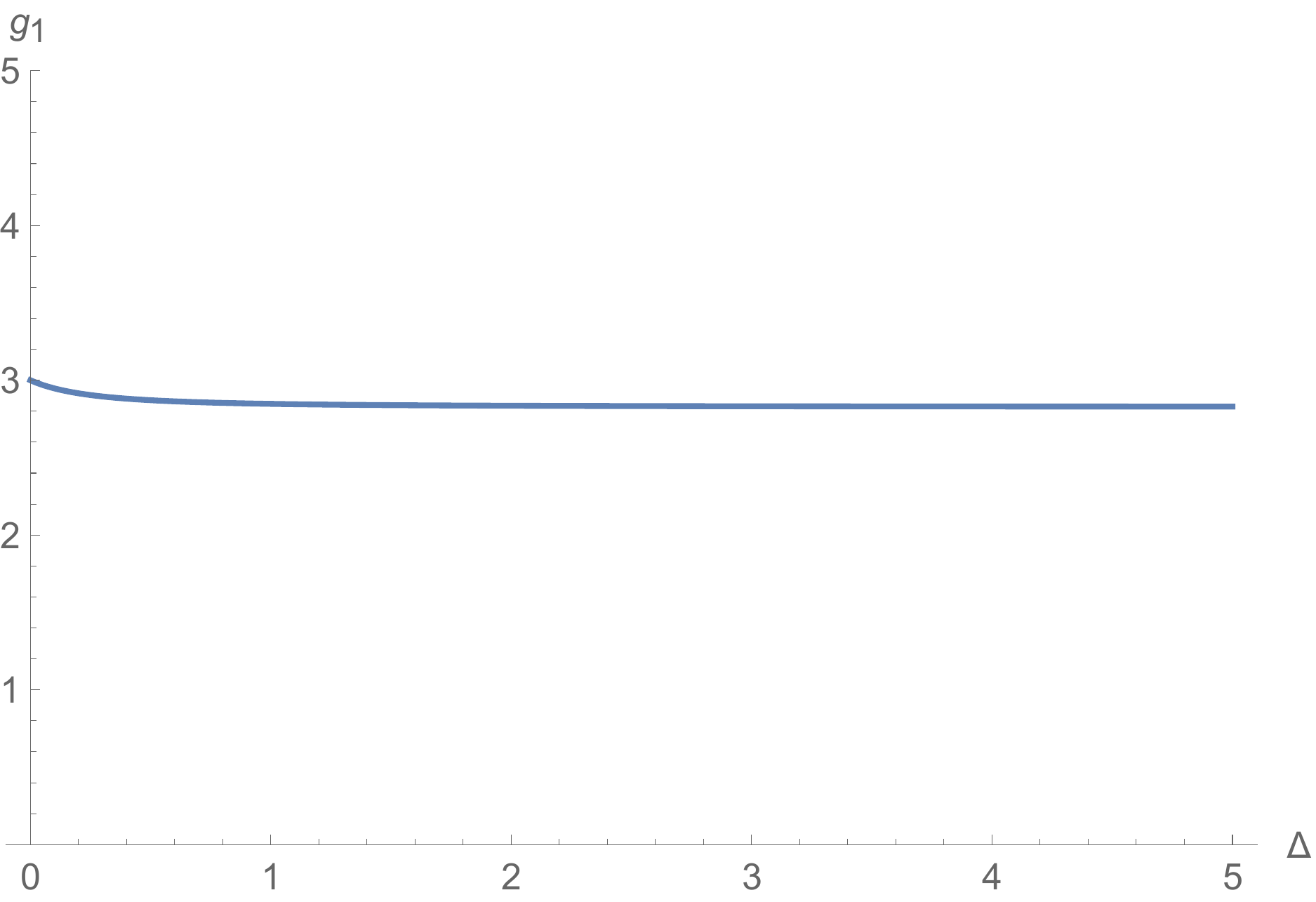}~~~~
\includegraphics[scale=0.25]{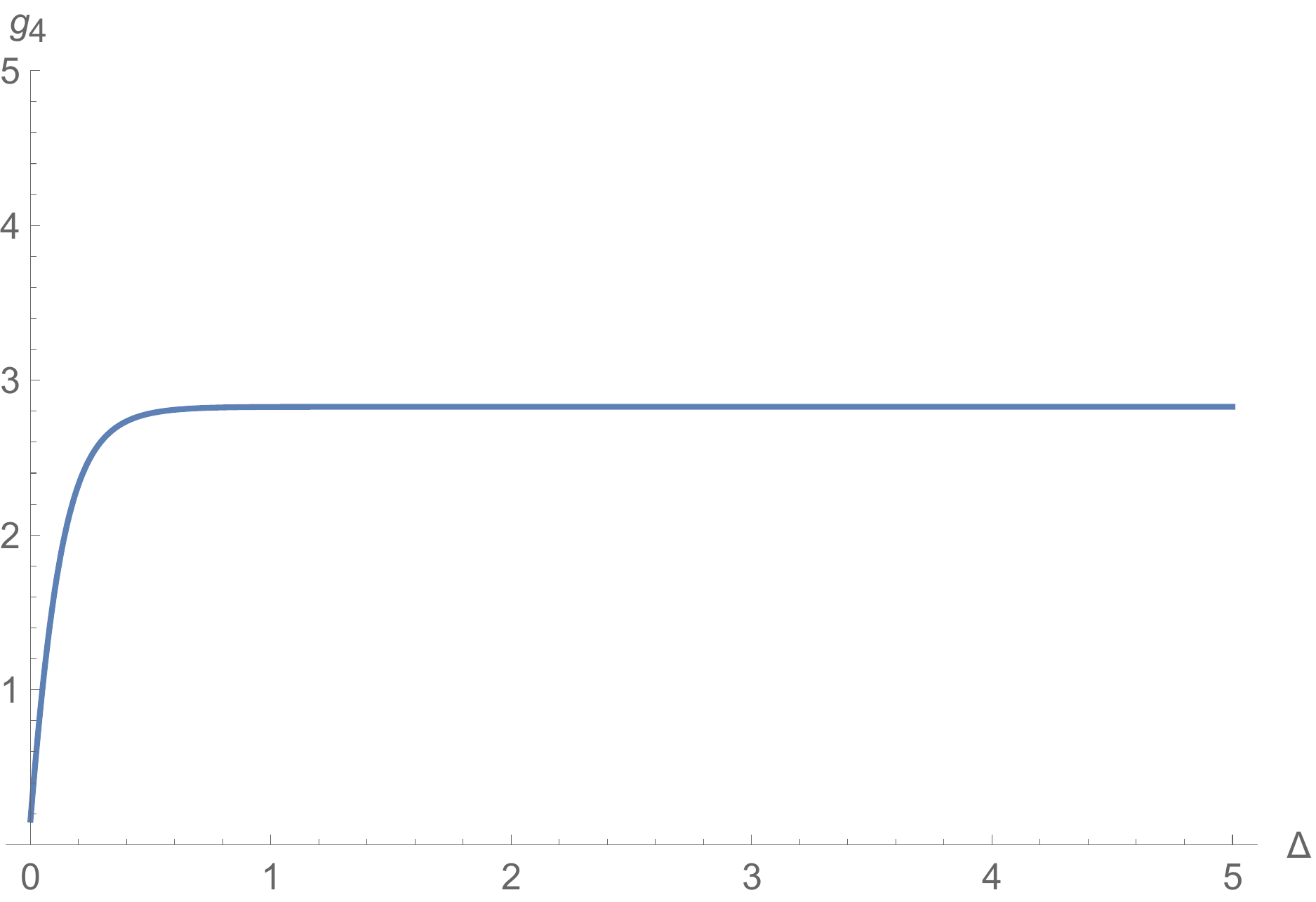}~~~~
\includegraphics[scale=0.25]{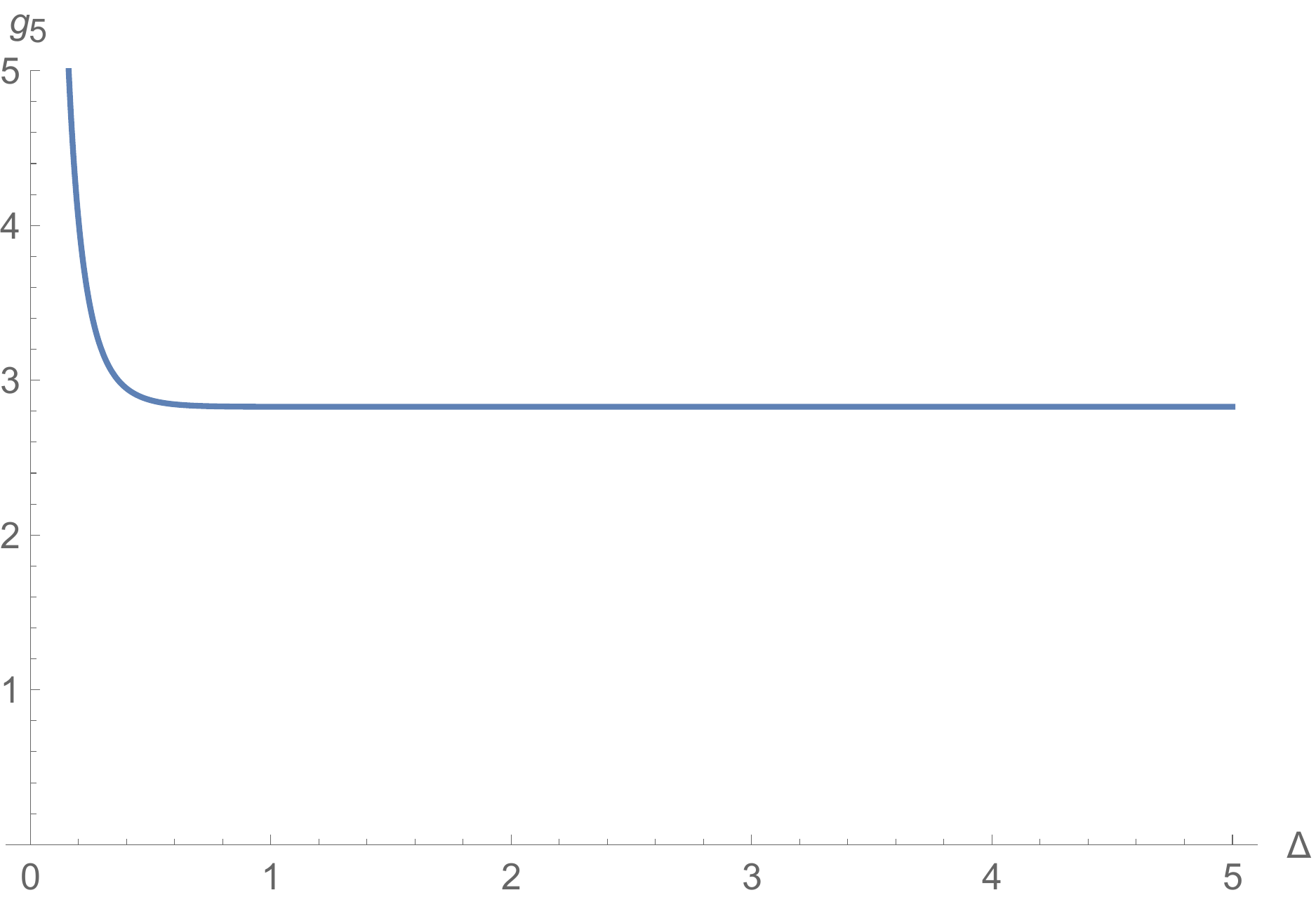}
\caption{The plots of the functions $g_1$, $g_4$, and $g_5$.}
\label{Pos1}	
\end{center}
\end{figure}

Motivated by the large $\Delta$ analysis, we would like to see whether the block vectors \eqref{TaylorBlock} in the Taylor scheme give rise to a cyclic polytope for general $\Delta$. We will proceed with our analysis iteratively in the dimension of the block vectors as in Section \ref{sec:cyclic}.
Let us first introduce a shorthand notation that will be used in the rest of this paper:
\begin{align}
\langle i_1, i_2,\cdots,i_{d+1}\rangle \equiv \epsilon_{I_1I_2\cdots I_{d+1}} G^{I_1}_{\Delta_{i_1}}\cdots G_{\Delta_{i_{d+1}}}^{I_{d+1}}\,.
\end{align}
For $d=1$, we would like to show
\eq\label{v1v2}
\langle 1,2\rangle>0,\quad \forall  \,\Delta_1<\Delta_2\,.
\eqe
This is equivalent to the statement that ${\bf G}_{\Delta_1}$ and ${\bf G}_{\Delta_2}$ can never be linearly dependent. Following the discussion in Section \ref{sec:cyclic}, this implies that
\eq\label{d1}
c_1+c_2G_\Delta^{1}=c_1+2c_2\Delta \alpha(\Delta)
\eqe
can not have two positive roots for any $c_i$s. Let's prove the statement by contradiction.
If \eqref{d1} has two positive roots, then the derivative must have at least a positive root.
 The derivative of \eqref{d1} is shown in Figure \ref{Pos1}, where we see that it is never zero for $\Delta>0$. Thus the absence of a positive root for $g_1=(G_\Delta^{1})'$, with $\,'=\frac{d}{d\Delta}$  then implies that  (\ref{v1v2}) holds.

Now let us move on to $d=2$, where we will show
\eq\label{v1v2v_3}
\langle 1,2,3\rangle>0,\quad \forall\,\Delta_1<\Delta_2<\Delta_3\,.
\eqe
Again this is true so long as the following function does not have three distinct positive roots:
\eq\label{d2}
c_1+c_2G_\Delta^{1}+c_3G_\Delta^{2}=c_1+2c_2\Delta \alpha(\Delta)+c_3\left(4\Delta(\Delta-1)+2\Delta\alpha(\Delta)\right)
\eqe
for any choice of the $c_i$'s. We again proceed by showing a contradiction.
Assuming \eqref{d2} has three positive roots for some $c_i$'s, then its first derivative
\eqa
c_2(G_\Delta^{1})'+c_3(G_\Delta^{2})'
=(G_\Delta^{1})'\left(c_2+c_3\frac{(G_\Delta^{2})'}{(G_\Delta^{1})'}\right)\,,
\eqae
 must have at least two positive roots. Since we have shown that $(G_\Delta^{1})'$ has no positive root, the second parentheses must have at least two positive roots. Recycling our argument, this implies that:
\eq
g_2\equiv \left(\frac{(G_\Delta^{2})'}{(G_\Delta^{1})'}\right)'\,,
\eqe
must have a at least a positive root.  However,  we have explicitly checked that the above function is never zero for $\Delta>0$, and hence the contradiction.
Thus we have shown (\ref{v1v2v_3}).

We can iteratively proceed for higher $d$. In summary the condition for positivity boils down to the following conditions for general $\Delta$:
\begin{align}
&d=1: ~g_1= (G_\Delta^{1})'>0,\quad d=2:~g_2= \left(\frac{(G_\Delta^{2})'}{(G_\Delta^{1})'}\right)'>0,\quad\\
&d=3: ~g_3= \left(\frac{\left(\frac{(G_\Delta^{3})'}{(G_\Delta^{1})'}\right)'}{\left(\frac{(G_\Delta^{2})'}{(G_\Delta^{1})'}\right)' }\right)'>0\,,\quad
 d=4:~g_4= \left(\left(\frac{\left(\frac{(G_\Delta^{4})'}{(G_\Delta^{1})'}\right)'}{\left(\frac{(G_\Delta^{2})'}{(G_\Delta^{1})'}\right)' }\right)'/\left(\frac{\left(\frac{(G_\Delta^{3})'}{(G_\Delta^{1})'}\right)'}{\left(\frac{(G_\Delta^{2})'}{(G_\Delta^{1})'}\right)' }\right)'\right)'>0,\cdots. \nonumber
\end{align}
We have numerically verified that up to $d=5$, these conditions hold true for general $\Delta>0$.
On the right of Figure \ref{Pos1}, we show the plot of the function $g_1,g_4$, $g_5$, which is always positive. Note that the positivity of $g_5$ is enough to guarantee all the lower orders, since (as we will talk about in more detail below), we can obtain the lower dimensional geometries simply by projecting through the infinity block, so that positive in $d$ dimensions immediately implies positivity for lower dimensions.  We show the plots for $d=1,d=4$ and $d=5$ to illustrate an interesting trend: for $d=1$ the functions are clearly positive. But for $d=4$ it gets ``close" to zero for small $\Delta$, while for $d=5$ it is monotonically increasing at low $\Delta$ and is still manifestly positive.

However starting at $d=6$, the analogous function $g_6$ {\it does} cross zero at very small $\Delta$. This does not by itself prove that some brackets will be negative--the condition we are using is sufficient but not necessary--and indeed randomly scanning the brackets for billions of choices of dimensions does not reveal any negativity. However we have managed to prove that indeed for sufficiently small $\Delta$'s, the brackets can indeed become negative. However, quite amazingly, the negativity occurs only when all seven block vectors in the bracket are small, $\Delta_i \lesssim 1/10$, and the most negative the brackets ever become never exceeds a magnitude $\sim 10^{-20}$!

Let us explain how we know the brackets become negative when all $\Delta$'s are small enough. We do this by studying the limit $\Delta \ll 1$ in more detail. Let us first consider the block vectors. To keep analytic control, lets first expand the block vectors around $z=0$,
\eq
z^{\Delta}\,_2F_1(\Delta,\Delta,2\Delta,z)=z^{\Delta}\left(\sum_{i=0}^{n_{c}} c_i(\Delta)z^i+\mathcal{O}(z^{n_c+1})\right)\,,
\eqe
where $n_c$ is the cut off of the expansion, and now $c_i(\Delta)$ is an analytic function in $\Delta$. Next, we re-expand the polynomial around $z=\frac{1}{2}$ to obtain the approximate block vector in the Taylor scheme ${\bf G}_{\Delta}$.

Now we compute the determinant using the approximate block vector. The cutoff $n_c$ can be determined by gradually increasing the cutoff until the determinant stabilize. In general this requires $n_c\sim 50$. Since we are interested in the regime where $\Delta_i\ll 1$, we can further Taylor expand the determinant in $\Delta_i$. Then to leading order we have
\eq
\langle \vec{G}_{\Delta_1},\vec{G}_{\Delta_2},\cdots,\vec{G}_{\Delta_d}\rangle=\alpha_d\prod_{i<j}(\Delta_j-\Delta_i)+\mathcal{O}(\Delta^{\frac{d(d-1)}{2}+1})\,.
\eqe
The subleading terms can be suppressed by considering arbitrary small $\Delta_i$s, and the sign of the determinant is given by $\alpha_d$. For $n_c=50$, we find the following $\alpha_d$:
\eq
\alpha_2=3,\,\alpha_3=10.6137,\,\alpha_4=58.5971,\,\alpha_5=12.8414,\, \alpha_6=689.048,\, \alpha_7=-951.43\;.
\eqe
Note that the values are of order 1. We see that for the $7\times7$ determinant we encounter a negative leading term.

From numerical experimentation at larger $d$ up to $d=20$, the pattern appears to be that all the brackets are positive unless at least $7$ of the $\Delta$'s are in this same range of $\Delta \lesssim 1/10$, and again the (negative) minors are always miniscule.

Curiously, while the full conformal blocks are not always positive, the
functions$\,_2F_1(\Delta, \Delta, 2 \Delta; z)$ do appear to be
positive. Indeed experimentally this appears to be the case for all
hypergeometric functions $_2F_1(a,b,c=a+b,z)$ with $a,b>0$. Statements of positivity of this sort about Hypergeometric functions do not appear to have been studied by mathematicians.

The tininess of the negative minors and the small dimensions involved means that the negativity of brackets for small $\Delta$ and large $d$ is irrelevant from any practical point of view. Nonetheless we find this phenomenon absolutely fascinating. After all we start with a problem with no parameters whatsoever--indeed the only numbers that show up are ``1" and ``2" in the hypergeometric functions--and yet we produce counter-examples to positivity with exponentially tiny brackets! This is an concrete example of a dream to generate ``exponentially large hierarchies from nothing".  The application of this observation to inspiring radically new solutions to the cosmological constant and hierarchy problems is left as an exercise for the interested reader.

\section{The Unitarity Polytope and the Crossing Plane}

\begin{figure}
\begin{center}
\includegraphics[scale=0.5]{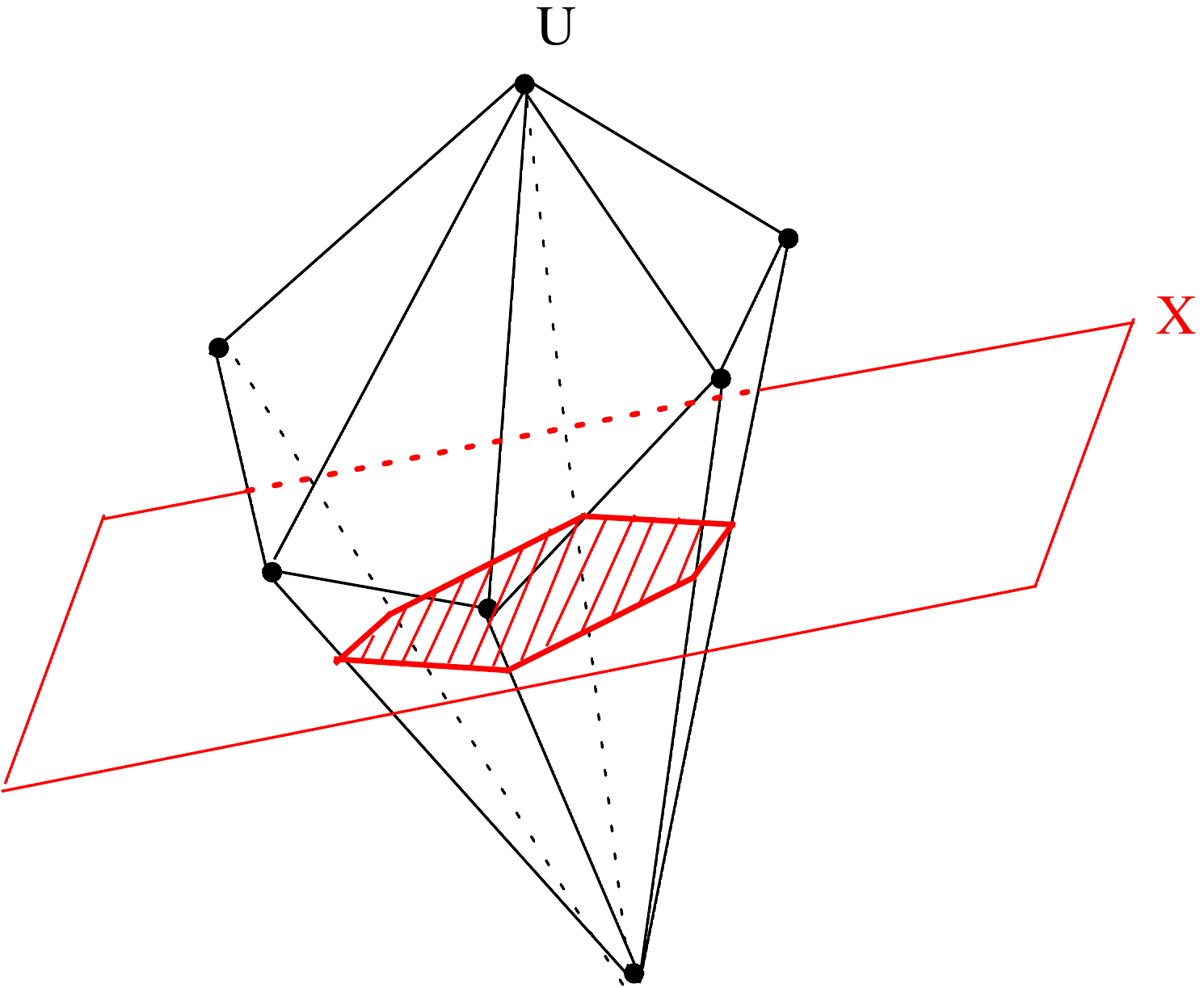}
\caption{A four-point function is consistent if it lies on the intersection between the crossing plane ${\bf X}[\Delta_\phi]$ and the unitarity  polytope ${\bf U}[\{\Delta_i\}]$.}
\label{intersect}
\end{center}
\end{figure}

To summarize our discussion so far,   the bootstrap constraints from unitarity and crossing symmetry can be phrased as follows.
Unitarity requires that the Taylor coefficients of the four-point function $F(z)$ expanded around $z=1/2$ has to lie inside a polytope spanned by the block vectors \eqref{TaylorBlock}.  We will call this polytope the unitarity polytope, denoted as ${\bf U}[\{\Delta_i\}]$.  Crossing symmetry, on the other hand, restricts the Taylor coefficients of $F(z)$ to lie on a half-dimensional plane ${\bf X}[\Delta_\phi]$ defined in Section \ref{sec:bootstrap}.  Here  $\Delta_\phi$ is the dimension of external operators.  A consistent CFT  four-point function must lie in the region defined by the intersection of the crossing plane ${\bf X}[\Delta_\phi]$ and the polytope ${\bf U}[\{\Delta_i\}]$ as in Figure \ref{intersect}.

To solve this intersection problem, we need to know what  the facets of ${\bf U}[\{\Delta_i\}]$ are.  As discussed Section \ref{sec:polytope}, this is a daunting task for general polytopes. Fortunately from Section \ref{sec:cyclic}, we see that the convex hull of the block vectors,  at least up to $d=5$, is  a cyclic polytope, for which all the faces are known.  Here $d$ is the number of Taylor coefficients we keep.   We will only obtain nontrivial constraints if $d$ is odd, i.e.\ $d=2N+1$, then the crossing plane ${\bf X}[\Delta_\phi]$ is $N$-dimensional, and the facets of the unitarity polytope ${\bf U}[\{\Delta_i\}]$ are
\eq\label{cfacets}
(0,i_1,i_1+1, i_2, i_2+1,\cdots, i_{N}, i_{N+1})\,\bigcup\,(i_1,i_1+1, i_2, i_2+1,\cdots, i_{N}, i_{N+1},\infty)\,,
\eqe
where each entry $i$ represents the block vector ${\bf G}_{\Delta_i}$ given in \eqref{TaylorBlock}. The lower dimension faces are obtained from the facets by knocking out one label at a time.

The geometry problem we want to solve is the intersection between an $N$-dimensional crossing plane ${\bf X}[\Delta_\phi]$ and a $(2N+1)$-dimensional unitarity polytope in $\mathbb{P}^{2N+1}$.
 Generically, an $N$-dimensional plane intersects with an $(N{+}1)$-dimensional face of the polytope at a point  in $2N{+}1$ dimensions. Such an $(N+1)$-dimensional face of the cyclic polytope takes the form
\eq\label{face}
(i_1,i_1+1, i_{2},\cdots,, i_{N{+}1})\,.
\eqe
From (\ref{Intersect}), the condition that the the crossing plane ${\bf X}[\Delta_\phi]$ intersect the above face \textit{in} the unitarity polytope ${\bf U}[\{\Delta_i\}]$ is
\begin{align}\label{Cintersect}
&\langle i_1+1, i_{2},\cdots,, i_{N{+}1},{\bf X}\rangle\,,\quad
-\langle i_1,i_{2},\cdots,, i_{N{+}1},{\bf X}\rangle \,,\quad
\langle i_1, i_1+1, i_{3},\cdots,, i_{N{+}1},{\bf X}\rangle\,,\notag\\
&\cdots\, , (-1)^{N}\langle i_1,i_1+1, i_{2},\cdots,, i_{N},{\bf X}\rangle\,,~~~~\text{all have the same sign}\,.
\end{align}
Thus, for a four-point function with external operators of dimension $\Delta_\phi$, \textit{a consistent CFT must contain $N+2$ operators in its spectrum such that (\ref{Cintersect}) holds.}

\begin{figure}
\begin{center}
\includegraphics[scale=0.8]{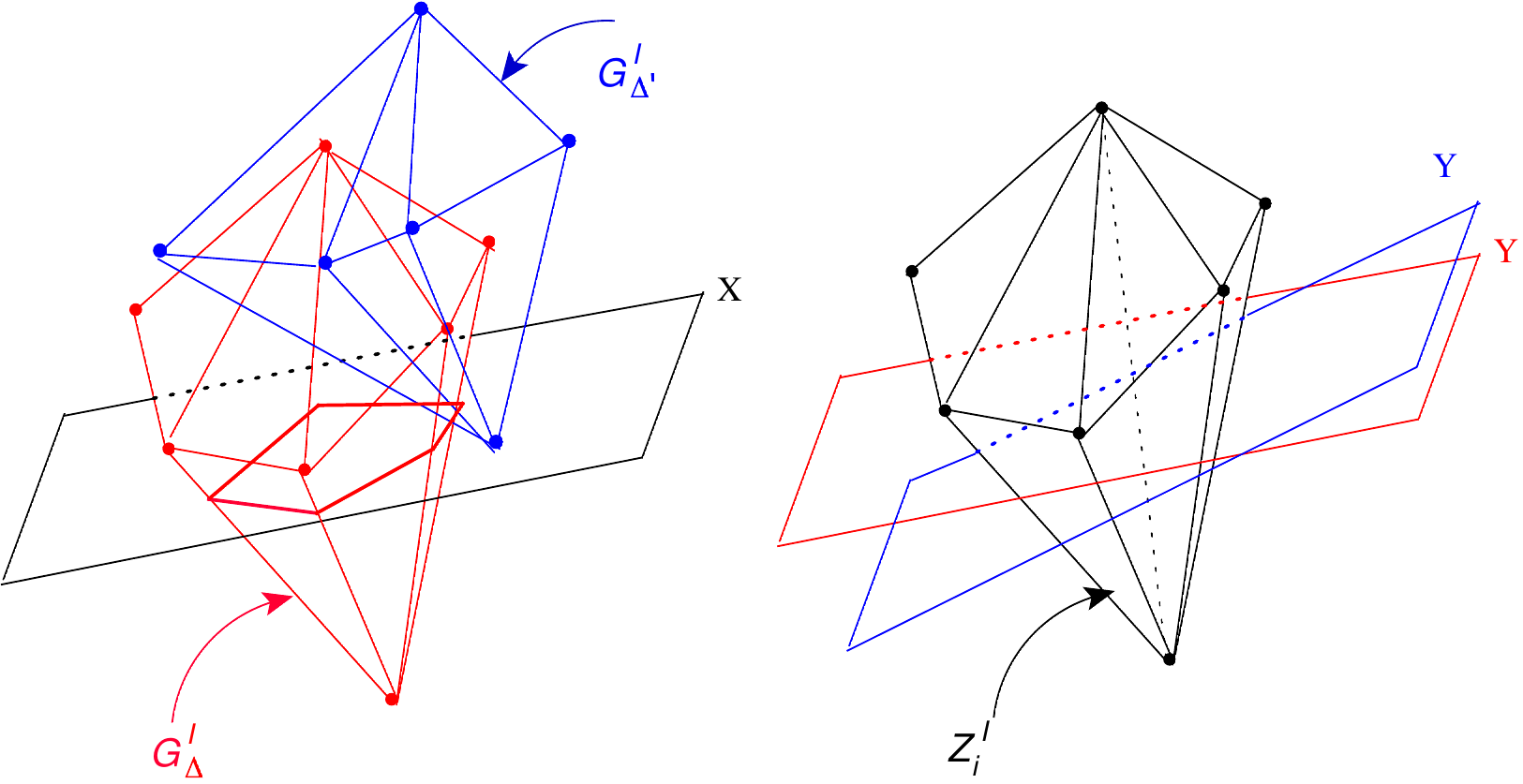}
\caption{The comparison between the geometry for CFT and the Amplituhedron. In the CFT case, we have a fixed crossing plane $\bf X$ that intersects with the unitarity polytope. The latter is determined by the spectrum of the CFT operators. In the Amplituhedron case, we have the polytope determined by the momentum twistors $Z^I_i$ held fixed, and we look for $k$-planes $\bf Y$, constrained by positivity/projected winding number conditions}.
\label{Compare}
\end{center}
\end{figure}

This picture of positive geometry is very reminiscent of the (tree) Amplituhedron for (tree) scattering amplitudes of $\mathcal{N}=4$ SYM~\cite{Arkani-Hamed:2013jha}. In that case, we have a polytope comprised of $(4{+}k)$-component vectors that are the bosonized super momentum twistors ($Z^I_i$), and a $k$-plane $Y_\alpha^I$,  with $\alpha=1,\cdots, k$ and $I=1,\cdots, 4+k$.\footnote{Here $k$ labels the helicity sector.  For example, the pure gluon amplitude  has $2{+}k$  negative helicity states with $k\geq 0$.} The Amplituhedron then puts a constraint on the plane $\bf Y$. In the original formulation, the $k-$plane $Y_{\alpha}^I$ must be expressible as $Y_\alpha^I = C_{\alpha a} Z_a^I$, where the $k \times n$ matrix $C$ is in the positive Grassmannian $G_+(k,n)$, with all positive minors. For $k=1$ the Amplituhedron is simply the cyclic polytope, while for higher $k$ it defines a generalization of polytopes into the Grassmannian. A more recent definition of the Amplituhedron is defined by characterizing the geometry obtained by projecting {\it  through} $\bf Y$. The data $Z_a$ becomes 4-dimensional after the projection, and is required to have a maximum ``winding number" ~\cite{Arkani-Hamed:2017vfh}. But in both definitions, the external data $Z_a^I$ is thought of as fixed, and we are interested in carving out an allowed space of $k$-planes $Y$ by various conditions. then defined as the set of planes $\bf Y$ which are in a general region where $\bf Y$ intersects the $Z$-polytope with the  prescribed winding number. For the CFT case discussed in this paper, we instead have the crossing plane $\bf X$ fixed by the external dimension $\Delta_\phi$, while the unitarity polytope varies with the spectrum and intersects with the former.  See  Figure \ref{Compare}.

\section{$N=1$: Three-Dimensional Polytope}\label{sec:1dgeometry}

We now begin studying in detail the geometry of the crossing plane intersecting the cyclic polytope. In the Taylor scheme, we keep  the first $(2N{+}1)$-th Taylor coefficients of the four-point function and the conformal block around $z=\frac{1}{2}$. The four-point function resides on the plane $\bf X$, which is $N$-dimensional, and conveniently parameterized as:
\eq\label{6dF}
{\bf F}=\left(\begin{array}{c}F^0 \\4\Delta_\phi F^0 \\ F^2 \\ \frac{16}{3}(\Delta_\phi-4\Delta_\phi^3)F^0+4\Delta_\phi F^2 \\ F^4 \\ \frac{64}{15}\Delta_\phi(32\Delta_\phi^4 -20\Delta_\phi^2+3)F^0-\frac{16}{3}\Delta_\phi(4\Delta^2_\phi-1)F^2+4\Delta_{\phi}F^4\\\vdots\end{array}\right)\in \mathbb{P}^{2N+1}
\,.
\eqe
Here $F^0,F^2, F^4, \cdots, F^{2N}$ parametrize the $N$-dimensional crossing plane $\bf X$ in $\mathbb{P}^{2N+1}$.
Using the rescaling freedom in $\mathbb{P}^{2N+1}$, we can set $F^0=1$.
As we increase $N$,  it suffices to restrict ourselves to the case of odd-dimensional polytopes. This is because in even dimensions the intersection problem merely constrains the new parameter $F^{2N}$, and we do not learn anything new compared to the analysis in the previous dimensionality.

With $N=1$, we are considering block vectors given by:
\eq
{\bf G}_{\Delta}=\left(\begin{array}{c}1 \\2\alpha \Delta\\ 4\Delta(\Delta-1)+2\Delta\alpha\\ \frac{8}{3}\alpha\Delta(\Delta^2-\Delta+1)\end{array}\right)\,,
\eqe
where again $\alpha= \,_2F_1(\Delta,\Delta+1 ,2\Delta,1/2)/\,_2F_1(\Delta,\Delta,2\Delta,1/2) $.
The unitarity polytope is a three-dimensional cyclic polytope,  positively spanned by the vectors ${\bf G}_{\Delta_i}$.  The two-dimensional facets consists of the following two sets:
\eq\label{1dwall}
(0,i, i+1)\,,\quad (i,i+1,\infty)\,.
\eqe
Here $i$ represents the vector ${\bf G}_{\Delta_i}$, $0$ is the identity operator ${\bf G}_{0}=(1,0,\cdots,0)$, and $\infty$ is ${\bf G}_{\infty}=(0,0,\cdots,1)$. The subscripts $i$ and $i{+}1$ label two operators $\Delta_i<\Delta_{i+1}$ with nothing in between.  

Note that here we assume that the last vertex of the cyclic
polytope is at $\Delta = \infty$, which is expected for
realistic CFT's. 
However for the geometry problem, with even a
putative finite spectrum of $\Delta$'s, we can always think of the
finite cyclic polytope as living inside a larger one where we add a
final vertex at infinity.

The crossing plane $\bf X$ is one-dimensional, which is represented by a $4\times 2$ matrix
\eq
\renewcommand{\kbldelim}{(}
\renewcommand{\kbrdelim}{)}
{\bf X}=  \kbordermatrix{
    & F^0 & F^2  \\
     & 1 & 0   \\
     & 4\Delta_\phi & 0   \\
     & 0 & 1\\
     &\frac{16}{3} (\Delta_\phi-4\Delta_\phi^3) & 4\Delta_\phi}\,.
\eqe
where the two columns are the vectors parameterized by $F^0$ and $F^2$.  The crossing plane $\bf X$ intersects with the face $(0,i,i+1)$ if and only if (see \eqref{Cintersect})
\eq\label{1did}
\langle {\bf X} , i ,i+1\rangle,~~ - \langle {\bf X}, 0, i+1\rangle,~~\langle {\bf X},0, i\rangle\,,~~~~~~\text{all have the same sign.}
\eqe
  Similarly the crossing plane $\bf X$ intersects with the face $(i,i+1, \infty)$ if and only if
\eq\label{1dinfty}
\langle {\bf X} , i ,i+1\rangle,~~ - \langle{\bf  X}, \infty, i+1\rangle,~~\langle {\bf X}, \infty, i\rangle\,~~~~~~\text{all have the same sign}\,.
\eqe
  Since  \eqref{1dwall} are all the faces of the three-dimensional polytope, the crossing plane intersects with the polytope iff \textit{either} one of the two conditions \eqref{1did} and \eqref{1dinfty} is satisfied.  Of course generically if one condition is satisfied
the other will not be; the only way $\bf X$ can intersect ${\bf U}[\{\Delta_i\}]$ on both kinds of
faces is if it only intersects ${\bf U}[\{\Delta_i\}]$ on the edge 
$(i, i+1)$ common to both
kinds of faces, forcing  $\langle {\bf X} ,i, i+1\rangle = 0$.

Conditions \eqref{1did} union \eqref{1dinfty} are the sufficient and necessary condition for the bootstrap problem to have a solution in this dimensionality.  To extract useful constraints from these conditions, it is often useful to derive necessary (but not necessarily sufficient) conditions by projecting the geometry to lower dimensions.  Specifically, we will project our three-dimensional polytope through one of its vertex to reduce it to a two-dimensional polytope.  The crossing plane, on the other hand, retains its dimensionality and is still a line.  We will discuss the projected problems and derive  constraints on the spectrum below.

\subsection{Projecting through the Identity $0$}

Let us start by projecting our three-dimensional cyclic polytope through the identity, ${\bf G}_0 = (1,0,0,0)$, to obtain a two-dimensional cyclic polytope in $\mathbb{P}^2$.  The crossing plane $\bf X$, on the other hand, is still one-dimensional after the projection.
In this way we have reduced the geometric question to asking whether a two-dimensional polygon (or a curve in the case of continuous spectrum)  intersects with a line or not.
The edges of the polygons are $(i , i+1) _{proj}$, which descend from $( 0,i,i+1)$ before the projection.  We use a subscript to remind the reader that these are facets of the projected geometry.

In practice, the projection through the identity 0 amounts to the following manipulation on the determinant $\langle 0,{\bf F}, i,i{+}1\rangle$,
\eqa\label{det0}
&&{\rm det}\left(\begin{array}{cccc}1 & 1 & 1 & 1 \\0 & 4\Delta_\phi & G^1_{\Delta_i} & G^1_{\Delta_{i{+}1}} \\0 & F^2 & G^2_{\Delta_i} & G^2_{\Delta_{i{+}1}} \\0 & \frac{16}{3}(\Delta_\phi{-}4\Delta_\phi^3){+}4\Delta_\phi F^2 & G^3_{\Delta_i} & G^3_{\Delta_{i{+}1}}\end{array}\right)={\rm det}\left(\begin{array}{ccc}  4\Delta_\phi & G^1_{\Delta_i} & G^1_{\Delta_{i{+}1}} \\ F^2 & G^2_{\Delta_i} & G^2_{\Delta_{i{+}1}} \\ \frac{16}{3}(\Delta_\phi{-}4\Delta_\phi^3){+}4\Delta_\phi F^2 & G^3_{\Delta_i} & G^3_{\Delta_{i{+}1}}\end{array}\right)\nonumber\\
&&\rightarrow \quad {\rm det}\left(\begin{array}{ccc}  1 & 1 & 1 \\ \frac{F^2}{4\Delta_\phi} & \frac{G^2_{\Delta_i}}{G^1_{\Delta_i}} & \frac{G^2_{\Delta_{i{+}1}}}{G^1_{\Delta_{i{+}1}}} \\ \frac{4}{3}(1{-}4\Delta_\phi^2){+} F^2 & \frac{G^3_{\Delta_i}}{G^1_{\Delta_i}} & \frac{G^3_{\Delta_{i{+}1}}}{G^1_{\Delta_{i{+}1}}}\end{array}\right)\;.
\eqae
In the last step we have rescaled the vectors by their leading positive  component $G^1_{\Delta_i}=2\Delta_i\alpha$.

If the crossing plane and the polytope intersect before the projection, they must again intersect after the projection, but not vice versa. The condition for the projected crossing plane to intersect the two-dimensional polygon is
\begin{align}\label{proj0}
\langle {\bf X} ,i\rangle_{proj}=\langle 0,{\bf X},i\rangle\,,~~~~~
-\langle {\bf X}  ,i+1\rangle_{proj}= -\langle 0,{\bf X},i+1\rangle\,~~~~\text{have the same sign}\,.
\end{align}
This condition is weaker than \eqref{1did} union \eqref{1dinfty} as it only requires the two geometric objects to intersect after the projection.  
 Indeed, note that our condition keeps only two of those in \eqref{1did},
and none of those in \eqref{1dinfty}. The absence of the conditions from \eqref{1dinfty} 
reflects the fact that the
edges $(i, \infty)$ and $(i+1, \infty)$ of the two-dimensional faces
$(i, i+1, \infty)$  all end up inside the two-dimensional polygon
after the projection through the identity.

\begin{figure}
\begin{center}
\includegraphics[scale=0.4]{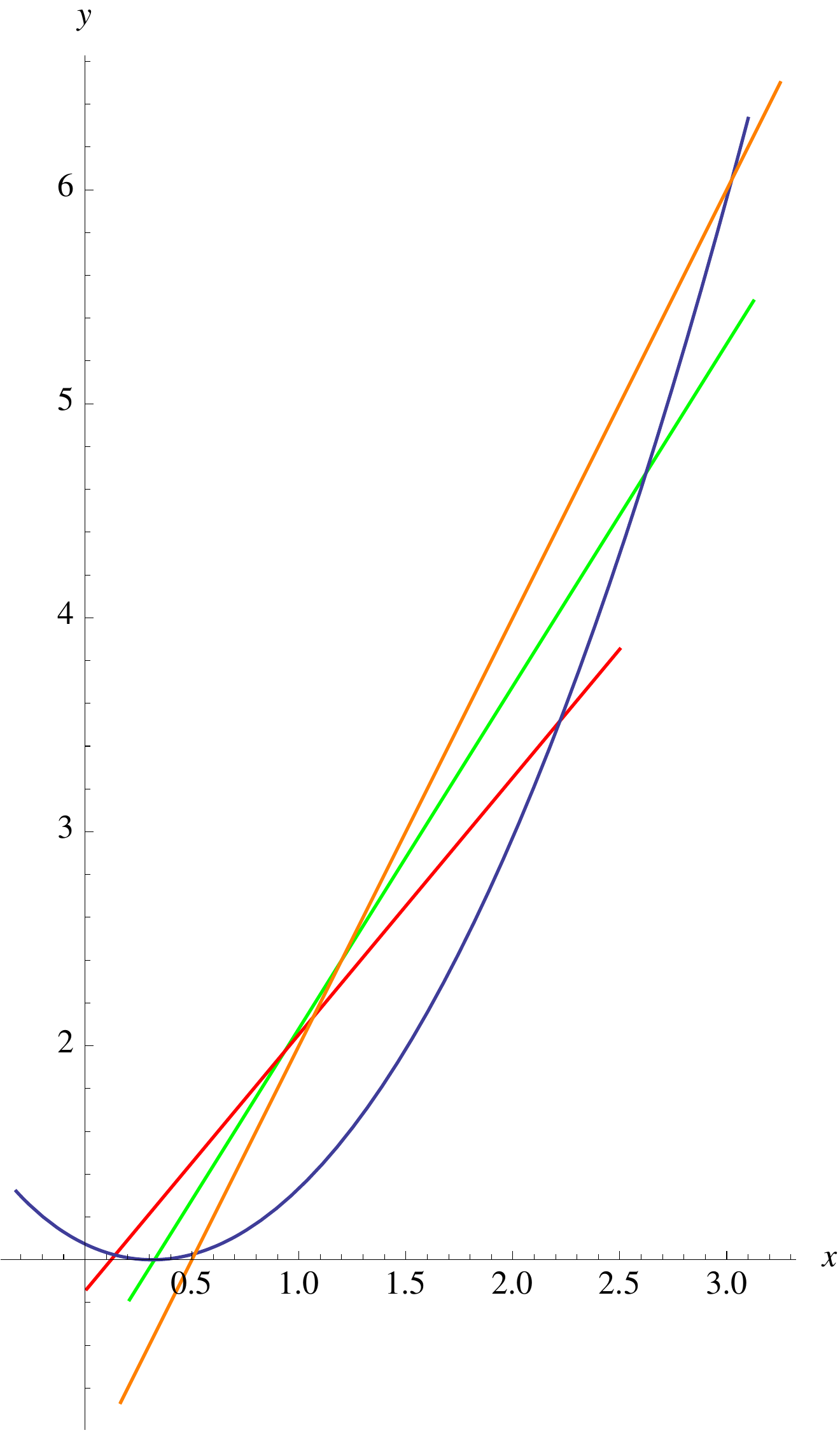}
\caption{The 2$d$ geometry obtained by projecting through the identity $0$. The blue curve is the block vectors parametrized by $y= G^3_\Delta/G^1_\Delta, \, x=G^2_\Delta/G^1_\Delta$ as in \eqref{det0}. The unitarity polytope is constructed from the convex hull of points on these curves. The red ($\Delta_\phi=0.3$) green ($\Delta_\phi=0.4$)  and orange ($\Delta_\phi=0.5$) lines are the crossing planes ${\bf X}[\Delta_\phi]$ with various external dimensions, parametrized as $y=\frac 43 (1-4\Delta_\phi^2)+F^2,\, x= {F^2\over 4\Delta_\phi}$.}
\label{2Dgeometry}
\end{center}
\end{figure}
\begin{figure}[h]
\begin{center}
\includegraphics[scale=0.6]{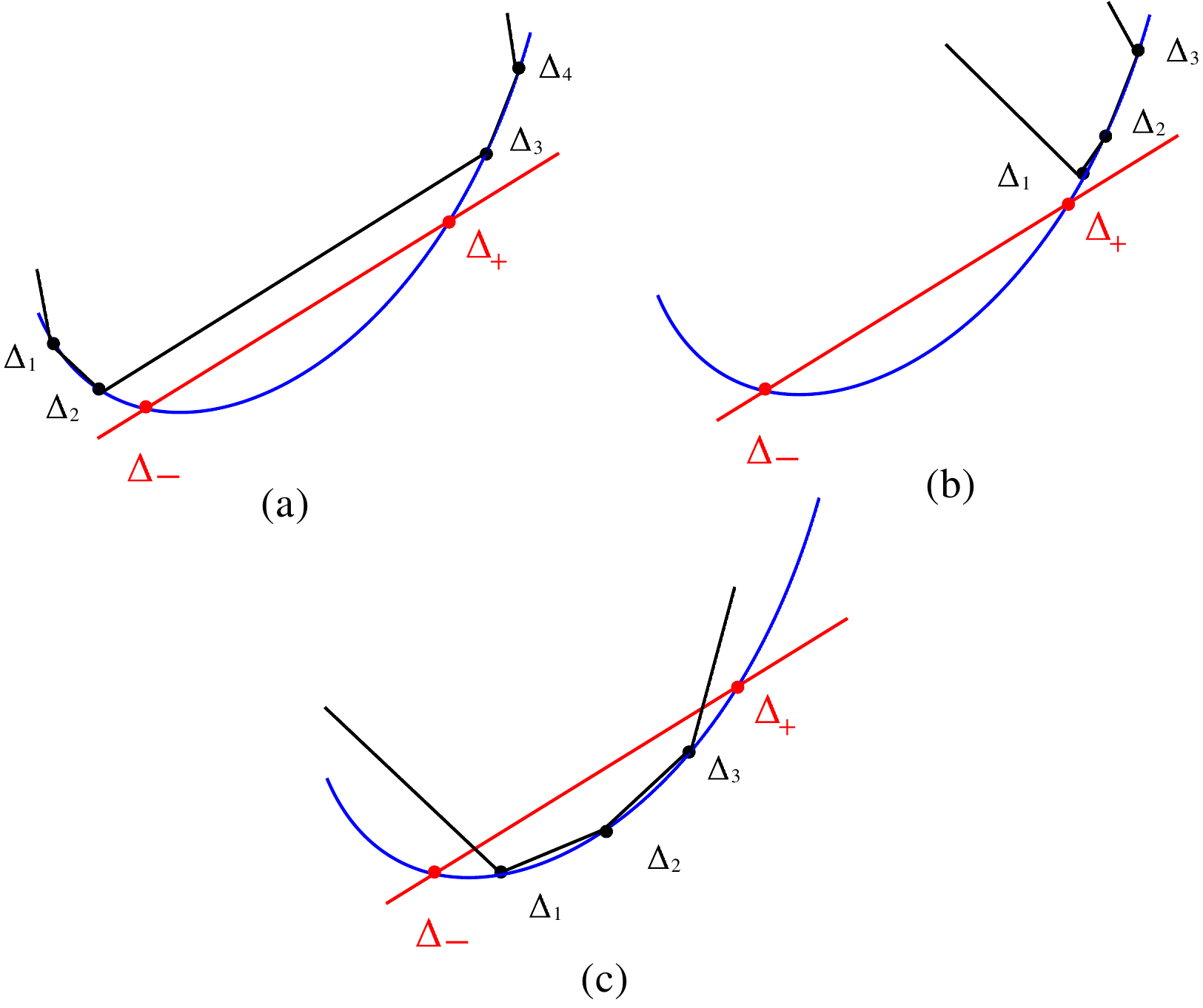}
\caption{In order for the unitarity polytope to intersect with the crossing plane, we must have operators lying between $\{\Delta_-,\Delta_+\}$. Case (a) and (b) demonstrates the cases where all the operators are outside this region, and hence cannot intersect with the crossing line. For case (c) it does intersects.  }
\label{Spec1}
\end{center}
\end{figure}

We can immediately say something interesting about the spectrum from \eqref{proj0}.  Geometrically, the block vectors form a curve, parametrized by $\Delta$, and the crossing plane is a line. We display the geometry in Figure \ref{2Dgeometry}. We see that generically the crossing plane intersects with the block curve at two points. These two points are the solution to the equation
\begin{align}
0&= \langle 0,{\bf X},\Delta\rangle= 4\Delta_\phi \left[
G^3_\Delta -4 \Delta_\phi G^2_\Delta
+\frac 43 (4\Delta_\phi^2-1)G^1_\Delta
\right]\notag\\
&= {32\over 3} \Delta_\phi \Delta
\left[
-6\Delta_\phi(\Delta-1) +\alpha(\Delta^2-\Delta +4\Delta_\phi^2 -3\Delta_\phi)
\right]\,.\label{XpmDef}
\end{align}
Since $\alpha= {\,_2F_1(\Delta,\Delta+1 ,2\Delta,1/2)\over\,_2F_1(\Delta,\Delta,2\Delta,1/2)}$ is a slowly varying function of $\Delta$ (see Figure \ref{AlphaPlot}), we can effectively treat it as a constant in a wide range of $\Delta$. From this approximation we see that there are two roots in $\Delta$ for this equation, denoted as $\Delta_\pm$ with $\Delta_+>\Delta_-$.  $\Delta_\pm$ are functions of $\Delta_\phi$.  The necessary condition \eqref{proj0} tells us that there must be a pair of operator such that either $\Delta_+$ or $\Delta_-$ lies in between. In other words, there must exist at least an operator with dimension $\Delta$ satisfying
\begin{align}
\Delta_- < \Delta <\Delta_+\,.
\end{align}
Examples for spectrums allowed and disallowed are shown in Figure \ref{Spec1}. This statement in particular implies that the gap in the spectrum between the identity and the lightest operator has to be smaller than $\Delta_+$.

\begin{figure}
\begin{center}
\includegraphics[scale=0.5]{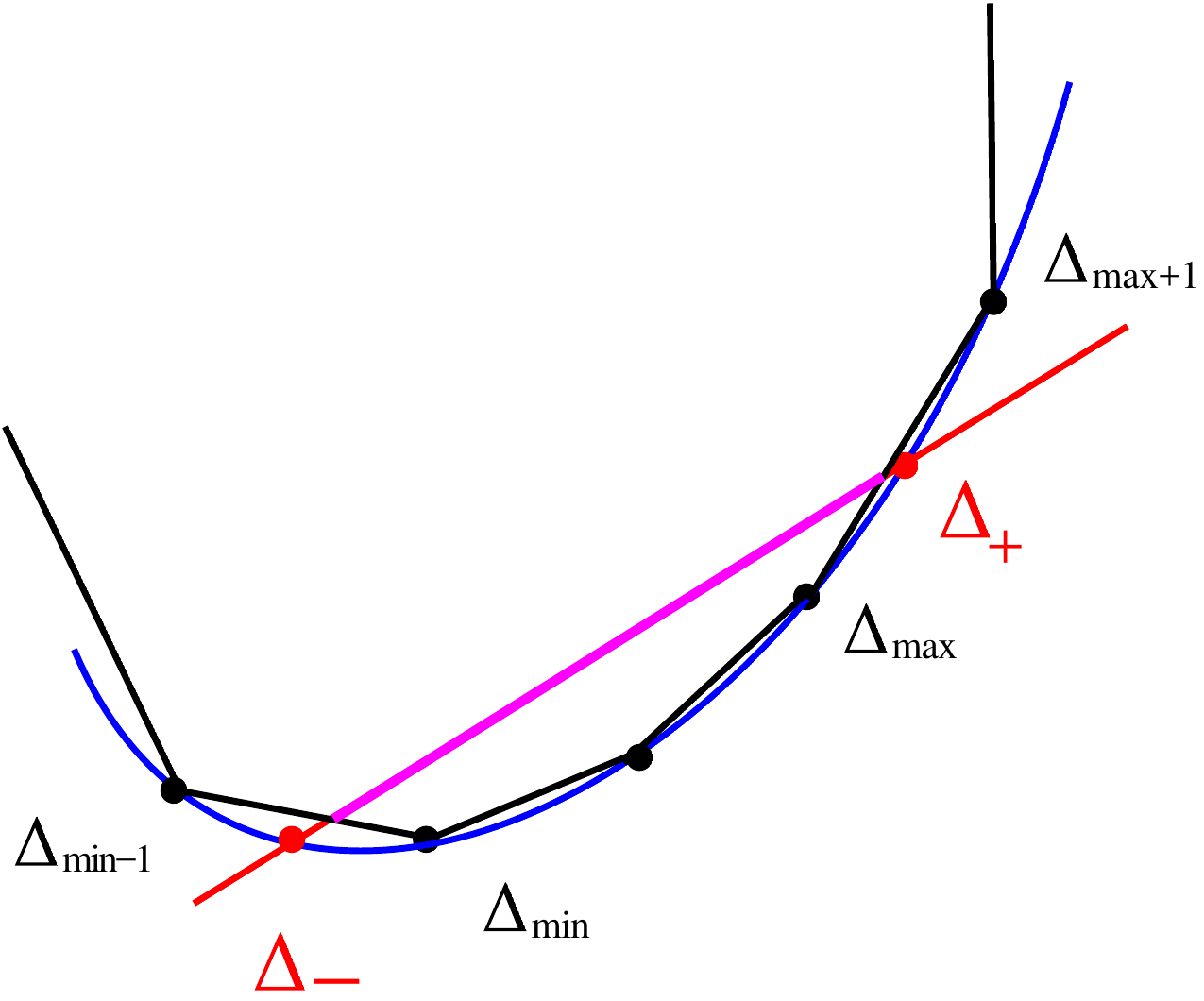}
\caption{For a putative spectrum the four point function is bounded by the operators that are the closest to  $\{\Delta_-,\Delta_+\}$.}
\label{FourptBd}
\end{center}
\end{figure}

Note that this analysis also gives us constraints on the four-point function for a given putative spectrum. Consider the part of the spectrum that lies between $\{\Delta_-,\Delta_+\}$, and denote the operator closest to $\Delta_-$ as $\Delta_{min}$, while that closest to $\Delta_{+}$ as $\Delta_{max}$. The four-point function $\bf X$ is then bounded by:
\eq
\langle  0,{\bf F}, \Delta_{min{-}1}, \Delta_{min}\rangle >0\quad \langle  0,{\bf F}, \Delta_{max}, \Delta_{max+1}\rangle >0\,.
\eqe
We illustrate this in Figure \ref{FourptBd}.
\subsection{Projecting through $\infty$}

We can derive another necessary condition by projecting through the infinity vertex ${\bf G}_\infty = (0,\cdots, 0,1)$.   Again after the projection, the polytope is two-dimensional with edges $(i,i+1)_{proj}$ and the crossing plane is an one-dimensional line.
The condition for the two to intersect in the projected geometry is
\begin{align}\label{projinfty}
\langle {\bf X} ,i\rangle_{proj}=\langle {\bf X},i, \infty\rangle\,,~~~~~
-\langle {\bf X}  ,i+1\rangle_{proj}=-\langle {\bf X},i+1,\infty\rangle\,~~~~\text{have the same sign}\,.
\end{align}
The equation
\begin{align}
0= \langle {\bf X}, \Delta, \infty\rangle=  -2 \alpha\Delta +4\Delta_\phi\,.
\end{align}
The equation has a unique positive solution in $\Delta$, denoted as $\Delta_*$.  This implies that there must be a pair of operators (possibly including the identity) sandwiching $\Delta_*$. This constraint is always satisfied in  a realistic CFT spectrum, since we have an identity operator to the left of $\Delta_*$, and infinity many operators near infinity.

\section{$N=2$: Five-Dimensional Polytope}
We now consider the five-dimensional polytope, where the block vectors has six components:
\eqa\label{5dbv}
{\bf G}_{\Delta}=\left(\begin{array}{c}1 \\ 2 \Delta\alpha \\ 4\Delta(\Delta-1)+2\Delta\alpha \\ \frac{8}{3}\Delta\alpha(\Delta^2{-}\Delta{+}1) \\ \frac{8}{3} \Delta(\Delta{-}1)(\Delta^2{-}\Delta{+}3) +4\Delta\alpha \\ \frac{16}{15}\left[\Delta\alpha(\Delta^2{-}\Delta{+}1)(\Delta^2{-}\Delta{+}6)-2\Delta^2(\Delta{-}1)^2\right]\end{array}\right)\,.
\eqae
 The crossing plane at this order is two dimensional, given by
\renewcommand{\kbldelim}{(}
\renewcommand{\kbrdelim}{)}
 \begin{align}
{\bf X}=  \kbordermatrix{
    & F^0 & F^2 &F^4 \\
     & 1 & 0  &0 \\
     & 4\Delta_\phi & 0 &0  \\
     & 0 & 1&0\\
     &{16\over3} (\Delta_\phi-4\Delta_\phi^3) & 4\Delta_\phi&0\\
     & 0 & 0&1\\
     & \frac{64}{15}\Delta_\phi(32\Delta^4_\phi-20\Delta_\phi^2+3) & {16\over3} (\Delta_\phi-4\Delta_\phi^3)&4\Delta_\phi
  }\,.
\end{align}
and the boundaries are
\eq
(0,i,i{+}1,j,j{+}1),\quad (i,i{+}1,j,j{+}1,\infty)\,.
\eqe
This implies that a putative spectrum is consistent only if there exists a set of four operators $(i,i+1,j,k)$ that intersects the crossing plane at a point, i.e.\
\eq\label{Triplets}
\langle {\bf X} ,i,i{+}1,j\rangle,\; -\langle {\bf X}, i{+}1,j,k\rangle,\;\langle {\bf X}, i,j,k\rangle,\; -\langle {\bf X}, i,i{+}1,k\rangle,\quad {\rm have\; the\; same \;sign}\,.
\eqe
Here $j,k$ could be $0$ or $\infty$.

Following our experience for the three-dimensional polytope, the interesting constraints arrises when we will view the geometry projected through $0$. In this case, we are considering the cyclic polytope with facets $(i,i{+}1,j,j{+}1)$. Since the crossing plane is still two dimensional, and it intersects the polytope if there exists three operators $(i,i{+}1, j)$ such that
\eq\label{5dsign}
\langle {\bf X}, i,i{+}1\rangle,\quad -\langle{\bf X}, i,j\rangle, \quad \langle{\bf X}, i{+}1,j\rangle\,,\quad\quad{\rm have\; the\; same \;sign}\,.
\eqe
Note that in the above, since we are projecting through $0$, each vector has five component as the leading piece is removed, and the brackets now correspond to taking the determinant of $5\times5$ matrices.

Now our geometry is four-dimensional. We can reduce the dimensionality by either projecting through, or projecting on to the two-dimensional crossing plane. Projecting through the crossing plane indicates that we are looking at the part of the geometry orthogonal to $\bf X$. As discussed previously, we can consider a $GL(3)$ transformation acting on the five component vectors such that the crossing plane $\bf X$ takes the following form
\eq
{\bf X}=\left(\begin{array}{ccc}1 & 0 & 0  \\0 & 1 & 0  \\ 0 & 0 & 1 \\0 & 0 & 0  \\0 & 0 & 0 \end{array}\right)\,.
\eqe
Acting on all block vectors with the same $GL(3)$ transformation, the orthogonal directions are simply the remaining two components. On the other hand, projecting on to the crossing plane means that we are considering the image of the unitarity polytope on the two-dimensional plane. In practice the two components of this projection, for points $(i,i{+}1,j)$, can be read off from the coefficient of $F^2$ and $F^4$ in the following determinant:
\eqa
\langle{\bf  F}, i, i+1, j\rangle&=&\det\left(\begin{array}{cccc}\frac{F^2}{4\Delta_\phi} & \tilde{G}^2_{\Delta_i} & \tilde{G}^2_{\Delta_{i+1}}  & \tilde{G}^2_{\Delta_j}  \\ \frac{4}{3}(1-4\Delta_\phi^2)+F^2 &  \tilde{G}^3_{\Delta_i} & \tilde{G}^3_{\Delta_{i+1}}  & \tilde{G}^3_{\Delta_j} \\ \frac{F^4}{4\Delta_\phi}   &  \tilde{G}^4_{\Delta_i} & \tilde{G}^4_{\Delta_{i+1}}  & \tilde{G}^4_{\Delta_j} \\   \frac{16}{15}(32\Delta_\phi^4-20\Delta_\phi^2+3)+\frac{4}{3}(1-4\Delta_\phi^2)F^2+F^4&  \tilde{G}^5_{\Delta_i} & \tilde{G}^5_{\Delta_{i+1}}  & \tilde{G}^5_{\Delta_j}\end{array}\right)\,.\nonumber\\
\eqae
where we've used the short hand notation $\tilde{G}^I_{\Delta}\equiv \frac{G^I_{\Delta}}{G^1_{\Delta}}$.

The two different projections will give us complementary information. The projection through $\bf X$ allows us to have a bird's eye view of all the block vectors, gaining information on how operators of a consistent spectrum must distribute on the curve of block vectors. Once the polytope intersect, the projection onto the crossing plane gives us information with regards to the neighboring operators as well as bounds on the four-point function itself.
\subsection{Projecting through $\bf X$}\label{sec:thruX}
We begin with projection through $\bf X$.
Instead of finding the $\Delta_\phi$-dependent $GL(3)$ transformation that separates the directions orthogonal to the crossing plane, we can  compute the following projection of the block vectors
\eq
\{\langle {\bf X}, 0, {\bf v}_1, \Delta \rangle, \;\;\langle {\bf X}, 0,{\bf v}_2, \Delta\rangle\} \,,
\eqe
where ${\bf v}_1,{\bf v}_2$ are some arbitrarily chosen auxiliary vectors that does not lie on the crossing plane $\bf X$.
The projected block vector is parametrized by these two coordinates on a two-dimensional plane.
By construction, the crossing plane $\bf X$ and the identity 0 are located at the origin on this two-dimensional plane.
The block vectors \eqref{5dbv} form a curve, starting from the origin and parameterized by $\Delta$.

\begin{figure}[h]
\begin{center}
\includegraphics[scale=0.3]{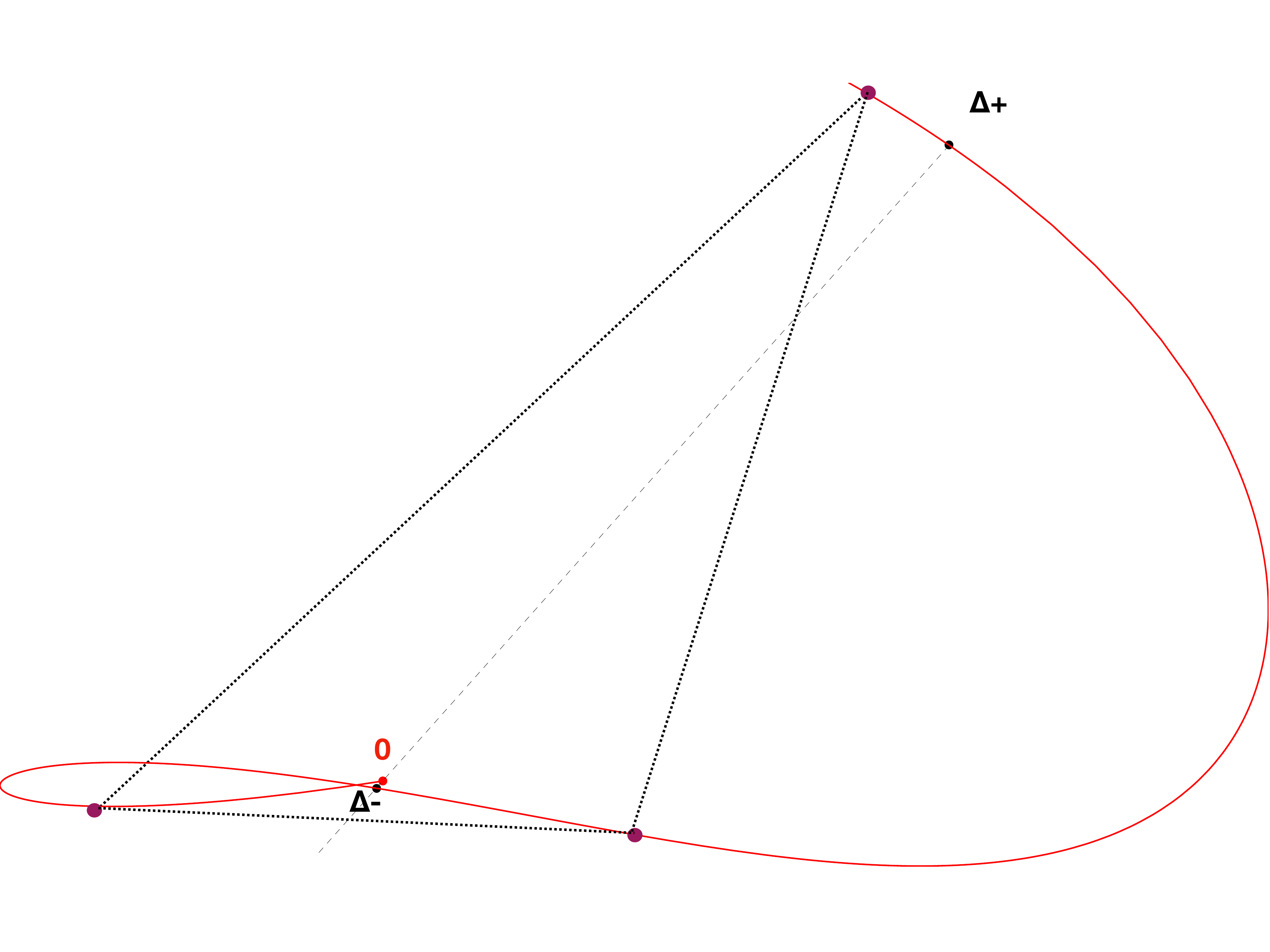}
\caption{Here we present the 2$d$ plot of the block vectors with its coordinates given by $(\langle {\bf X}, 0, {\bf v}_1, i\rangle, \;\langle {\bf X}, 0,{\bf v}_2, i\rangle)$, where ${\bf v}_1=(0, 1, 6, 15, 2, {-}2)$ and ${\bf v}_2=(-1, -2, -\frac{1}{4}, \frac{1}{3}, 1, 0)$. The external dimension that defines the crossing plane was chosen to be $\Delta_\phi=1.5$. Notice that $(0, \Delta_-,\Delta_+)$ are collinear. The purple dots correspond the position of three operators, labeled $(i,i{+}1,j)$, that forms a triangle that incloses the origin $0$. }
\label{CurveExp1}
\end{center}
\end{figure}

The two points $\Delta_\pm$,  defined as the solutions to (\ref{XpmDef}), play a special role in this geometry. 
In particular, they are collinear with the origin. To see this  note that since (\ref{XpmDef}) tells us that $\langle {\bf X}_1,{\bf X}_2, 0,\Delta_{\pm}\rangle=0$, where ${\bf X}_1,{\bf X}_2$ are the two vectors of the crossing ``line" in the $N=1$ problem, this implies that the three-dimensional block vector for $\Delta_{\pm}$ can be spanned by ${\bf X}_1,{\bf X}_2$, and the identity. This then tells us that the four-component determinants $\langle {\bf X}_1, 0,\Delta_-,\Delta_+\rangle=\langle{\bf  X}_2, 0,\Delta_-,\Delta_+\rangle=0$. Now explicitly expanding the five component determinant $\langle {\bf X} ,0 ,\Delta_+, \Delta_-\rangle$, one finds that it can be written as a linear combination of $\langle{\bf X}_1, 0,\Delta_-,\Delta_+\rangle$, $\langle {\bf X}_2, 0,\Delta_-,\Delta_+\rangle$ and $\langle {\bf X}_1, {\bf X}_2, 0,\Delta_{\pm}\rangle$, which all vanishes. Thus
\eq
\langle {\bf X} ,0 ,\Delta_+, \Delta_-\rangle=0\;.
\eqe
In other words, the line between $\Delta_{\pm}$ must pass through the origin on this two-dimensional plane. This is illustrated in Figure \ref{CurveExp1} (where  the external dimension is chosen to be $\Delta_\phi=1.5$), where the two points $\Delta_{\pm}$ are on the  \textit{opposite} sides of the origin.  
 Rather amusingly, the vertex $\Delta_-$ is always much closer to the origin compared to $\Delta_+$.

\begin{figure}
\begin{center}
\includegraphics[scale=0.33]{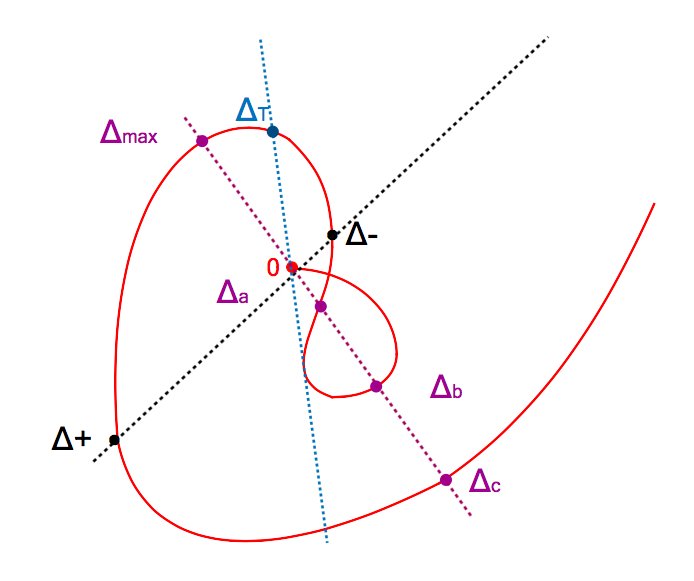}\includegraphics[scale=0.35]{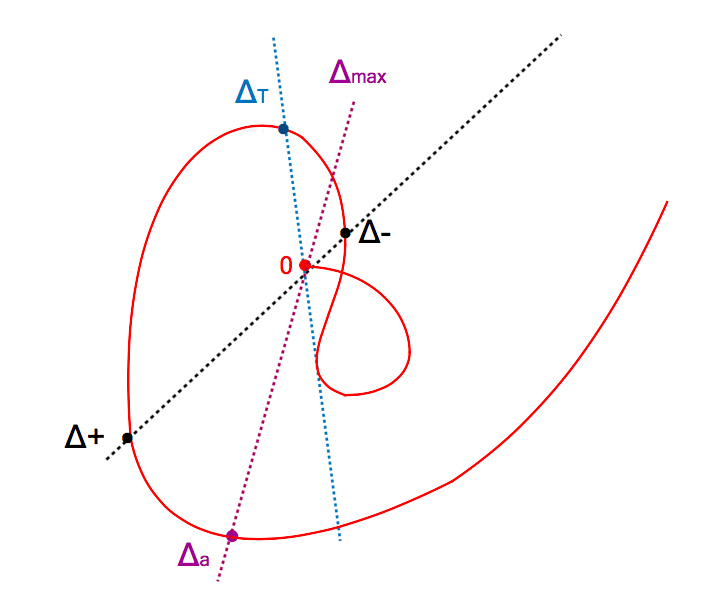}
\caption{There are two possible configurations of the spectrum depending on where $\Delta_{max}$ lies with respect to $\Delta_T$, which is the point where the line that passes through the origin and tangent to the curve below $\Delta_-$, intersects with the block curve. On the LHS we have $\Delta_{max}>\Delta_T$, for which we conclude that either $\Delta_{max{+}1}$ is between $\Delta_+$ and $\Delta_c$, or that there is a light operator between $\Delta_{a}, \Delta_b$. On the RHS for $\Delta_{max}<\Delta_T$ we must have $\Delta_{max{+}1}$ in between $\Delta_+$ and $\Delta_a$.}
\label{CurveExp4}
\end{center}
\end{figure}

\begin{figure}
\begin{center}
\label{fig:D1D2}
\includegraphics[width=2.5in]{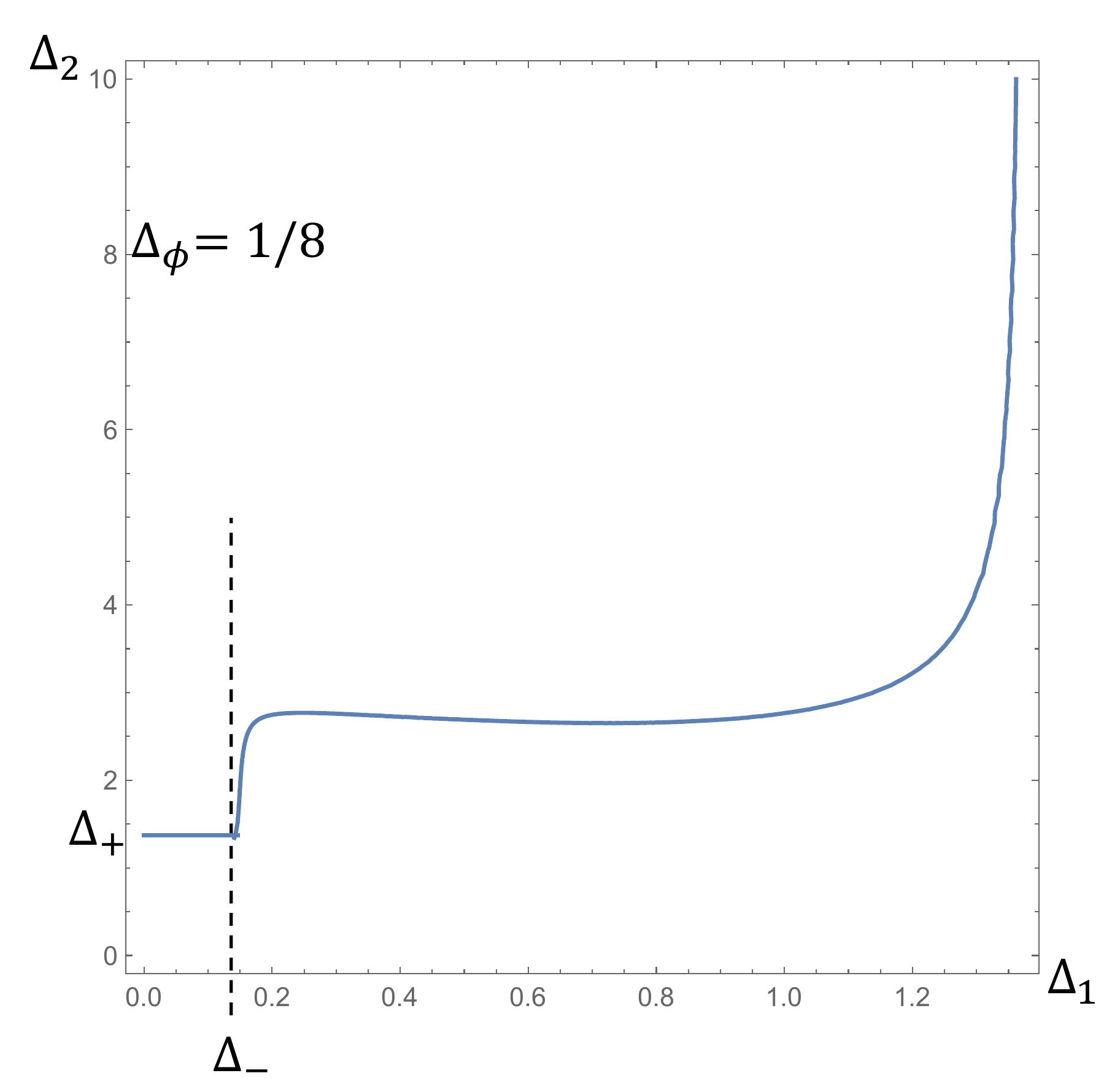}
\caption{Constraints on the first two primary operators $(\Delta_1,\Delta_2)$. The region below the blue line is allowed.   Before the dashed line, $\Delta_1<\Delta_-$ and hence $\Delta_2$ must be smaller than $\Delta_+$. After the dashed line $\Delta_-<\Delta_1$, then $\Delta_2$ must be below the curve $\langle {\bf X}, 0,\Delta_1,\Delta_2\rangle=0$. Finally for $\Delta_1>\Delta_+$ there are no solutions and is ruled out. }\label{fig:D1D2}
\end{center}
\end{figure}

The criterion for the consistency of a spectrum is that there exists a set of three operators such that (\ref{5dsign}) is satisfied, i.e.\
\eq
\langle{\bf  X}, i ,i{+}1\rangle,\quad \langle{\bf X}, i{+}1,j\rangle,\quad \langle{\bf X},j, i\rangle\quad {\rm have \; the\;same \;sign}\,.
\eqe
This means that $\bf X$ is on the same side of the boundaries $(i,i{+}1)$, $(i{+}1,j)$ and $(j,i)$,  i.e.\ the origin must be in the convex hull of the three block vectors.
In other words, \textit{there have to be
three points on the block curve, forming a triangle that encloses the origin.}
 See Figure \ref{CurveExp1} for an example.
 Note that since $\Delta_+, \Delta_-$ are collinear with the origin,  this criterion automatically requires that there is at least one operator  between $\Delta_+$ and $\Delta_-$, which is the consistency condition from the $N=1$ geometry  in Section \ref{sec:1dgeometry}.

The above criterion imposes global constraints on the spectrum.
Let the dimension of the heaviest operator  between $\Delta_\pm$ be  $\Delta_{max}$. 
Since $\Delta_\pm$ are   on the opposite sides of the origin, we can  draw a line that is tangent to the curve before $\Delta_-$ and passes through the origin. This line  intersects  the curve in between $\Delta_{\pm}$, which we denote as $\Delta_T$, as shown in Figure \ref{CurveExp4}.\footnote{Figure \ref{CurveExp4}, and similarly for Figure \ref{fig:ffeg}, has been deformed from the realistic plot (e.g.\ Figure \ref{CurveExp1}) to make it more visible.}
 Now if $\Delta_{max}>\Delta_T$, then we have the same conclusion as before, either there exists a $\Delta_{max{+}1}$ in between $\Delta_{+}$ and $\Delta_c$, or there is a light operator between $\Delta_a$ and $\Delta_b$. However if  $\Delta_{max}<\Delta_T$, the line between $\Delta_{max}$ and the origin only intersects the block curve above $\Delta_+$, i.e.\ $\Delta_{b,c}$ no longer exists. Thus we conclude that we \textit{must} have $\Delta_{max{+}1}$ above $\Delta_+$ and $\Delta_a$.

Finally we consider the constraints for the first two primaries $\Delta_1$ and $\Delta_2$ in the spectrum. For $\Delta_1 <\Delta_-$, we know that $\Delta_2$ cannot be greater than $\Delta_+$, otherwise there will be no operators between $\Delta_{\pm}$.  Similarly, the region with $\Delta_1>\Delta_+$ is also ruled out by the same argument. For $\Delta_1$  between $\Delta_-$ and $\Delta_+$,  $\Delta_2$ cannot be above the contour plot of $\langle {\bf X}, 0,\Delta_1,\Delta_2\rangle=0$, otherwise there does not exist a triplet of operators $\Delta_i,\Delta_{i+1}, \Delta_j$ satisfying \eqref{Triplets}. Combining all these statements, we carve out the space of consistent 1$d$ CFT for the first two primaries $(\Delta_1,\Delta_2)$ in Figure \ref{fig:D1D2}.

\subsection{Projecting onto $\bf X$}

As before, we start with the geometry projected through the identity 0. The cross section of the four-dimensional polytope by the crossing plane $\bf X$ is a two-dimensional polygon.   The polygon gives a bound on the actual values of the four-point function, parametrized by $F^0,F^2,F^4$ as in \eqref{6dF}.

\begin{figure}
\centering
\includegraphics[width=.7\textwidth]{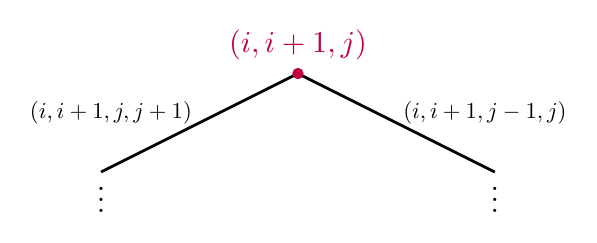}
\caption{The combinatoric rule of the polygon on $\bf X$ inherited from the cyclic geometry. An edge is given by four numbers of the form $(i,i+1,j,j+1)$. Two edges intersect at a vertex of the form $(i,i+1,j)$ if they share three numbers.}\label{fig:rule}
\end{figure}
\begin{figure}[h!]
\centering
\subfloat[]{
\includegraphics[width=.5\textwidth]{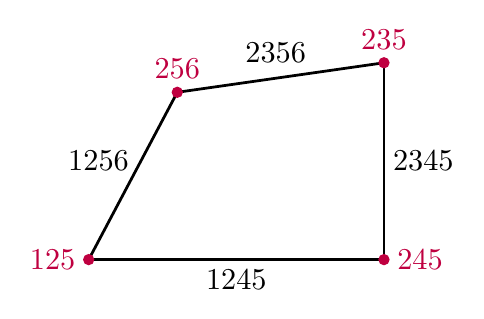}
}
\subfloat[]{
\includegraphics[width=.5\textwidth]{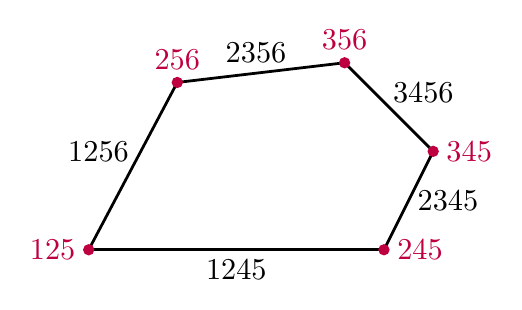}
}
\caption{The projection of the cyclic polytope on the two-dimensional crossing plane $\bf X$ is a polygon.   The edges and vertices of the polygon are constrained by the cyclic geometry \eqref{polygonrule}. Here we show two examples allowed by the combinatorics. }\label{fig:polygoneg}
\end{figure}
The shape of the polygon is determined by the spectrum  through some interesting combinatorics inherited from the cyclic polytope. The edges of the polygon come from the facets of the four-dimensional cyclic polytope, which are of the form $(i,i+1,j,j+1)$.
A vertex $(i,i+1,j)$, on the other hand, is the intersection of two edges who share exactly three numbers.
See Figure \ref{fig:rule} for an illustration of this combinatoric rule.
A triplet $(i,i+1,j)$ is an actual vertex of the polygon if the corresponding facet intersects with the crossing plane, i.e.\ if \eqref{5dsign} is true.
To recap,
\begin{align}\label{polygonrule}
\text{Edges}:~~(i,i+1,j,j+1)\,,~~~~~\text{Vertices}:~~(i,i+1,j)\,.
\end{align}
For example, in Figure \ref{fig:polygoneg} we give two examples of polygons allowed by the above combinatoric rule.

\begin{figure}[h]
\centering
\subfloat[]{\label{fig:triangle1}
\includegraphics[width=.55\textwidth]{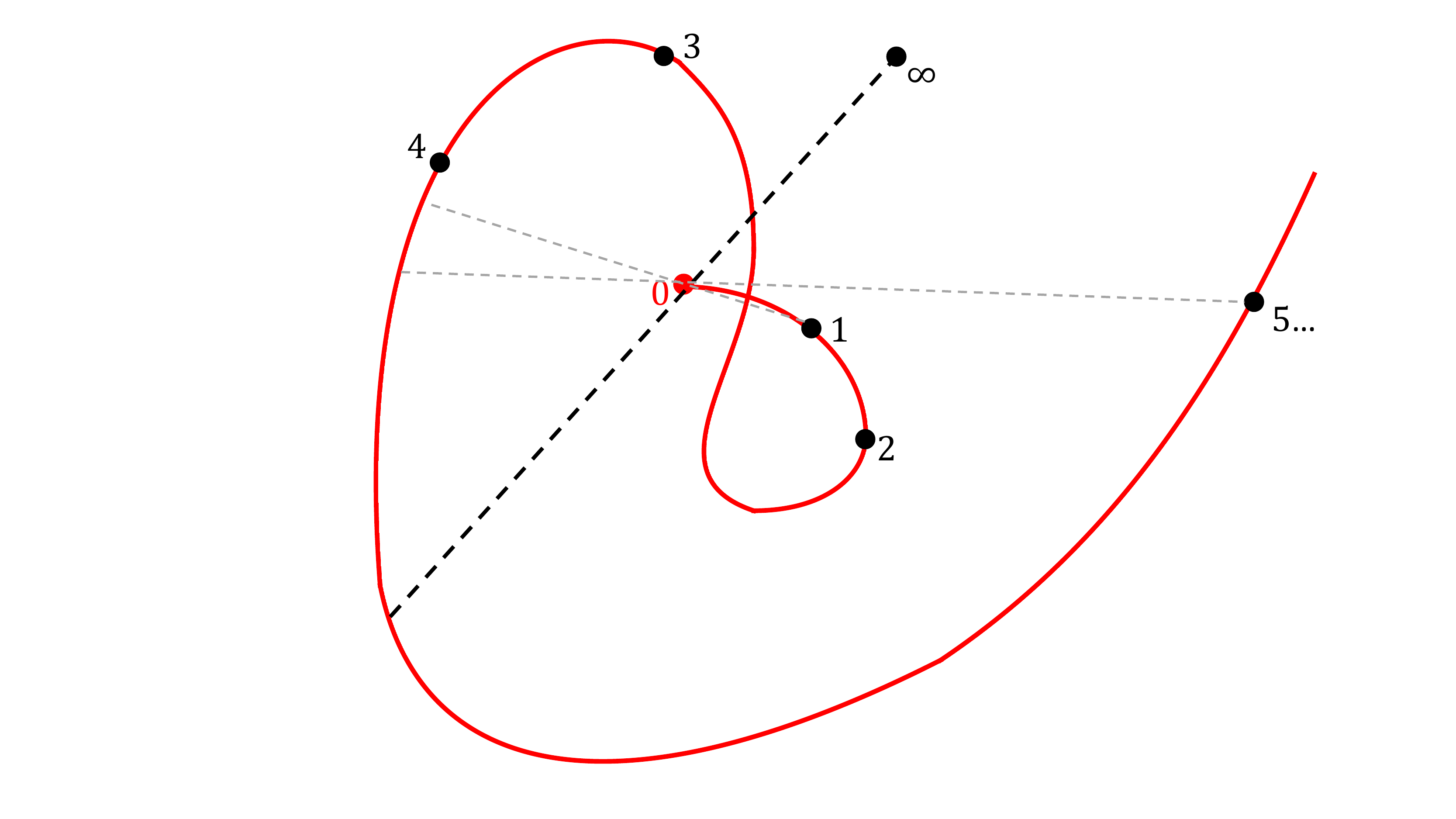}
}
\subfloat[]{\label{fig:triangle2}
\includegraphics[width=.45\textwidth]{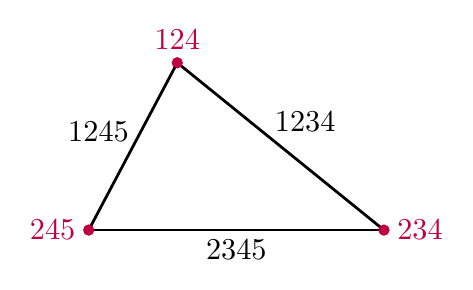}
}
\\
\subfloat[]{\label{fig:pentagon1}
\includegraphics[width=.55\textwidth]{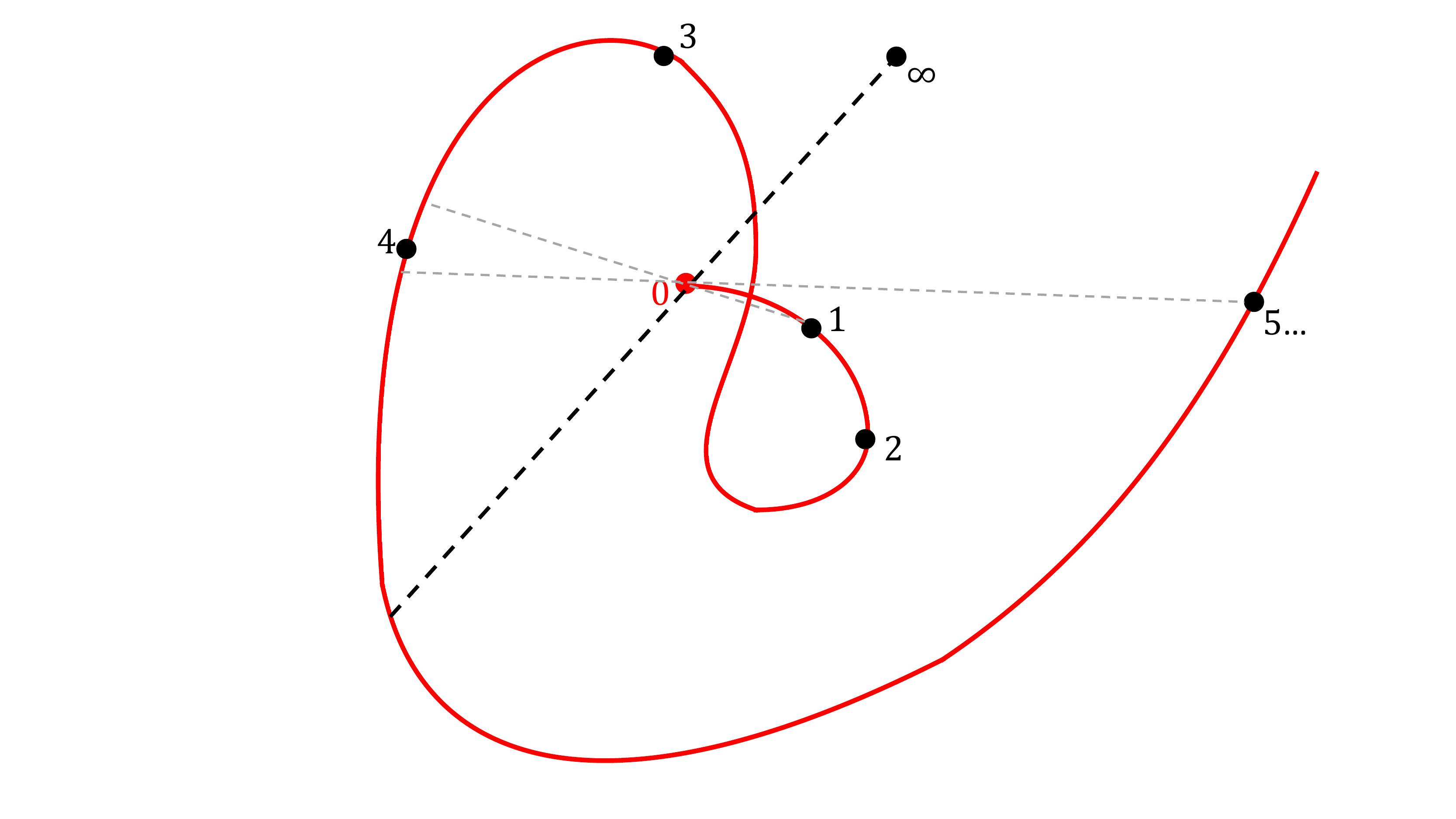}
}
\subfloat[]{\label{fig:pentagon2}
\includegraphics[width=.45\textwidth]{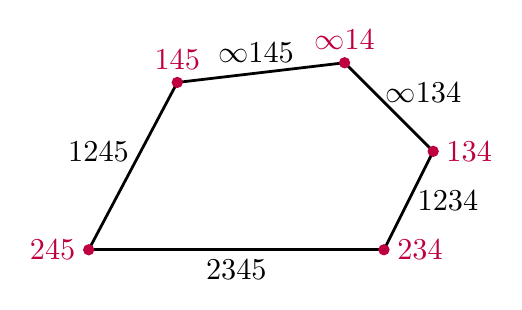}
}
\\
\centering
\subfloat[]{\label{fig:infinigon1}
\includegraphics[width=.55\textwidth]{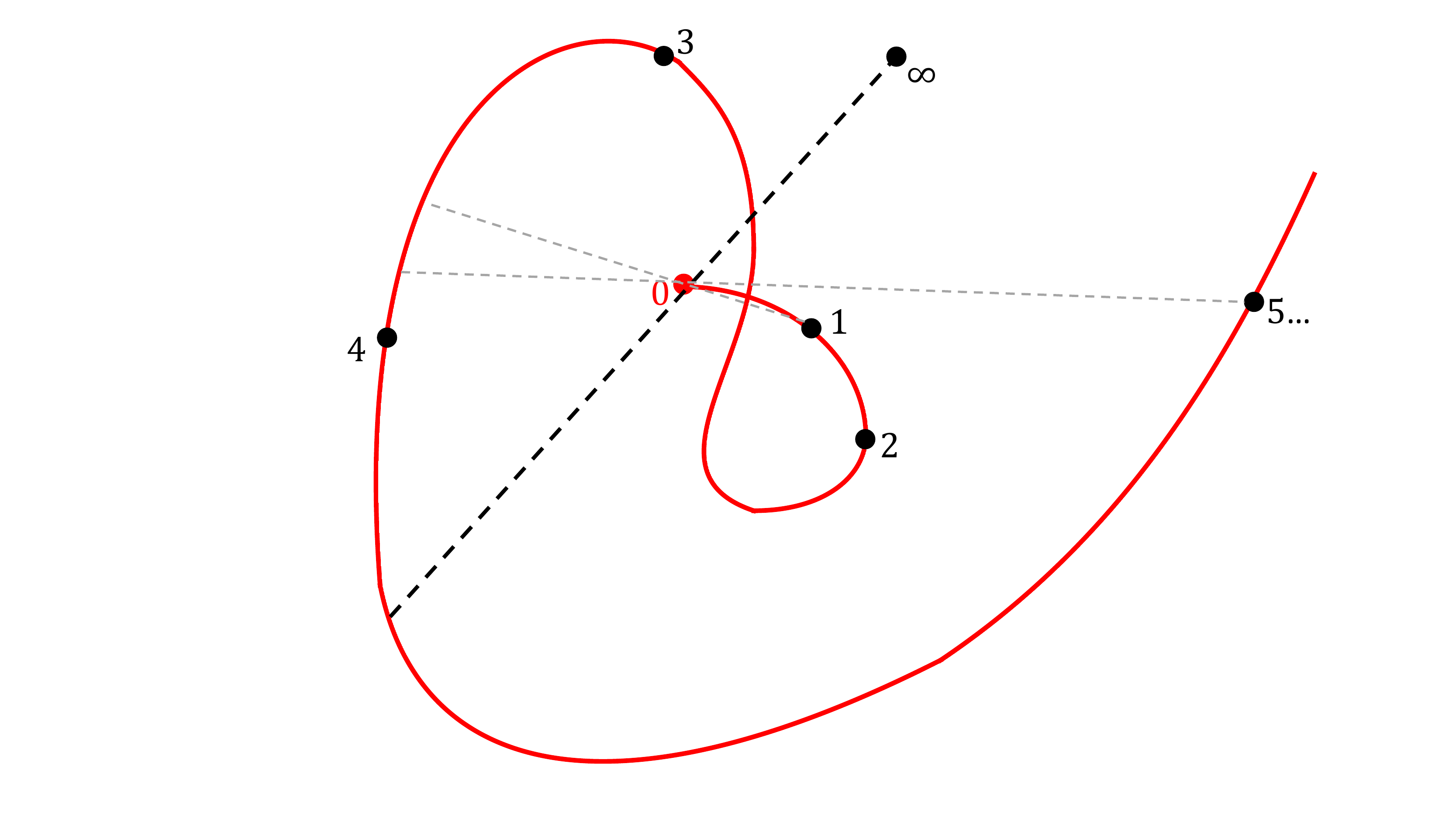}
}
\subfloat[]{\label{fig:infinigon2}
\includegraphics[width=.45\textwidth]{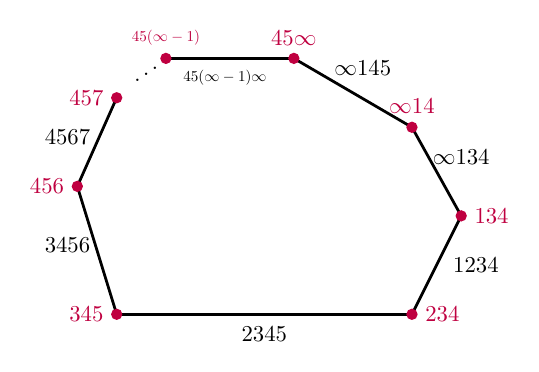}
}
\caption{An illustration of how the polygon on the crossing plane changes its shape as we deform the spectrum.    Left: The polytope projected through the crossing plane and the identity 0 as discussed in Section \ref{sec:thruX}.  The references vectors $v_i$'s are chosen as before.  Right: The polygon from the polytope projected onto the crossing plane $\bf X$, parametrized by $F^2/F^0$ and $F^4/F^0$.  }\label{fig:ffeg}
\end{figure}

We illustrate how the polygon on the crossing plane $\bf X$ changes its shape as we vary the spectrum in Figure \ref{fig:ffeg}.  The  polytopes projected through $\bf X$ and the identity $0$ are shown on the left.   The red curve on the left is the trajectory of the block vector \eqref{5dbv} as we vary $\Delta$. The number $i$  on the red curve label where the block vector for the $i$-th operator of dimension $\Delta_i$ is located at.  In the figure we show the first 5 operators but we assume the spectrum continues to infinity.   As we vary the dimension $\Delta_4$ of the fourth operator, we see that the polygon changes from a triangle to a polygon, and then to a polygon with infinitely many edges. 
To make the plot more visible, Figure \ref{fig:ffeg}  is not drawn to scale.

\begin{figure}[h!]
\centering
\includegraphics[width=.6\textwidth]{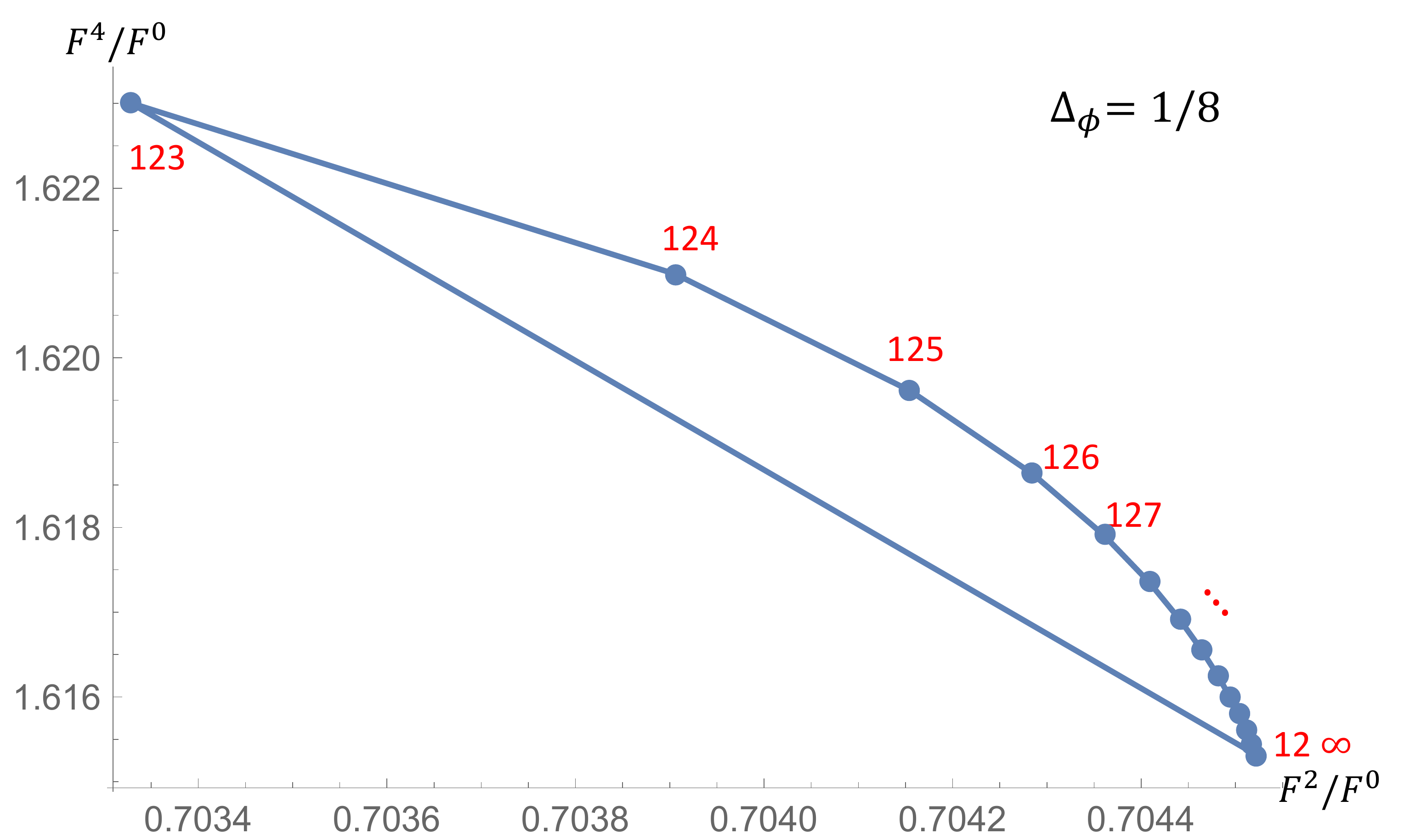}
\caption{The cyclic polytope projected onto the crossing plane for the free fermion spectrum $\Delta_i  =2\Delta_\phi+ 2i-1$ ($i=1,2,\cdots$). The dimension $\Delta_\phi$ of the external operator (the fermion field) is shown in the plot.}\label{fig:freefermion}
\end{figure}

As a concrete example, the polygon on the crossing plane for the generalized free fermion is shown in Figure \ref{fig:freefermion}.   The region inside the polygon, which is amazingly thin and tiny, is the allowed four-point function constrained by the bootstrap equation.

\section{Recursive Increase of Resolution}

We have completely investigated the bootstrap geometry problem up to the $N=2$ case.
This is the case where the crossing plane $\bf X$ is two-dimensional in $\mathbb{P}^5$, or stated non-projectively, three-dimensional in six dimensions.
We have seen how, as we go to the higher-dimensional problem, our
``resolution" on CFT data increases, and we further carve out the space of
allowed operator dimensions. In this section we discuss how this recursive
procedure increases resolution more systematically.

As before, it is natural to work with a $(2N + 2)$ dimensional space of
Taylor coefficients for the four-point function.  The crossing plane ${\bf
X}$ is a half-dimensional $(N+1)$-plane in $(2N+2)$ dimensions. We project
through the $\bf X$ plane as well as the identity block vector ${\bf G}_0$ to end
up with a geometry in $(2N+2) - (N+1) - 1 = N$ dimensions. As we have
discussed, the CFT data $\left\{\Delta\right\}$ is consistent if and only if
there are $(N+1)$ $\Delta_i$ such that the simplex with vertices $\left
\{\Delta_{i_1}, \cdots, \Delta_{i_{N+1}} \right \}$ contains the origin.

Now, we would like to systematically relate the solution of this problem with
some $N$, to the one where we increase $N \to (N+1)$. We have already
highlighted an obvious geometric relation between the two problems: we know
that if we project the $(N+1)$-dimensional geometry through the infinity
vertex ${\bf G}_\infty$, then we go back to the $N$
dimensional problem. This motivates us to incorporate the infinity block into
our thinking about the geometry problem at hand.

Indeed there is a very natural way to do this, using the following trivial
but important piece of basic linear algebra and geometry. Suppose ${\bf v}_1,
\cdots, {\bf v}_{D+1}$ are $(D+1)$ vectors in $D$ dimensions, such that the
origin is contained inside the vertices of the simplex defined by them, i.e.\,
such that $c_1 {\bf v}_1 + \cdots c_{D+1} {\bf v}_{D+1} = 0$ for some $c_i >
0$. Then the claim is that for an arbitrary vector ${\bf w}$, there is some
collection of $D$ of the $(D+1)$ ${\bf v}$s, such that the a positive linear
combination of $\left\{{\bf v}_{i_1}, \cdots, {\bf v}_{i_D}, {\bf w} \right
\}$ {\it also} contains the origin.

The proof is trivial. By  $GL(D)$ transformations and positive rescalings, we can always bring the vectors
${\bf v}_1, \cdots, {\bf v}_{D+1}$ to the form ${\bf v}_i = {\bf e}_i$ for
$i=1, \cdots, D$, and ${\bf v}_{D+1} = - ({\bf e}_1 + \cdots  + {\bf e}_D)$.
Here the ${\bf e}_i$ are unit vectors  with non-vanishing components in the
$i$-th slot, and we have ${\bf v}_1 + \cdots {\bf v}_D + {\bf v}_{D+1} = 0$.
Now, we can expand any ${\bf w}$ in the basis of ${\bf e}_i$ as ${\bf w} =
w_1 {\bf e}_1 + \cdots w_D {\bf e}_D$, and let us consider the equation $c_1
{\bf v}_1 + \cdots c_D {\bf v}_D + c_{D+1} {\bf v}_{D+1} + {\bf w} = 0$,
which implies $c_i = c_{D+1} -w_i $. If all the $w_i$ are negative, we simply put $c_{D+1} = 0$ and $c_i = - w_i
> 0$. If instead some of the $w_i$ are positive, let $w_{i_*}$ be the largest
of all the $w$'s. Then we can set $c_{D+1} = w_*$, and $c_{i_*}=0$, then all
the other $c_i$ are $c_i = c_{D+1} - w_i = w_{i_*} - w_i > 0$. Thus in all
cases, we have found some positive $c_i$ for which $c_{i_1} {\bf v}_{i_1} +
\cdots c_{i_D} {\bf v}_{i_D} + {\bf w} = 0$, as claimed.

With this simple fact in hand, let us return to our CFT geometry problem, and
observe that being able to find $(N+1)$ vectors $\left\{\Delta_{i_1}, \cdots,
\Delta_{i_{N+1}} \right\}$ containing the origin, is completely equivalent to
demanding that there are just $N$ $\Delta$'s which, together with infinity,
contain the origin. On the one hand, if $\left\{\Delta_{i_1}, \cdots,
\Delta_{i_N}, \infty \right\}$ contains the origin, we have good CFT data
(with $\Delta_{i_{N+1}} = \infty)$. On the other hand, we have just seen that
for {\it any} set of $(N+1) \Delta$'s containing the origin, there are $N$ of
them for which the set $\left\{ \Delta_{i_1}, \cdots \Delta_{i_N}, \infty
\right \}$ contains the origin.

Thus, the set of legal CFT data in the $N$ dimensional geometry is associated
with solving for the space $S_N$ of $\left \{\Delta_1, \cdots, \Delta_N
\right \}$ such that $\left \{\Delta_1, \cdots, \Delta_N, \infty \right \}$
contains the origin; we must then demand that the CFT spectrum contains $N$
operators with dimensions inside this set $S_N$.

Let us choose some particular $\left\{\Delta_1, \cdots, \Delta_N \right \}$
inside $S_N$. Define $T_N[\Delta_1, \cdots, \Delta_N]$ to be the set
of all $\Delta_{N+1}$'s such that $\left \{\Delta_1, \cdots,
\Delta_N,\Delta_{N+1} \right \}$ contains the origin. Obviously $T_N$ is not
empty since by definition as $\Delta_{N+1} \to \infty$, $\left \{\Delta_1,
\cdots, \Delta_N, \infty \right \}$ contains the origin. Note that the union
of $S_N$ and $T_N$ is a way of labelling {\it every} possible simplex that
contains the origin, with $N$ $\Delta$'s such that $\left \{\Delta_1, \cdots,
\Delta_N, \infty \right \}$ contains the origin, and the possible ranges for
the $\Delta_{N+1}$ that could along with these $N$ $\Delta$'s.

Now, let's go up to $(N+1)$ dimensions. We are now interested in the space $S_{N+1}$, i.e.\ we are looking for some $\left \{
\Delta^\prime_1, \cdots, \Delta^\prime_N, \Delta^\prime_{N+1}, \infty \right
\}$ such that
\begin{align}
\begin{split}
&\langle {\bf X}0 \Delta_1^\prime \cdots \Delta_N^\prime
\Delta^\prime_{N+1} \rangle > 0\,,\\
& \langle {\bf X }0
\Delta^\prime_1 \,\cdots \Delta^\prime_N \infty \rangle < 0\,, \cdots, \langle
{\bf X }0 \Delta_2^\prime \cdots \Delta^\prime_{N+1} \infty \rangle > 0\,.
\end{split}
\end{align}
 Note that the conditions in the second line simply demand that $\left \{\Delta^\prime_1, \cdots,
\Delta^\prime_{N+1} \right \}$ is a legal simplex in the $N$-dimensional
problem, since projecting through ${\bf G}_\infty$ sends us back to the $N$ dimensional geometry.

Thus, we learn that $\left\{\Delta^\prime_1, \cdots, \Delta^\prime_{N+1}
\right \}$ must actually be in $S_N \cup T_N$. But importantly, we have one more
constraint on  $\Delta^\prime_{N+1}$!  We must have that $\langle {\bf X }0
\Delta^\prime_1 \cdots, \Delta^\prime_N \Delta^\prime_{N+1} \rangle$ has the
opposite sign as $\langle {\bf X } 0\Delta^\prime_1 \cdots \Delta^\prime_N
\infty \rangle$. This is crucial. Recall that $T_N$ was always non-empty
since $\Delta_{N+1}$ could always go to $\infty$. But the above
condition forbids this. Indeed, $\Delta^\prime_{N+1}$ must be smaller than
the largest root in the variable $\Delta$ of the function $f[\Delta;\Delta_1,
\cdots, \Delta_N] = \langle {\bf X} 0\Delta_1 \cdots \Delta_N \Delta
\rangle$.

So, let us finally define $U_N[\Delta_1, \cdots, \Delta_N]$ to be the set of
all $\Delta$'s such that $\langle {\bf X}0 \Delta_1 \cdots \Delta_N \Delta
\rangle$ has the opposite sign as $\langle {\bf X}0 \Delta_1 \cdots,
\Delta_N \infty \rangle$. Again, this set $U_N$ is manifestly bounded from
above. The recursive expression relating the set of legal CFT data in $(N+1)$
dimensions to the set in $N$ dimensions:
\begin{equation}
S_{N+1} = S_N \cup \left[T_N \cap U_N \right]
\end{equation}
Note that $T_N \cap U_N$ may be empty, in which case we are ruling out parts
of $S_N$ that were consistent in $N$ dimensions but which are seen to be
inconsistent with the increased resolution of the $(N+1)$-dimensional
problem.

In this way, we can recursively build up the space of allowed $\Delta$'s. In
$N$ dimensions we must have some $\left \{ \Delta_1, \cdots, \Delta_N \right
\}$ in some specified ranges. Given any allowed $\left \{\Delta_1, \cdots,
\Delta_N \right \}$ in $N$ dimensions, there is some {\it manifestly bounded
from above}, and possible empty,  range for $\Delta_{N+1}$'s. This manifests
the way in which the space of possible operator dimensions is further carved
out as we systematically step to higher and higher dimensions.

\section{Outlook}

Our investigations in this paper have only scratched the surface of what
appears to be a rich set of connections between the geometry and
combinatorics of total positivity and the conformal bootstrap program. Even
sticking with the $d=1$ CFT geometry associated with the $SL(2,\mathbb{R})$ conformal
blocks, there are a  number of open avenues for future investigation.

To begin with, we would like to have a much better understanding of the
positivity we have experimentally observed for conformal blocks. In
particular, we would like to again draw attention to the extraordinary
``fine-tuning" that arises seemingly out of the platonic thin-air of
conformal blocks: minors of block vectors are {\it almost} always positive,
except when $7$  or more dimensions are smaller than $\sim 10^{-1}$, and the
most negative the minors ever become is about $\sim 10^{-20}$! It would be
fascinating to better understand this phenomenon better. And might there be a
slightly different basis of conformal blocks for which the positivity is
exact?

The exploration of total positivity properties involving hypergeometric
functions is also interesting from a purely mathematical point of view. The
conformal blocks \eqref{block} are $G_\Delta(z) = z^\Delta\, _2F_1(\Delta, \Delta, 2
\Delta; z)$. Of course, as we discussed the class of function $z^\Delta$
enjoys our total positivity properties. Interestingly, the
functions$\,_2F_1(\Delta, \Delta, 2 \Delta; z)$ {\it also} appear to be
totally positive; indeed experimentally this appears to be the case for all
hypergeometric functions $_2F_1(a,b,c=a+b,z)$ with $a,b>0$, a fact not studied to our knowledge in the mathematical literature.
It would be
interesting to find a conceptual proof for this statement, which might give
some inroads into understanding some exact statement about positivity for the
blocks.

Turning to the classification of CFT data, is it possible to more explicitly
carry out the recursive procedure for increasing resolution as more Taylor
coefficients are kept? Also, the intersection of the unitarity polytope and
the crossing plane is in general an interesting polytope, whose facet
structure is controlled by the combinatorics of cyclic polytopes, as we
illustrated in our low-dimensional examples. But we worked in the simplest
cases of one and two-dimensional geometries where convex polytopes have a
universal shape--line segments and convex polygons--while higher-dimensional
polytopes are much more interesting and intricate. Is it possible to
characterize all the polytope shapes that can arise in the ${\bf U} \cap {\bf
X}$ problem in a nice combinatorial way?

Perhaps the most interesting set of questions from both the physical and
positive-geometrical points of view have to do with understanding of the
bootstrap problem with higher-dimensional conformal symmetry. As we have
seen, the most obvious notions of positivity are associated with some
ordering: we have block vectors $G_{\Delta}(z)$, and both the ``parameter"
$\Delta$ and the one-dimensional space labelled by $z$ have a natural
ordering. The situation is more interesting in higher dimensions, where we
have blocks that depend on two variables $(z, \bar{z})$, and two parameters
$(\Delta,s)$ for dimension and spin. Is there a natural extension of
positivity involving this pair of two-dimensional spaces $(z, \bar{z})$ and
$(\Delta, s)$ in a non-trivial way? We close by making some preliminary and
elementary observations on this problem, beginning with the bootstrap for
$d=2$ CFTs.

In two dimensions operators can be labelled by $h, \bar{h}$ labelled to spin
and dimension by $s = h - \bar{h}\in\mathbb{Z}$ and $\Delta = h + \bar{h}$.
The $d=2$ global conformal blocks are simply given as the product of $d=1$ blocks
via $G_{\Delta, s}(z, \bar{z}) = G_h (z) G_{\bar{h}}(\bar{z})$. The
four-point function can then be written as
\begin{equation}
F(z, \bar{z}) = \sum_{h, \bar{h}} p_{h, \bar{h}} G_h(z) G_{\bar{h}}(\bar{z})\,, \, ~~~\, p_{h,\bar{h}} \geq 0\,,~~~h - \bar{h} = s\in \mathbb{Z}\,.
\end{equation}
Note that the non-trivial correlation between $h, \bar{h}$ is enforced by the
requirement that the spin $h -
\bar{h} = s$ is an integer.

But we can begin with a simpler problem: we consider the same expansion, but
relax the requirement on the spin. Said another way, in the real 2$d$
CFT we can think of the sum over all $(h,\bar{h})$ but with $p_{h, \bar{h}} =
0$ unless $h - \bar{h}=s$ is integer, but we can consider the looser
restriction placing no conditions other than the positivity of
$p_{h,\bar{h}}$. Clearly, it is natural to call this the ``direct product" of
the geometries associated with the $z, \bar{z}$ problems in $d=1$.

The notion of the ``direct product" of polytopes is a natural one, though not
much studied in the mathematical literature.
Some general properties can be
easily summarized. Let us consider vertices $X_a^i$ and $Y_A^I$ of two
projective polytopes $P,Q$ in $(N+1)$ and $(M+1)$ dimensions. Will be define
the direct product polytope $(P \times Q)$, living in the $(N+1) \times
(M+1)$ direct product space, to be the convex hull of the direct product of
the vertices, as $F^{iI} = \sum_{a,A} p_{aA} (X_a^i Y_A^I)$.   Now, suppose the
facets of $P$ are $w_i$ and those of $Q$ are $W_I$. What can we say about the
faces of $P \times Q$?

Obviously, $(w_i W_I)$ bound $(P \times Q)$, since trivially $(w_i W_I) F^{i
I} \geq 0$.
Moreover, $(w_i W_I)$ is actually guaranteed to
be one of the faces of the $(P \times Q)$. The reason is that we can find
more than $(N+1) \times (M+1)$ vertices of $(P \times Q)$ that lie on $(w_i
W_I)$.
Indeed,  there are at least $(N+1)$ different $X_a^i$ that lie on $w_i$ and at
least $(M+1)$ different $Y_A^I$ that lie on $W_I$. So at least these $(N+1)
\times (M+1)$ vertices $X_{a_i}^i Y_{A_I}^I$ lie on $(w_i W_I)$.
Now since $(w_iW_I)$ bounds $P\times Q$ and has at least $(N+1)\times (M+1)$ vertices, it must be a face of $(P
\times Q)$.

In general, there may be further faces of $(P \times Q)$ that are not of this
form. But with a little extra work, it is possible to prove the following
nice fact. A {\it simplicial} polytope is one where the neighborhood of every
vertex locally looks like a simplex, with the smallest number of faces
meeting at the vertex. Then, if $P,Q$ are both simplicial, {\it all} the
faces of $P \times Q$ are the direct products $w_i W_I$ of the faces of
$P,Q$. Cyclic polytopes are simplicial, so we do know the face structure of
the loosened geometry problem for 2$d$ CFTs, where we relax the condition
$p_{h, \bar{h}} = 0$ for $h - \bar{h}$ not equal an integer.

It is interesting to see how the real CFT polytope sits inside the direct
product polytope. Let the conformal weights of the CFT be $(h_i,\bar{h}_i)$ with $i$ labeling different global conformal primaries.
Then, the vertices of the direct product polytope are
$G_{h_i}(z)
G_{h_j}(\bar{z})$ for all $i,j$,
 but for the CFT
polytope we are only keeping the diagonal vertices $G_{h_i}(z)
G_{\bar{h}_i}(\bar{z})$.

Now, of course the faces of the direct product polytope still bound the CFT
polytope we are interested in. But in general, there aren't enough vertices
of the CFT polytope on these faces for them to correspond to faces of the CFT
polytope.
In practice, as we have seen the cyclic polytope constraints are already very
restrictive, so even these ``looser" direct product polytopes, that bound the
real CFT polytopes,are likely highly constraining on CFT data.

This is likely the most we can say directly following from the ideas in this
paper, without addressing the bigger challenge of determining the positive
geometry intrinsic to general CFTs in higher dimensions. 
Another application of the positive geometry is to the modular bootstrap of the torus partition function in two-dimensional CFT.  There the analogs of the block vectors are built from the Virasoro characters, which after a $GL$ transform  literally lie on a moment curve.  Again in this case all the techniques presented in the current paper can be directly applied.  
We leave the direct
exploration of these fascinating problems to future work.

 \section*{Acknowledgements}

We are grateful to  J. Kaplan, P. Galashin,  Y.-H. Lin, R. Mahajan, D. Stanford for enlightening discussions.
N.A-H. is  supported by DOE grant de-sc0009988. Y-t Huang is supported by  MoST Grant No. 106-2628-M-002-012-MY3.
S.H.S. is supported by the National Science Foundation grant PHY-1606531 and the Roger Dashen Membership.

\appendix

\section{Scalar Conformal Blocks in General Dimensions}\label{app:block}

In this appendix we give a self-contained  derivation of conformal blocks in general spacetime dimension, focusing on the case when all the operators are scalars.  We will review a Lorentzian integral formula for  conformal blocks \cite{Polyakov:1974gs,Czech:2016xec,Kravchuk:2018htv} and give a closed form expression in even dimensions.  For simplicity, we will focus on the case when the four external operators are all scalars and have the same dimension $\Delta_\phi$. We will also assume that the intermediate operator is a scalar  with dimension $\Delta$.  This conformal block will be denoted as $G_{\Delta,0}(z,\bar z)$ where the cross ratios are defined as
\begin{align}
z\bar z = {x_{12}^2 x_{34}^2 \over x_{13}^2 x_{24}^2 }\,,~~~~(1-z)(1-\bar z) = {x_{14}^2 x_{23}^2 \over x_{13}^2 x_{24}^2 }\,.
\end{align}

The defining properties of of the conformal blocks are (1) It is conformal invariant. (2) It is an eigenvector of the quadratic Casimir of the conformal group. (3) For small $z,\bar z$, $G_{\Delta,0}(z,\bar z)$ goes like $(z\bar  z)^{\Delta/2}$.  If we only impose the first two conditions, then the function $\Psi_{\Delta,0}$, defined via the following Euclidean integral
\begin{align}\label{EuclideanInt}
{1\over |x_{12}|^{2\Delta_\phi} |x_{34}|^{2\Delta_\phi}} \Psi_{\Delta,0}(z,\bar z) = {1\over |x_{12}|^{2\Delta_\phi -\Delta} |x_{34}|^{2\Delta_\phi-(d-\Delta)}}\int_{\mathbb{R}^d} d^d x_5 { 1\over |x_{15}|^\Delta  |x_{25}|^\Delta |x_{35}|^{d-\Delta} |x_{45}|^{d-\Delta}}
\end{align}
will do the job.   Here $|\cdot|$ is the Euclidean norm and the integral in $x_5$ is over the whole $\mathbb{R}^d$. This is usually known as the shadow representation of the conformal partial wave $\Psi_{\Delta,0}(z,\bar z)$. The latter is not quite the conformal block $G_{\Delta,0}(z,\bar z)$.  Rather, $\Psi_{\Delta,0}(z,\bar z)$ is a linear combination of $G_{\Delta,0}$ and $G_{d-\Delta,0}$.   The two terms have different monodromy around $z=\bar z=0$, which can be used to extract the conformal block \cite{SimmonsDuffin:2012uy}.

\begin{figure}
\centering
\includegraphics[width=.6\textwidth]{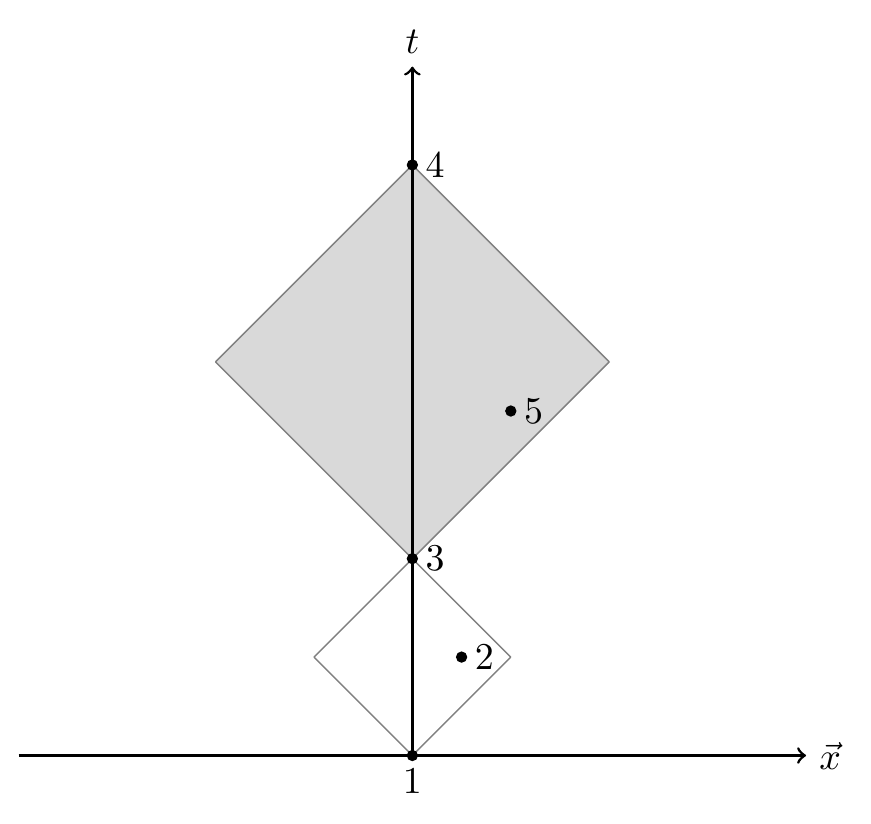}
\caption{The gray area is the integration region for  $x_5^\mu$ in the Lorentzian integral formula \eqref{LorentzianInt} for the conformal block.}\label{fig:causal}
\end{figure}

There is a simple Lorentzian variation of \eqref{EuclideanInt} that will directly yield a formula that obeys all three properties.   Let  $x_i^\mu \in \mathbb{R}^{1,d-1}$ with $i=1,\cdots ,4$.  We will choose  a particular causal structure such that $x^\mu_{i+1}$ is in the future lightcone of $x_i^\mu$ for $i=1,2,3$ (see Figure \ref{fig:causal}).
  The conformal block then admits the following Lorentzian integral representation \cite{Polyakov:1974gs,Czech:2016xec,Kravchuk:2018htv}
\begin{align}\label{LorentzianInt}
G_{\Delta,0}(z,\bar z) = \mathcal{N}{ |x_{12}|^{\Delta} \over|x_{34}|^{-d+\Delta}}\int_{x_3<x_5<x_4} d^d x_5 { 1\over |x_{15}|^\Delta  |x_{25}|^\Delta |x_{35}|^{d-\Delta} |x_{45}|^{d-\Delta}}\,,
\end{align}
where we only integrate $x_5$ in the intersection between the future lightcone of $x_3$ and the past lightcone of $x_4$, which is a conformally invariant region in Lorentzian signature.  Here $|\cdot|$ is the Lorentzian norm in $\mathbb{R}^{1,d-1}$.  $\mathcal{N}$ is a normalization constant to be fixed later.  Furthermore, the integral is non-singular as we take $x_1\to x_2$, hence the righthand side above reproduces the correct small $z,\bar z$ behavior.

In the following we will explicitly evaluate the Lorentzian integral in one dimension and even spacetime dimensions.  By conformal invariance, we can choose the four external points to be at $x_1 = (t=0,\vec x=0)$, $x_3 = (t=1, \vec x=0)$, $x_4=(t=\infty, \vec x=0)$ with  $x_2$  in the intersection between the future lightcone of $x_1$ and the past lightcone of $x_3$.   The causal structure is shown in Figure \ref{fig:causal}.

\subsection{One Dimension}

In one dimension, let $\chi \equiv x_5-x_3= x_5 - 1$.  The Lorentzian integral reduces to
\begin{align}
G_\Delta(z) =  {\cal N} z^\Delta \int_0^\infty d\chi {1\over (\chi+1)^\Delta (\chi +1-z)^\Delta \chi^{1-\Delta}}\,.
\end{align}
Using the integral representation of  the Gauss hypergeometric function
\begin{align}\label{2F1}
\,_2F_1(a,b,c,z)  = {\Gamma(c) \over \Gamma(b)\Gamma(c-b) }
\int_0^\infty du  {1\over u^{b-c+1} (u+1)^{c-a} (u+1-z)^a}\,,
\end{align}
we obtain the 1$d$ conformal block \eqref{block} used throughout the main text of the paper:
\begin{align}
G_\Delta(z) =  z^\Delta\,_2F_1(\Delta/2,\Delta/2,\Delta, z)\,.
\end{align}
Here we have chosen $\cal N$ to normalize the block to start from $z^\Delta$ in the small $z$ expansion.

\subsection{Two Dimensions}

Let us parametrize $x_5$ as $x_5^\mu=  ( t =\tau+1,x^1=  \rho)$ and  $x_2$ as $x_2^\mu  =  (t=T,  x^1 = X)$.  The cross ratios are $z=T+X$ and $\bar z=T-X$. In  two dimensions the Lorentzian integral takes the form
\begin{align}
&G_{\Delta,0}(z,\bar z) = \mathcal{N}  (z\bar z)^{\Delta\over2}\int_0^\infty d\tau\,   \int_{0}^\tau  d\rho \,\left[
 { 1\over   [ (\tau+1)^2  -\rho^2 ]^{\Delta \over2}
  [ (\tau+1 -T)^2   - ( \rho-X)^2]^{\Delta\over2}
  [ \tau^2 - \rho^2 ]^{d-\Delta\over2} }
  \right.\notag\\
  &\left.+
   { 1\over   [ (\tau+1)^2  -\rho^2 ]^{\Delta \over2}
  [ (\tau+1 -T)^2   - ( \rho+X)^2]^{\Delta\over2}
  [ \tau^2 - \rho^2 ]^{d-\Delta\over2} }
  \right]\,.
\end{align}
Let us define $u=\tau +\rho $ and $v=\tau-\rho$.  We can then write
\begin{align}
&G_{\Delta,0}(z,\bar z) ={ \mathcal{N} \over2} (z\bar z)^{\Delta\over2}
\int_0^\infty du\,   \int_0^udv \,\left[
 { 1\over   [ (u+1)(v+1) ]^{\Delta \over2}
  [ (u+1 -z)(v + 1 -\bar z))]^{\Delta\over2}
(  uv)^{d-\Delta\over2} }\right.\notag\\
&\left.
+{ 1\over   [ (u+1)(v+1) ]^{\Delta \over2}
  [ (u+1 -\bar z)(v + 1 -z))]^{\Delta\over2}
(  uv)^{d-\Delta\over2} }
\right]\,.
\end{align}
Since the integrand is symmetric in exchanging $u$ with $v$, we can replace the integration range as $0<u<\infty$ and $0<v<\infty$ at the price of a factor $1/2$.  It follows that the the integral factorizes to products of terms like \eqref{2F1}, and we reproduce the standard scalar $SL(2,\mathbb{C})$ conformal block in two dimensions \cite{Ferrara:1974ny,Dolan:2000ut,Dolan:2003hv,Dolan:2011dv}:
\begin{align}
G_{\Delta,0} (z,\bar z) = (z\bar z)^{\Delta\over2} \,_2F_1(\Delta/2,\Delta/2, \Delta,z)\,_2F_1(\Delta/2,\Delta/2,\Delta,\bar z)\,,
\end{align}
where we have chosen the normalization constant $\cal N$ such that the leading term is $(z\bar z)^{\Delta\over2}$ in the small $z,\bar z$ expansion.

\subsection{Even Dimensions}

Let us parametrize the integration point  $x_5$ as $x_{53}^t =\tau$ and $|\vec x_{53}|=\rho$.  The second external point  $x_2$ will be parametrized as $x_2^\mu  =  (t=T,  x^1 = X, x^\perp=0)$.  The cross ratios are $z=T+X$ and $\bar z = T-X$.  The Lorentzian formula  \eqref{LorentzianInt} is then
\begin{align}
&G_{\Delta,0}(z,\bar z) = \mathcal{N}V_{d-1}  (z\bar z)^{\Delta\over2} \int_0^\infty d\tau\,   \int_0^\tau \rho^{d-2} d\rho \int_{0}^\pi d\theta\sin^{d-3}\theta \,\notag\\
&\times
 { 1\over   [ (\tau+1)^2  -\rho^2 ]^{\Delta \over2}
  [ (\tau+1 -T)^2   - ( \rho^2  -2\rho X\cos\theta+X^2 )]^{\Delta\over2}
  [ \tau^2 - \rho^2 ]^{d-\Delta\over2} }\,,
\end{align}
where $V_D={2\pi ^{D+1\over2} \over \Gamma({D+1\over2})}$ is the volume of a $D$-dimensional sphere.
Let us first focus on the $\theta$ integral:
\begin{align}
& \int_{0}^\pi d\theta\sin^{d-3}\theta
 {1\over
   [ (\tau+1 -T)^2   - ( \rho^2  -2\rho X\cos\theta+X^2 )]^{\Delta\over2}}
  = \sum_{n=0}^{d-4\over2}  (-1)^n{{d-4\over2} \choose n} I_{2n}({\Delta/2})\,,
   \end{align}
where we have defined $u=\tau+\rho$ and $v=\tau -\rho$ and
\begin{align}
I_m(\alpha)\equiv   \int_{-1}^1 dy  { y^{m}
   \over
   [  (\tau+1 -T)^2   - ( \rho^2  -2\rho Xy+X^2 )]^{\alpha}
   }\,.
\end{align}

Let us compute the integral $I_m(\alpha)$. First we note that
\begin{align}
 \int_{-1}^1dy {1 \over (y +D)^\alpha} = {1\over \alpha-1}
 \left[ {1\over (D-1)^{\alpha-1}  }  -{1\over (D+1)^{\alpha-1}}\right]  \,.
\end{align}
It follows that
\begin{align}
& \int_{-1}^1dy {y^m \over (y +D)^\alpha}
 = \int_{-1}^1dy {( y+D-D)^m \over (y +D)^\alpha}
 =\sum_{k=0}^m  (-1)^k {m \choose k}D^k  \int_{-1}^1 dy \,
 {1\over (y+D)^{\alpha-m+k}}\notag\\
 =&
 \sum_{k=0}^m  (-1)^k {m \choose k}  {D^k\over \alpha- m+k-1}
 \left[
 {1\over (D-1)^{\alpha-m+k-1} }  - {1\over (D+1)^{\alpha-m+k-1} }
 \right]
  \,.
\end{align}

Back to $I_m(\alpha)$, we then have
\begin{align}
I_m(\alpha) =
&{1\over (2\rho X)^{ m+1}}
\sum_{k=0}^m(-1)^k {m\choose k}{1\over \alpha- m +k-1}
\left[ (\tau+1-T)^2 -(\rho^2 +X^2) \right]^k   \notag\\
&\times
\left[
{1\over [  (u -\bar z +1)(v- z +1)]^{\alpha-m+k-1}  }
-{1\over [  (u -z +1)(v-\bar z +1)]^{\alpha-m+k-1}  }
\right]
\end{align}
We would like to write the integrand as a sum of powers of  $(u-\bar z+1)$ and $(v-z+1)$, so that the final integral can be done using \eqref{2F1}.  To achieve this, we rewrite
\begin{align}
&(\tau+1-T)^2 -(\rho^2 +X^2)
 = \frac 12 (u -\bar z +1)(v- z +1)
+\frac 12 (u -z +1)(v-\bar z +1)
\end{align}
We then have
\begin{align}
&I_m(\alpha) = {1\over (2\rho X)^{m+1} }
\sum_{k=0}^m{ (-1)^k\over 2^k} { m \choose k}
{1\over \alpha-m+k-1}\notag\\
&\times
\sum_{\ell=0}^k
{k \choose \ell}
\left\{
{ [  (u -z +1)(v-\bar z +1)]^{\ell}\over [  (u -\bar z +1)(v- z +1)]^{\alpha-m-1 +\ell}  }
-{ [  (u -\bar z +1)(v- z +1)]^\ell\over [  (u -z +1)(v-\bar z +1)]^{\alpha-m-1-\ell}  }
\right\}
\end{align}
We should further do the binomial expansion in the numerator by writing $u-z+1 = u-\bar z +1  -  (z-\bar z)$ and so on.  We therefore arrive at
\begin{align}
&I_m(\alpha) = {1\over (2\rho X)^{m+1} }
\sum_{k=0}^m{ (-1)^k\over 2^k} { m \choose k}
{1\over \alpha-m+k-1}\\
&\times
\sum_{\ell=0}^k
{k \choose \ell}
\sum_{p_1,p_2=0}^\ell
(-1)^{p_1}
{\ell \choose p_1}
{\ell \choose p_2}
\left\{
{(z-\bar z)^{p_1+p_2}\over   (u -\bar z +1)^{\alpha-m-1 +p_1}(v- z +1)^{\alpha-m-1 +p_2}  }
-(z\leftrightarrow \bar z)
\right\}\,.\notag
\end{align}
Note that $I_{2n}(\alpha)$ is an even function in $\rho$.

Let us now return to the conformal block:
\begin{align}
&G_{\Delta,0}(z,\bar z)  = \mathcal{N}{ V_{d-1}\over 2^{d-1}}(z\bar z)^{\Delta\over2}
 \int_0^\infty du \int_0^u dv { (u-v)^{d-2}
  \over (u+1)^{\Delta\over2}  (v+1)^{\Delta\over2}   u^{d-\Delta\over2} v^{d-\Delta\over2} }
  \sum_{n=0}^{d-4\over2} (-1)^n {{d-4\over2}  \choose n} I_{2n}(\Delta/2) \,.
\end{align}
Since $I_{2n}(\alpha)$ is even under the exchange $u\leftrightarrow v$, we can replace the integration  range to be $0<u<\infty$ and $0<v<\infty$ at the price of a factor $1/2$:
\begin{align}
&G_{\Delta,0}(z,\bar z)  =\mathcal{N}{ V_{d-1}\over 2^{d}}(z\bar z)^{\Delta\over2}
 \int_0^\infty du \int_0^\infty dv { (u-v)^{d-2}
  \over (u+1)^{\Delta\over2}  (v+1)^{\Delta\over2}   u^{d-\Delta\over2} v^{d-\Delta\over2} }
  \sum_{n=0}^{d-4\over2} (-1)^n {{d-4\over2}  \choose n} I_{2n}(\Delta/2) \notag\\
  &= {\cal N}{V_{d-1}\over 2^d} (z\bar z)^{\Delta\over2}\sum_{n=0}^{d-4\over2}(-1)^n {{d-4\over2} \choose n}
  {2^{2n+1}\over (z-\bar z)^{2n+1} }
  \sum_{k=0}^{2n} {(-1)^k\over 2^k}  {2n\choose k} {1\over \Delta/2-2n+k-1}
\notag\\
&\times  \sum_{\ell=0}^k { k\choose \ell}
  \sum_{p_1,p_2=0}^\ell
  (-1)^{p_1} {\ell \choose p_1}{\ell\choose p_2}
  \notag\\
  & \times\int_0^\infty du \int_0^\infty dv
  \Big[
  {(z-\bar z)^{p_1+p_2}(u-v)^{d-2-2n-1}
  \over
   (u+1)^{\Delta\over2}  (v+1)^{\Delta\over2}   u^{d-\Delta\over2} v^{d-\Delta\over2}
   (u-\bar z+1)^{\Delta/2 - 2n -1+p_1}
   (v-z +1)^{\Delta/2 -2n-1+p_2}
  }\notag\\
  &-(z\leftrightarrow \bar z)\Big] \,.
\end{align}
The $u,v$ integrals can be done by expanding  $(u-v)^{d-2-2n-1}$ and using  \eqref{2F1}. We therefore arrive at the following final expression for the even-dimensional scalar conformal block as a finite sum of the hypergeometric functions:
\begin{align}\label{evenblock}
&G_{\Delta,0}(z,\bar z) = {\cal N}
 {V_{d-1}\over 2^{d}} (z\bar z)^{\Delta\over2}
 \sum_{n=0}^{d-4\over2}
   \sum_{k=0}^{2n}
    \sum_{\ell=0}^k
      \sum_{p_1,p_2=0}^\ell
        \sum_{p=0}^{d-3-2n}
  {1\over (z-\bar z)^{2n+1} }
 (-1)^{n+k+p_1+p}  {2^{2n+1-k}\over \Delta/2-2n+k-1}
\notag\\
&\times {{d-4\over2} \choose n}
{2n\choose k}{ k\choose \ell}
 {\ell \choose p_1}{\ell\choose p_2}
{d-3-2n \choose p}
\\
  & \times {\Gamma({ -d+\Delta\over2} +1+p+p_1)\Gamma({d+\Delta\over2} -2-2n-p)
  \Gamma({d+\Delta\over2} - 2n -2+p_2-p)\Gamma({-d+\Delta\over2} +1+p)
  \over
  \Gamma(\Delta-2n-1+p_1)  \Gamma(\Delta-2n-1+p_2)}
 \notag\\
 &\times
  \Big[ (z-\bar z)^{p_1+p_2}
  \,_2F_1(\Delta/2 -2n-1+p_1, { -d+\Delta\over2}+1 +p+p_1 , \Delta-2n-1+p_1, \bar z)\notag\\
  &\times
  \,_2F_1(\Delta/2 -2n-1+p_2 , {d+\Delta\over2} -2n-2+p_2-p ,\Delta-2n-1+p_2,z)   -(z\leftrightarrow \bar z)\Big]\,.\notag
  \end{align}

 Let us fix the overall normalization constant ${\cal N}$.  To do this, we only need to compute the integral at $z=\bar z=0$:
 \begin{align}
 {1\over {\cal N}}&=V_{d}\int_0^\infty d\tau \int_0^\tau d\rho{\rho^{d-2}\over
 [(\tau+1)^2 -\rho^2 ]^{\Delta}  (\tau^2-\rho^2)^{d-\Delta\over2}
 }
 \notag\\
 &=
 {V_{d}\over2^d}
 \sum_{k=0}^{d-2}{d-2\choose k}(-1)^k
 \int_0^\infty du{1\over(u+1)^\Delta u^{{d-\Delta\over2} -(d-2)+k} }
  \int_0^\infty dv {1\over (v+1)^\Delta v^{{d-\Delta\over2} -k} }\notag\\
&  ={V_{d}\over2^d}
 \sum_{k=0}^{d-2}{d-2\choose k}(-1)^k
 {\Gamma(  {-d+\Delta\over2}  +k+1)^2
 \Gamma(  {d+\Delta\over2} -k-1)^2
 \over \Gamma(\Delta)^2} \,.
 \end{align}

In four dimensions, $d=4$, various sums in  \eqref{evenblock} collapse and we reproduce the standard scalar conformal block \cite{Dolan:2000ut,Dolan:2003hv,Dolan:2011dv}
\begin{align}
&G_{\Delta,0}(z,\bar z)= {2(\Delta-1) \over (\Delta-2)} {(z\bar z)^{\Delta\over2}\over z-\bar z}
\left[
\,_2F_1({\Delta\over2}-1 , {\Delta\over2}  , \Delta-1, z)
\,_2F_1 ({\Delta\over2}-1 , {\Delta\over2}-1 ,\Delta-1 ,\bar z)
-(z\leftrightarrow \bar z)
\right] \notag\\
=& {z\bar z\over z-\bar z}   \left[
z^{\Delta\over2}  \bar z^{ {\Delta-2\over2}}
\,_2F_1({\Delta\over2 },{\Delta\over2} , \Delta,z)
\,_2F_1({\Delta\over2}-1,{\Delta\over2}-1,\Delta-2,\bar z) -(z\leftrightarrow \bar z)
\right]\,.
\end{align}

In six dimensions we have also checked that \eqref{evenblock} agrees numerically with the known expression in \cite{Dolan:2000ut}.

\bibliography{PosCFT}
\bibliographystyle{utphys}
\end{document}